\documentclass [12pt]{article}

\usepackage{bm}
\usepackage{float}
\restylefloat{table}
\restylefloat{figure}
\usepackage{shuffle}

\usepackage{graphicx,amssymb,amsbsy,amsfonts,amssymb,amsmath}
\usepackage[usenames, dvipsnames]{xcolor}
\usepackage{calc}
\usepackage{url}

\allowdisplaybreaks
\usepackage{array}
\usepackage{bbm}
\usepackage{slashed}

\numberwithin{equation}{section}

\newif\ifdraft
\drafttrue
\newif\ifpreprint
\preprinttrue

\restylefloat{figure}
\allowdisplaybreaks

\newcommand{\req}[1]{(\ref{#1})}
\def\vev#1{\langle #1 \rangle}

\def\bet{\beta}
\def\fc#1#2{\frac{#1}{#2}}
\def\h{\frac{1}{2}}
\newcommand{\nwc}{\newcommand}
\nwc{\ba}  {\begin{array}}
\nwc{\ea}  {\end{array}}
\nwc{\bdm} {\begin{displaymath}}
\nwc{\edm} {\end{displaymath}}

\nwc{\bea} {\begin{equation}\ba{lcl}}
\nwc{\eea} {\ea\end{equation}}

\nwc{\be} {\begin{equation}}
\nwc{\ee} {\end{equation}}

\nwc{\bda} {\bdm\ba{lcl}}
\nwc{\eda} {\ea\edm}

\nwc{\bc}  {\begin{center}}
\nwc{\ec}  {\end{center}}

\nwc{\ds}  {\displaystyle}

\nwc{\nn} {\nonumber}
\nwc{\nnn} {\nonumber \vspace{.2cm} \\ }
\nwc{\ra}{\rightarrow}
\nwc{\lra}{\longrightarrow}
\def\lf{\left}\def\ri{\right}

\nwc{\p} {\partial}
\nwc{\Tr}{{\rm Tr}}
\def\IN{{\bf N}}

\def\ap{\alpha'}

\def\th{\theta}
\def\F{{_{q+1}F_q}}
\def\eps{\epsilon}
\def\z{\zeta}
\def\al{\alpha}
\def\bet{\beta}
\def\fa{{\frak a}}
\def\fb{{\frak b}}
\def\Oc{{\cal O}}

\begin{document}

\title{\textbf{Differential Equations, Associators, \\ 
and Recurrences for  Amplitudes}  \\ }

\medskip
\author{\large Georg Puhlf\"urst and  
Stephan Stieberger\\[2cm]}
\date{}

\maketitle

\vspace{-2.5cm}
\centerline{\it  Max--Planck--Institut f\"ur Physik}
\centerline{\it Werner--Heisenberg--Institut}
\centerline{\it 80805 M\"unchen, Germany}

\vspace{3cm}
\begin{abstract}
We provide new methods to straightforwardly obtain compact and analytic expressions 
for $\eps$--expansions of functions appearing in both field and string theory amplitudes. 
An algebraic method is presented to explicitly solve for recurrence relations connecting 
different $\eps$--orders of a power series solution in $\eps$  of a differential equation. 
This strategy  generalizes the usual iteration by Picard's method. 
Our tools are demonstrated for generalized hypergeometric functions.
Furthermore, we match  the $\eps$--expansion  of specific generalized hypergeometric functions with the underlying Drinfeld associator with proper Lie algebra and monodromy representations. 
We also apply our tools for computing $\epsilon$--expansions for solutions to generic first--order Fuchsian equations (Schlesinger system).
Finally, we set up our methods to systematically get compact and explicit $\alpha'$--expansions of tree--level superstring amplitudes to any order in $\ap$.

\end{abstract}
\bigskip\bigskip
\medskip
\vspace{-0.35cm}

\begin{flushright}
{\small  MPP--2015--150}
\end{flushright}


\thispagestyle{empty}

\newpage
\setcounter{tocdepth}{2}
\tableofcontents
\break

\section{Introduction}

Scattering amplitudes describe the interactions of physical states and play an important role to determine physical observables measurable at colliders. 
Moreover, the computation of perturbative scattering amplitudes is of considerable interest both in quantum field--theory and string theory to reveal their underlying  hidden symmetry structure and mathematical framework. The latter specifies the module of functions describing the amplitudes. These functions typically depend on the data of the external particles like  their momenta, masses
and scales. In field--theory the amplitudes are given by Feynman integrals over loop momenta. 
The generic functions describing Feynman integrals
are iterated integrals, elliptic functions and perhaps generalizations thereof.
On the other hand, in string theory amplitudes are given by integrals over vertex operator positions and the
moduli space of the underlying world--sheet describing the interaction process of string states.

Amplitudes in field--theory  very often can be described by certain differential equations 
or systems thereof with a given initial value problem subject to physical conditions 
\cite{Kotikov:1991pm}.
On the other hand, partial differential equations based on Lie algebras appear in the context of conformal field theory underlying the symmetries on the string world--sheet \cite{Knizhnik:1984nr,Bernard:1987df}.
The differential equations (of regular singular points or generic divisors) capture essential information on singular limits of the amplitudes. These features are directly related to properties
of the underlying string world--sheet.

In field--theory at a given loop order amplitudes depend on the parameter $\eps$ describing the dimensional regularization of the underlying Feynman integral with the space--time dimension $D=4-2\eps$ as regularization parameter. One is interested in the Laurent series
in $\eps$ of a Feynman integral.
On the other hand, perturbative string amplitudes depend on the string tension 
$\ap$ accounting for the infinitely many heavy string modes of masses $M_{\rm string}^2\sim\ap^{-1}$. In this case one is interested in their power series expansion w.r.t. $\ap$. Each term in 
this $\ap$--expansion is typically described by ${\bf Q}$--linear combinations of iterated integrals
of the same weight multiplied by  homogeneous polynomials in kinematic invariants \cite{Motivic,Broedel:2014vla}.

Although the field--theory parameter $\eps$ and string tension $\ap$ are of completely different origin the functional dependence 
of amplitudes w.r.t. to these parameters leads to qualitative similar structures in both field-- and string theory.
Therefore, in practice finding  explicit and analytic results for power series expansions w.r.t. these parameters is of fundamental importance.
Obtaining in a fully systematic way a closed and compact
 expression for a given order in $\eps$ or $\ap$, which does not rely on its lower orders to be computed in advance and which is straightforwardly applicable, is one main task 
in this work.

The differential equations underlying the amplitude integral give rise  to recurrence relations connecting different orders of $\epsilon$  of a power series Ansatz in $\eps$.
We present an algebraic method to systematically solve for such recurrence relations stemming from  differential equations with non--linear coefficients, i.e. coefficients of different orders in $\epsilon$.
This procedure gives an explicit solution to the recurrence relation providing for each order in 
$\epsilon$ a closed, compact and analytic expression.
This strategy  generalizes the usual iteration by Picard's method, which for a given order in $\eps$
requires information on all lower orders.

The novelty and importance of our recurrence method can be summarized as follows:
\begin{itemize}
\item We can straightforwardly construct power series expansions in $\epsilon$ as solutions of higher order differential equations with coefficients polynomial in $\epsilon$.

\item We obtain analytic, explicit and compact expressions at each order in epsilon
in terms of iterated integrals without the necessity to compute  their lower orders.

\item Generically  multiple polylogarithms (MPLs) appear only {\it linearly} and not any powers thereof.

\item We can find finite solutions to differential equations at their regular singular points.
 Singularities do not appear at any point in our calculations, thus no regularization is necessary.

\end{itemize}

The module of hypergeometric functions is ubiquitous both in computing tree--level string amplitudes~\cite{Oprisa:2005wu} and in the evaluation of Feynman diagrams cf. e.g. \cite{Smirnov}.
Therefore, finding an efficient procedure to determine the $\ap$-- or $\eps$--expansion of these functions is an important problem and will be addressed in this work.
Their underlying higher order differential equations lead to recurrence relations, which we explicitly solve.
Our method provides closed expressions for $\eps$--expansions of this big family of functions.

The differential equations for generalized hypergeometric functions are Fuchsian differential equations with regular singular points at $0,1$ and $\infty$. Hence, by properly
assigning the Lie algebra and monodromy representations of the Knizhnik--Zamolodchikov
(KZ) equation their underlying fundamental solutions can be matched with the $\eps$--expansion  of specific  generalized hypergeometric functions.
More concretely, for $x\!=\!1$ we obtain a connection between the $\eps$--expansion and the underlying Drinfeld associator, while for generic $x$ a relation is established between the 
$\eps$--expansion
and one fundamental solution of the underlying KZ equation. This way we obtain a very elegant way of casting the 
full $\epsilon$--expansion of a generalized hypergeometric function into the form of the 
Drinfeld associator.
We also set up and apply our tools for computing $\epsilon$--expansions of solutions to generic first order Fuchsian equations (Schlesinger system) and give explicit solutions by solving the underlying recurrences.

The organization of this work is as follows. In section 2 we introduce and discuss linear homogeneous recurrence relations and their generic solutions. The latter depends 
on the initial values only. In section 3 after introducing some basics in generalized 
hypergeometric functions we introduce integral operators for  iterated integrals 
and we set up the recurrence relations
 for a power series Ansatz in $\epsilon$ solving the underlying hypergeometric differential equation. Then, we explicitly solve for these recursion relations providing compact and closed expressions for any order in the $\eps$--expansion of generalized hypergeometric functions. 
In section 4 we establish the connection between generalized hypergeometric functions and the Drinfeld associator. We first rewrite the hypergeometric differential equation as first order Fuchsian differential equation. The latter is matched with the Lie algebra and monodromy representation of the underlying KZ equation thereby relating its fundamental solutions to 
 hypergeometric functions with specific boundary conditions at the regular singular points.
Furthermore, we also discuss $\epsilon$--expansions of solutions to generic first order Fuchsian  equations and apply our technique to solve recurrence relations for this type of differential equations.
In section 5 we apply our results to $\ap$--expansions of open superstring amplitudes
yielding explicit expressions for any order in $\ap$.
Finally, in Section 6 we present some other applications of our results. Generically, the latter are not given in terms of a minimal basis of multiple zeta values (MZVs). As a consequence, cyclic symmetry in the
kinematic invariants is non--trivially fulfilled. As a result general MZV identities are generated, e.g. the sum theorem and generalizations thereof.
The functions, which enter the all--order expansions and the MZV identities follow from a combinatorial approach.
They can be related to hypergeometric functions, binomial coefficients and the (generalized) Fibonacci numbers.

\section{Recurrence relations}

One essential step of this work are $n$--th order linear homogeneous recurrence relations
\begin{align}\label{2recrel}
w_k=\sum\limits_{\alpha=1}^{n}c_{\alpha}\ w_{k-\alpha}\ ,
\end{align}
with constant coefficients $c_i$ and initial values $w_k=\bar{w}_k$ for $k=0,1,\ldots,n-1$.
In general the coefficients do not commute: $c_ic_j\neq c_jc_i$. 

In subsection 2.1 useful definitions and notations for  non--commutative
coefficients will be introduced.
The solution, i.e. a formula that expresses all $w_k$ ($k\geq n$) in terms of the initial values 
$\bar{w}_k$ only, will be presented in subsection 2.2.

\subsection{The generalized operator product}

A simple example is the following second order recurrence relation
\begin{align}\label{2example}
 w_k=c_1w_{k-1}+c_2w_{k-2}\ ,
\end{align}
with initial values $w_0=1$ and $w_1=c_1$. For $k=5$ this gives:
\begin{align}\label{2order5}
 w_5=c_1^5+c_1^3c_2+c_1^2c_2c_1+c_1c_2c_1^2+c_2c_1^3+c_1c_2^2+c_2c_1c_2+c_2^2c_1\ .
\end{align}
It can be related to the integer partitions of $5$, which use only 2 and 1:
\begin{align}
 5=1+1+1+1+1=2+1+1+1=2+2+1\ .
\end{align}
Denoting how often 1 appears in a partition by $j_1$ and the number of 2's by $j_2$, then each of these
three partitions can be identified by a product $c_1^{j_1}c_2^{j_2}$, which for the case of interest are $c_1^5$, $c_1^3c_2$ and $c_1c_2^2$.
All these terms appear on the r.h.s. of \req{2order5}. The remaining terms in \req{2order5}  are permutations of these three products. Let us introduce the  bracket $\{c_1^{j_1},c_2^{j_2}\}$ for the sum of all 
possible distinct permutations of factors $c_i$, each one appearing $j_i$ times ($i=1,2$).
For example $j_1=1$,  $j_2=2$ yields the following sum of three products:
\begin{align}\label{2exa}
\{c_1,c_2^2\}= c_1c_2^2+c_2c_1c_2+c_2^2c_1\ .
\end{align}
With this bracket we can now  write $w_5$ more compact as:
\begin{align}\label{2solve5}
 w_5=\sum\limits_{j_1+2j_2=5}\{c_1^{j_1},c_2^{j_2}\}\ .
\end{align}
The sum over non--negative integers $j_1$ and $j_2$ represents all the integer partitions. It turns out, that the generalization
of \req{2solve5} solves the recurrence relation \req{2example}:
\begin{align}\label{2solvek}
 w_k=\sum\limits_{j_1+2j_2=k}\{c_1^{j_1},c_2^{j_2}\}\ .
\end{align}
Before the solution of the more general recurrence relation \req{2recrel} and its proof are discussed, in the following subsection a proper
definition and some basic properties of a generalized version of the bracket  $\{c_1^{j_1},c_2^{j_2}\}$ 
is given.

\subsubsection{Definition}
\def\mathbb#1{\bf #1}

The object
\begin{align}\label{2gop}
\left\{c_1^{j_1},c_2^{j_2},\ldots,c_n^{j_n}\right\}
\end{align}
is defined as the sum of all the
\begin{align}\label{2multi}
\binom{\sum\limits_{\alpha=1}^n j_{\alpha}}{j_1, j_2, \ldots, j_n} 
\end{align}
possible distinct permutations of non--commutative factors $c_i$, each one appearing $j_i$
times (with $j_i\in\IN \cup\{0\}$,\ $i=1,2,...,n$). The non--negative integers $j_i$ are referred to  indices
and the factors $c_i$ to arguments of the generalized operator product \req{2gop}. For example:
\begin{align}\label{2ex0}
\left\{c_1,c_2\right\}&=c_1 c_2+c_2 c_1\ ,\\
\{c_1^{2},c_2,c_3\}&=c_1^2c_2c_3+c_1c_2c_1c_3+c_1c_2c_3c_1+c_2c_1^2c_3+c_2c_1c_3c_1+c_2c_3c_1^2\nonumber\\
&+c_1^2c_3c_2+c_1c_3c_1c_2+c_1c_3c_2c_1+c_3c_1^2c_2+c_3c_1c_2c_1+c_3c_2c_1^2\ .\label{2ex}
\end{align}
For the case of two arguments the object \req{2gop} was used in \cite{Rumaenien} to solve a second order recurrence relation with
non--commutative coefficients. There is a useful recursive definition\footnote{The same formula with $c_\alpha$ to the right of the generalized operator product
holds as well.} for \req{2gop} as:
\begin{align}\label{2def}
 \left\{c_1^{j_1},c_2^{j_2},\ldots,c_n^{j_n}\right\}=\sum\limits_{\substack{\alpha=1\\j_{\alpha}\neq0}}^{n}c_\alpha
 \left\{c_1^{j_1},c_2^{j_2},\ldots,c_{\alpha}^{j_{\alpha}-1},\ldots,c_n^{j_n}\right\}+\prod\limits_{\beta=1}^{n}\delta_{0 j_\beta}\ .
\end{align}
The product of Kronecker deltas gives a non--vanishing contribution in case all indices $j_1,\ldots,j_n$ are zero. Without this product an inconsistency would occur: 
for $j_1=\ldots=j_n=0$ the sum on the r.h.s. of \req{2def} becomes zero, while the l.h.s. should be one.
Furthermore we have\footnote{Since in \req{2gop} the order of the arguments  is irrelevant, 
we write identities, such as the first line of \req{2defadd}, without loss of generality for the first arguments
only.}:
\begin{align}
\begin{split}\label{2defadd}
 \left\{c_1^{0},c_2^{j_2},\ldots,c_n^{j_n}\right\} &=\left\{c_2^{j_2},\ldots,c_n^{j_n}\right\},\\
 \{c^j\}&=c^j\ ,\\
\text{and especially }\{c^0\}&=1\ .
\end{split}
\end{align}
The definition \req{2def} together with eqs. \req{2defadd} allows to decrease step by step the
indices and the number of arguments. This way \req{2gop} can be written in terms of non--commutative
products. For instance applying \req{2def} twice to all generalized operator products on the l.h.s. of \req{2ex} yields:
\begin{align}
 \{c_1^{2},c_2,c_3\}&=c_1\{c_1,c_2,c_3\}+c_2\{c_1^2,c_3\}+c_3\{c_1^2,c_2\}\nonumber\\
 &=c_1^2\{c_2,c_3\}+c_1c_2\{c_1,c_3\}+c_1c_3\{c_1,c_2\}+c_2c_1\{c_1,c_3\}+c_2c_3c_1^2\nonumber\\
 &+c_3c_1\{c_1,c_2\}+c_3c_2c_1^2\ .
\end{align}
Applying \req{2def} once again or using \req{2ex0} gives the r.h.s. of \req{2ex}. The definition of \req{2gop} can be extended to negative
integer indices as:
\begin{align}\label{2negative}
 \{c_1^{j_1},c_2^{j_2},\ldots,c_n^{j_n}\}=0\ , \ \ j_1<0\ .
\end{align}
This extension turns out to be useful, when the indices of generalized operator products include summation indices. It allows to reduce the
conditions for the summation regions. E.g. the condition $j_{\alpha}\neq0$ in the sum on the r.h.s. of \req{2def}
can be dropped with this extension.

To prove that \req{2def} gives indeed all distinct permutations, it is sufficient to show that:
\begin{enumerate}
	\item the number of terms equals \req{2multi},
	\item there are no identical terms
	\item and all terms contain each non--commutative factor $c_i$ exactly $j_i$ times.
\end{enumerate}
The third point is obviously fulfilled. The second one is also quite obvious, since every summand of the sum
in \req{2def} starts with a different factor. Using the definition again yields, that all terms coming from the
same summand and therefore have the same first factor, have a different second factor and so on. The first
point is also true, since the number of terms on the r.h.s. of \req{2def} is
\begin{align}
\sum\limits_{\alpha=1}^{n}\binom{-1+\sum_{\beta=1}^{n}j_{\beta}}{j_1, j_2, \ldots,j_{\alpha}-1,\ldots, j_n}\ .
\end{align}
The above expression can easily  be transformed into \req{2multi} using the definition of the multinomial coefficient in terms of
factorials.

The generalized operator product \req{2gop} is closely related to the shuffle product:
\begin{align}\label{2shuffle}
\left\{c_1^{j_1},c_2^{j_2},\ldots,c_n^{j_n}\right\}=\underbrace{c_1\ldots c_1}_{j_1}\shuffle
\underbrace{c_2\ldots c_2}_{j_2}\shuffle\ldots\shuffle\underbrace{c_n\ldots c_n}_{j_n}\ .
\end{align}
However, the notation on the l.h.s. is more compact, in particular for the applications in the following sections.

\subsubsection{Basic properties}

With the definition \req{2def} the following basic properties can easily be proven.
Factors $a$ which commute with all arguments, i.e. $c_{i} a=a c_{i}$ can be factorized:
\begin{align}\label{2commute}
\left\{(a c_1)^{j_1},c_2^{j_2},\ldots,c_n^{j_n}\right\}=a^{j_1}\left\{c_1^{j_1},c_2^{j_2},\ldots,c_n^{j_n}\right\}\ .
\end{align}
Identical arguments can be combined as:
\begin{align}\label{2identical}
\left\{c_1^{j_1},c_1^{j_2},c_3^{j_3},\ldots,c_n^{j_n}\right\}=\left\{c_1^{j_1+j_2},c_3^{j_3},\ldots,c_n^{j_n}\right\}
\binom{j_1+j_2}{j_1}\ .
\end{align}
The binomial coefficient\footnote{In order not to conflict with \req{2negative}, we use $\binom{j_1+j_2}{j_1}=0$ for 
$(j_1<0)\lor (j_2<0)$.}
ensures, that the number of terms is the same on both sides. While sums can be treated according to
\begin{align}\label{2sum}
\{c_1+c_2,c_3^{j_3},\ldots,c_n^{j_n}\}=\{c_1,c_3^{j_3},\ldots,c_n^{j_n}\}+\{c_2,c_3^{j_3},\ldots,c_n^{j_n}\}\ ,
\end{align}
one has to be careful, when such arguments appear with exponents greater than one. Before these
cases are discussed, note that the generalized operator product can be used for a generalized version of the binomial theorem,
which is also valid for non--commutative quantities $c_1$ and $c_2$:
\begin{align}\label{2binth}
 (c_1+c_2)^j=\sum\limits_{j_1+j_2=j}^{}\left\{c_1^{j_1},c_2^{j_2}\right\}\ ,
\end{align}
with non--negative integers $j$, $j_1$ and $j_2$.
Applying naively  this relation to arguments of \req{2gop} leads to inconsistencies. E.g.:
\begin{align}
\left\{(c_1+c_2)^2,c_3\right\}&\stackrel{?}{=}\left\{\sum\limits_{\alpha=0}^{2}\left\{c_1^{2-\alpha},c_2^{\alpha}
\right\},c_3\right\}\nonumber\\
&=\left\{c_1^2,c_3\right\}+\left\{c_2^2,c_3\right\}+\left\{c_1c_2,c_3\right\}+\left\{c_2c_1,c_3\right\}\ .\label{2bin1}
\end{align}
Eq. \req{2sum} is used in the last step. Using instead the definition \req{2def}, gives:
\begin{align}
\left\{(c_1+c_2)^2,c_3\right\}&=(c_1+c_2)(c_1+c_2)c_3+(c_1+c_2)c_3(c_1+c_2)\nonumber\\
&+c_3(c_1+c_2)(c_1+c_2)\ .\label{2bin2}
\end{align}
Comparing \req{2bin1} and \req{2bin2} shows that $c_1c_3c_2+c_2c_3c_1$ is missing in \req{2bin1}. To avoid this problem, 
one simply has to ignore the inner curly brackets when applying \req{2binth} to arguments of \req{2gop}. Hence, the following relation is consistent:
\begin{align}
\{(c_1+c_2)^j,c_3^{j_3},\ldots,c_n^{j_n}\}=\sum\limits_{j_1+j_2=j}^{}\{c_1^{j_1},c_2^{j_2},c_3^{j_3},\ldots,c_n^{j_n}\}\ .
\end{align} 
This can be easily generalized to multinomials:
\begin{align}\label{2multigop}
\{(c_1+c_2+\cdots+c_n)^j,c_{n+1}^{j_{n+1}},\ldots\}
=\sum\limits_{j_1+j_2+...+j_n=j}^{}\{c_1^{j_1},c_2^{j_2},\ldots,c_n^{j_n},c_{n+1}^{j_{n+1}},\ldots\}\ .
\end{align}
Besides these basic properties, there are more intricate identities satisfied by generalized operator products.
They will be  discussed in section 6.

\subsection{Solution}

The $n$--th order linear homogeneous recurrence relation \req{2recrel} is solved by:
\begin{align}\label{2solution}
 w_k=\sum\limits_{\alpha=0}^{n-1}\sum\limits_{j_1+2j_2+\cdots+nj_n=k-n-\alpha}\left\{c_1^{j_1},c_2^{j_2},\ldots,c_n^{j_n}\right\}
 \sum\limits_{\beta=\alpha+1}^nc_{\beta}\bar{w}_{n-\beta+\alpha}\ ,\ \ k\geq n\ .
\end{align}
Note, that the r.h.s. of \req{2solution} only contains initial values $\bar{w}_l$. 
The second sum is over $n$--tuples of non--negative integers $j_1,\ldots,j_n$ which solve the 
equation:
\begin{align}
 \sum\limits_{\gamma=1}^n\gamma j_{\gamma}=k-n-\alpha\ .
\end{align}
In the following we shall prove by  induction that \req{2solution} solves \req{2recrel}. The regions $2n> k \geq n$ and 
$k\geq 2n$ are discussed separately. 
The first region is required to prove the base case $k=2n$ of the induction for 
$k\geq 2n$. The induction for  $ 2n> k \geq n$ has the two base cases $k=n$ and $k=n+1$. In the first 
case $k=n$, the only  non--zero 
contribution comes from $\alpha=j_1=j_2=\ldots=j_n=0$:
\begin{align}\label{2basecase1}
 w_n=\sum\limits_{\beta=1}^nc_{\beta}\ \bar{w}_{n-\beta}\ .
\end{align}
The second case $k=n+1$ has two parts. One
with $\alpha=1,~j_1=j_2=\ldots=j_n=0$ and the other with $j_1=1,~\alpha=j_2=\ldots=j_n=0$:
\begin{align}
 w_{n+1}&=c_1\sum\limits_{\gamma=1}^{n}c_{\gamma}\bar{w}_{n-\gamma}+\sum\limits_{\beta=2}^nc_{\beta}\bar{w}_{n+1-\beta}\nonumber\\
 &=c_1w_n+\sum\limits_{\beta=2}^nc_{\beta}\bar{w}_{n+1-\beta}\nonumber\\
 &=\sum\limits_{\beta=1}^nc_{\beta}w_{n+1-\beta}\ .\label{2basecase2}
\end{align}
Eq. \req{2basecase1} is used in the second line of \req{2basecase2}. 
Both cases \req{2basecase1} and \req{2basecase2} are in agreement with the eq.  \req{2recrel}
and the initial conditions. The recursive definition \req{2def} of the generalized operator product is particularly useful for the inductive step:
\begin{align}
 w_k&=\sum\limits_{\alpha=0}^{n-1}\sum\limits_{j_1+2j_2+\cdots+nj_n=k-n-\alpha}
 \sum\limits_{\gamma=1}^{n}c_{\gamma}\left\{c_1^{j_1},\ldots,c_{\gamma}^{j_{\gamma}-1},
 \ldots,c_n^{j_n}\right\} \sum\limits_{\beta=\alpha+1}^nc_{\beta}\bar{w}_{n-\beta+\alpha}\nonumber\\
 &+\sum\limits_{\alpha=0}^{n-1}\sum\limits_{j_1+2j_2+\cdots+nj_n=k-n-\alpha}\prod\limits_{\gamma=1}^n
 \delta_{0j_\gamma}\sum\limits_{\beta=\alpha+1}^nc_{\beta}\bar{w}_{n-\beta+\alpha}\ .\label{2indstep}
\end{align}
Shifting $j_{\gamma}\rightarrow j_{\gamma}+1$ on the r.h.s. of the first line gives:
\begin{align}
 &\sum\limits_{\gamma=1}^{n}c_\gamma\sum\limits_{\alpha=0}^{n-1}\sum\limits_{j_1+2j_2+\cdots+nj_n=k-n-\alpha-\gamma}
 \left\{c_1^{j_1},\ldots,c_n^{j_n}\right\}
 \sum\limits_{\beta=\alpha+1}^nc_{\beta}\bar{w}_{n-\beta+\alpha}\nonumber\\
 &=\sum\limits_{\gamma=1}^{n}c_\gamma \cdot\begin{cases}
     w_{k-\gamma} & \text{for } k-\gamma\geq n \\[3mm]
     0 & \text{else.} 
   \end{cases}\nonumber\\
 &=\sum\limits_{\gamma=1}^{\min(n,k-n)}c_\gamma w_{k-\gamma}\ .\label{2line1}
\end{align}
The second line of \req{2indstep} is non--zero only if there is a solution for $k-n-\alpha=0$. Inserting the
upper bound $\alpha\leq n-1$ of the first sum, gives the condition $k<2n$.
Combining both lines of \req{2indstep} for this region yields
\begin{align}\label{2line12}
 w_k=\sum\limits_{\gamma=1}^{k-n}c_{\gamma}w_{k-\gamma}+\sum\limits_{\beta=k-n+1}^{n}c_\beta\bar{w}_{k-\beta}\ ,
\end{align}
which is identical to \req{2recrel}. In the region $k\geq2n$ the second line of \req{2indstep} becomes zero
and the upper bound in \req{2line1} is $n$, since $k-n\geq n$. This also results in \req{2recrel}.

Finally, it is easy to prove that the inhomogeneous recurrence relation ($k\geq n$)
\begin{align}
 w_k^{\text{(inh)}}=\sum\limits_{\alpha=1}^nc_{\alpha}w_{k-\alpha}^{\text{(inh)}}+d_k
\end{align}
is solved by
\begin{align}\label{2inhsol}
 w_k^{\text{(inh)}}=w_k+\sum\limits_{\alpha=n}^k\ \sum\limits_{j_1+2j_2+\ldots+nj_n=k-\alpha}\{c_1^{j_1},c_2^{j_2},\ldots,c_n^{j_n}\}\ d_\al\ ,
\end{align}
where $w_k$ is the solution of the corresponding homogeneous recurrence relation and $d_k$ is a 
inhomogeneity, that depends on $k$.

\section{Expansion of generalized hypergeometric functions}

The generalized Gauss function or generalized hypergeometric function ${}_pF_{p-1}$
is given by the power series \cite{Slater}
\be\label{Hyper}
{}_pF_{p-1}(\vec{a};\vec{b};z)\equiv {}_pF_{p-1}\lf[{a_1,\ldots,a_p\atop b_1,\ldots,b_{p-1}};z\ri]=\sum_{m=0}^\infty\ 
\fc{\prod\limits_{i=1}^{p}(a_i)^m}{\prod\limits_{j=1}^{p-1}(b_j)^m}\ \fc{z^m}{m!}\ \ \ ,\ \ \ p\geq1\ ,
\ee
with some parameters $a_i,b_j\in{\bf R}$ and with the Pochhammer (rising factorial) symbol:
$$(a)^n=\fc{\Gamma(a+n)}{\Gamma(a)}=a\ (a+1)\ldots(a+n-1)\ .$$
The series \req{Hyper} 
enjoys the $p$--th order differential equation (with $y={}_pF_{p-1})$
\be\label{3hgdgl}
\theta\ (\theta+b_1-1)\ (\theta+b_2-1)\ldots(\theta+b_{p-1}-1)y-z\ 
(\theta+a_1)\ (\theta+a_2)\ldots(\theta+a_p) y=0\ ,
\ee
with the differential operator:
\be\label{differential}
\theta=z\ \fc{d}{dz}\ .
\ee
Furthermore, the function \req{Hyper} satisfies:
\begin{align}
 \begin{split}
\label{3hgrel}
 \left (\theta + a_i \right ){}_pF_{p-1}(\vec{a};\vec{b};z) &= a_i\;{}_pF_{p-1}\left[ 
 \begin{matrix}
 a_1,\dots,a_i+1,\dots,a_p \\
 b_1,\dots,b_{p-1}
 \end{matrix} ;z\right] ,\\[2mm]
\left (\theta+ b_j - 1 \right ){}_pF_{p-1}(\vec{a};\vec{b};z) &= (b_j - 1)\; {}_pF_{p-1}
\left[
 \begin{matrix}
 a_1,\dots,a_p \\
 b_1,\dots,b_j-1,\dots,b_{p-1} 
 \end{matrix} ;z \right],\\
\fc{d}{dz}\; {}_pF_{p-1}(\vec{a};\vec{b};z) &= \frac{\prod\limits_{i=1}^p a_i}{\prod\limits_{j=1}^{p-1} b_j}
\; {}_pF_{p-1}\left[ 
 \begin{matrix}
 a_1+1,\dots,a_p+1 \\ b_1+1,\dots,b_{p-1}+1 
 \end{matrix} ;z \right]\ .
 \end{split}
\end{align}
The $\eps$-- or $\alpha'$--expansion of a given generalized hypergeometric function
\begin{align}\label{3hg}
 {}_pF_{p-1}\left[
\begin{matrix} 
m_1+\alpha'a_{1},\ldots,m_p+\alpha'a_{p} \\ 
n_1+\alpha'b_{1},\ldots,n_{p-1}+\alpha'b_{p-1} 
\end{matrix};z\right],~~m_i,n_j\in{\bf Z}\ 
\end{align}
is expressible in terms of generalized polylogarithms with coefficients, that are ratios of polynomials. 
For each $p$ it is sufficient to derive the expansion for one set of integers $\vec{m}$ and $\vec{n}$ only.
By using eqs. \req{3hgrel} any function $_p F_{p-1}(\vec{a};\vec{b};z)$ can be expressed as
a linear combination of other functions $_p F_{p-1}(\vec{m}+\vec{a};\vec{n}+\vec{b};z)$ with parameters that differ from the original ones by an integer shift
 and the first $p-1$ derivatives thereof \cite{Takayama}. 

In this section we shall  present and apply our new technique to solve for recurrences to compute the $\alpha'$--expansion
of \req{3hg}. 
To warm up and for completeness we begin with the case of the $p=2$ hypergeometric function:
\begin{align}\label{32f1}
{}_2F_1\left[
\begin{matrix} 
-\alpha'a,~\alpha'b\\ 
1+\alpha'b \end{matrix}
;z\right]=\sum\limits_{k=0}^{\infty}u_k(z)(\alpha')^k\ .
\end{align}
However, we shall mainly be concerned with the  case $p=3$. For the latter the expansion takes the form: 
\begin{align}\label{33f2}
 {}_3F_2\left[
\begin{matrix} 
\alpha'a_{1},~\alpha'a_{2},~\alpha'a_{3} \\ 
1+\alpha'b_{1},~1+\alpha'b_{2} \end{matrix}
;z\right]=\sum\limits_{k=0}^{\infty}v_k(z)\ (\alpha')^k\ .
\end{align}
Finally, the general case $p$ will be elaborated at the end of this section.

To obtain all--order expressions for $v_k(z)$ and $u_k(z)$ the differential equations in $z$ for the
functions \req{3hg} are used. Combining eqs. \req{3hgrel} gives the differential equation \req{3hgdgl}.
Inserting the expansions \req{32f1} and \req{33f2} into \req{3hgdgl} leads to the following recursive differential equations for $u_k(z)$ and $v_k(z)$,
respectively:
\begin{align}
0&=(z-1)\ \theta^2u_k(z)+\left[\ z\ (b-a)-b\ \right]\ \theta u_{k-1}(z)-z\ ab\ u_{k-2}(z)\ ,\\[2mm]
0&=(z-1)\ \theta^3v_k(z)+\left[\ z(a_1+a_2+a_3)-b_1-b_2\ \right]\ \theta^2v_{k-1}(z)\nonumber\\
  &+\left[\ z\ (a_1a_2+a_2a_3+a_3a_1)-b_1b_2\ \right]\theta v_{k-2}(z)+z\ a_1a_2a_3\ v_{k-3}(z)\ .
\end{align}
To solve these equations we transform them into recurrence relations. This is achieved by replacing
the derivatives and integrations in the formal solution of these differential equations by differential
and integral operators, respectively. In addition, the boundary conditions have to be respected.
The recurrence relations have the form \req{2recrel}, where the non--commutative coefficients $c_i$ represent differential and integral operators.

Applying recurrences to expand hypergeometric functions was
first proposed in the field--theory context in \cite{Kalmykov:2006hu,Kalmykov}. In \cite{Boels} the recurrence relations have been used to calculate the expansions \req{33f2} and \req{32f1} order by order. However, with the general solution
\req{2solution} the all--order expansions can now  systematically be constructed and straightforwardly be given in closed form.

In the next subsection we shall discuss our notation. Then, in the subsequent subsections we shall present and solve the recurrence relations for $u_k(z)$, $v_k(z)$ and for general $p$.

\subsection{Integral operators, multiple polylogarithms and multiple zeta values}

We introduce the integral operators
\begin{align}
\begin{split}
I(1)\ f(z)&=\int_0^z \frac{dt}{1-t}\ f(t)\ ,\\[3mm]
\text{and }~~I(0)\ f(z)&=\int_0^z \frac{dt}{t}\ f(t)\label{3intop}
\end{split}
\end{align}
to define the following multiple integral operator
\begin{align}
I(\underbrace{0,\ldots,0}_{n_1-1},1,\ldots,\underbrace{0,\ldots,0}_{n_d-1},1):=I(0)^{n_1-1}I(1)I(0)^{n_2-1}I(1)\ldots I(0)^{n_d-1}I(1)\ ,\label{311}
\end{align}
with the multiple index $\vec{n}=(n_1,n_2,\ldots,n_d)$.
Acting with the operator \req{311} on\footnote{In the sequel we will not write
this $1$.} the constant function $1$ yields MPLs:
\be\label{3MPL}
I(\underbrace{0,\ldots,0}_{n_1-1},1,\ldots,\underbrace{0,\ldots,0}_{n_d-1},1)\ 1=
{\cal L}i_{\vec n}(z)\ \equiv {\cal L}i_{n_1,\ldots,n_d}(z,\underbrace{1,\ldots,1}_{d-1})=\sum_{0<k_d<\ldots<k_1}\fc{z^{k_1}}{k_1^{n_1}\cdot\ldots\cdot
k_d^{n_d}}\ .
\ee
For $z=1$ the MPLs become MZVs
\begin{align}\label{3mzv}
 \zeta(\vec{n})={\cal L}i_{\vec{n}}(1)=\left.I(\underbrace{0,\ldots,0}_{n_1-1},1,\ldots,\underbrace{0,\ldots,0}_{n_d-1},1)\ 1\right|_{z=1}\ ,
\end{align}
with the following definition of MZVs:
\be
\zeta(\vec{n})\equiv \zeta(n_1,\ldots,n_d)=\sum_{0<k_d<\ldots<k_1}k_1^{-n_1}\cdot\ldots\cdot k_d^{-n_d}\ .
\ee
Both for MPLs and for MZVs the weight $w$ is defined as the sum of all indices:
\begin{align}\label{3weight}
 w=n_1+n_2+\ldots+n_d\ .
\end{align}
Using the representations in terms of integral operators the weight is equivalent to the total number of integral
operators. The depth $d$ is defined as the number of indices, i.e.:
\begin{align}\label{3depth}
 d=\text{dim}(\vec{n})\ .
\end{align}
In terms of integral operators, this is the number of operators $I(1)$. For example
\begin{align}
I(0)I(1)I(1)={\cal L}i_{2,1}(z)\stackrel{z=1}{\rightarrow}\zeta(2,1)
\end{align}
are weight $w=3$ and depth $d=2$ MPLs and MZVs, respectively.
It is useful to combine both products of integral operators \req{3intop} and the differential operator $\theta$ defined in \req{differential} into the shorter form:
\begin{align}
I(p_1)I(p_2)\ldots I(p_n)=I(p_1,p_2,\ldots,p_n)\ , \ \  p_i\in\{0,1,\theta\}\ , \ \ I(\theta)\equiv\theta\ .
\end{align}
Up to boundary values the differential operator $\theta$ is the inverse of $I(0)$:
\begin{align}\label{3inverse}
 I(\theta,0)f(z)=I(0,\theta)f(z)=f(z)\ .
\end{align}

The results of the following sections will often contain sums of the form:
\begin{align}\label{3summzv}
 \sum\limits_{\cdots}\zeta(\vec{n})\ .
\end{align}
Above, the dots may represent conditions for the weight $w$, the depth $d$, specific indices $n_i$ or other
quantities referring to the MZVs $\zeta(\vec{n})$ in the sum. The sum runs over all sets of positive integers $\vec{n}$, that
satisfy these requirements. It is understood that $n_1>1$. For example the sum of all MZVs of weight $w=5$ and depth $d=2$ 
is represented as:
\begin{align}
 \sum\limits_{\substack{w=5\\d=2}}\zeta(\vec{n})=\zeta(4,1)+\zeta(3,2)+\zeta(2,3)\ .
\end{align}
Further conditions could include the first index $n_1$ or the number of indices $d_1$, which equal one:
\begin{align}
 \sum\limits_{\substack{w=5;~d=2\\n_1\geq3;~d_1=0}}\zeta(\vec{n})=\zeta(3,2)\ .
\end{align}
Obviously $d_1=0$ is equivalent to $n_i\geq2$ ($i=1,\ldots,d$). More general $d_i$ is defined as the number of indices, which
equal $i$, so that $d=\sum_i d_i$.
In some cases a weighting $\omega$ is included, which can depend on the indices $\vec{n}$ or other
quantities. For example the following sum has $\omega=d_1$:
\begin{align}
 \sum\limits_{\substack{w=6\\d=3}}\zeta(\vec{n})d_1=2\zeta(4,1,1)+\zeta(3,2,1)+\zeta(3,1,2)+\zeta(2,3,1)+\zeta(2,1,3)\ .
\end{align}
In our notation the well known sum theorem \cite{Granville} reads 
\begin{align}\label{3sumth}
 \sum\limits_{\substack{w=a\\d=b}}\zeta(\vec{n})=\zeta(a)\ ,
\end{align}
which means, that for given weight and depth the sum of all MZVs equals the single zeta value (depth one
MZV) of that weight (independent of the given depth). The same notation is used for MPLs.

Some sums use multiple indices $\vec{\alpha}=(\alpha_1,\alpha_2,\ldots,\alpha_a)$:
\begin{align}
 \sum\limits_{\vec{\alpha}}f(\alpha_1,\alpha_2,\ldots,\alpha_a)=
 \sum\limits_{\alpha_1}\sum\limits_{\alpha_2}\ldots
 \sum\limits_{\alpha_a}f(\alpha_1,\alpha_2,\ldots,\alpha_a)\ .
\end{align}
In this context it is necessary to distinguish between functions $f(\alpha_1,\alpha_2,\ldots,\alpha_a)$, which depend on
elements of multi--indices, and functions $g(\vec{\alpha})$, which have multi--indices as arguments:
\begin{align}\label{notation}
 \sum\limits_{\vec{\alpha}}g(\vec{\alpha})=\sum\limits_{\alpha_1}g(\alpha_1)
 \sum\limits_{\alpha_2}g(\alpha_2)\ldots\sum\limits_{\alpha_a}g(\alpha_a)\ .
\end{align}
The latter only occur in combination with multi--index sums and they represent functions $g(\alpha_i)$, which have only one element
as argument. The summation regions for the elements $\al_i$  follow from the summation region of the 
vector $\vec\al$  in a natural way.
All indices of these sums are non--negative integers ($\alpha_i\geq0$). The sum of all elements of a multi--index $\vec{\alpha}$
is written as $|\vec{\alpha}|=\alpha_1+\alpha_2+\ldots+\alpha_d$.
This notation is especially used for weightings in sums of MZVs \req{3summzv}. It is understood, that the number of
elements of the multi--indices equals the depth of the corresponding MZVs.

\subsection{Second order recurrence relation}

As a warm up in this subsection the solution of the recurrence relation for the 
coefficients $u_k(z)$ of the expansion \req{32f1} is calculated. This result  is already known \cite{D}.
The relation reads
\begin{align}\label{32ndrecrel}
u_k(z)=c_1 u_{k-1}(z)+c_2 u_{k-2}(z)\ ,
\end{align}
with
\begin{align}\label{32ndcoeff}
\begin{split}
c_1&=-a I(0,1,\theta)-b I(0)\ ,\\
c_2&=-a b I(0,1)\ ,
\end{split}
\end{align}
and the initial values $u_0=1$ and $u_1=0$. According to \req{2solution} the solution is:
\begin{align}\label{32ndsol}
 u_k(z)=-ab\sum\limits_{j_1+2j_2=k-2}\left\{(-aI(0,1,\theta)-bI(0))^{j_1},(-ab I(0,1))^{j_2}\right\}I(0,1)\ .
\end{align}
Eq. \req{3MPL} implies, that the final expression only contains the integral operators $I(0)$ and $I(1)$. 
Therefore the first step should be to eliminate the differential operator $I(\theta)$. This is
achieved by the relation 
\req{3inverse} and the following identity:
\begin{align}\label{3rmvtheta}
\left\{I(0,\vec{p}_1,\theta)^{j_1},I(0,\vec{p}_2,\theta)^{j_2},\ldots,I(0,\vec{p}_n,\theta)^{j_n}\right\}
=I(0)\left\{I(\vec{p}_1)^{j_1},I(\vec{p}_2)^{j_2},\ldots,I(\vec{p}_n)^{j_n}\right\}I(\theta)\ .
\end{align}
The removal of $I(\theta)$ works because every argument starts with an $I(0)$ and ends with an  $I(\theta)$.
The vectors $\vec{p}_i$ are arbitrary sequences of the elements $\{0,1,\theta\}$. With the relations \req{2commute}, 
\req{2multigop} and \req{3rmvtheta} the result \req{32ndsol} can be transformed to:
\begin{align}\label{32ndstep1}
 u_k(z)=\sum\limits_{\alpha=1}^{k-1}(-1)^{k+1}a^{k-\alpha}b^{\alpha}\sum\limits_{\beta}(-1)^{\beta}I(0)
 \left\{I(1)^{k-\alpha-1-\beta},I(0)^{\alpha-1-\beta},I(1,0)^{\beta}\right\}I(1)\ .
\end{align}
An identity, which will be discussed in section 6, allows to simplify the generalized operator product and the sum over $\beta$ to arrive at:
\begin{align}\label{32ndstep2}
 u_k(z)=\sum\limits_{\alpha=1}^{k-1}(-1)^{k+1}a^{k-\alpha}b^{\alpha}I(0)^{\alpha}I(1)^{k-\alpha}=
 \sum\limits_{\alpha=1}^{k-1}(-1)^{k+1}a^{k-\alpha}b^{\alpha}\;{\cal L}i_{(\alpha+1,\{1\}^{k-\alpha-1})}(z)\ .
\end{align}
In the final step eq. \req{3MPL} has been used to express the result in terms of MPLs. Therefore, the hypergeometric
function \req{32f1} can be written as:
\begin{align}\label{32ndresult}
{}_2F_1\left[
\begin{matrix} 
-\alpha'a,~\alpha'b\\ 
1+\alpha'b \end{matrix}
;z\right]=1-\sum\limits_{k=2}^{\infty}(-\alpha')^k\sum\limits_{\alpha=1}^{k-1}a^{k-\alpha}b^{\alpha}
\;{\cal L}i_{(\alpha+1,\{1\}^{k-\alpha-1})}(z)\ .
\end{align}
Of particular interest is the case $z=1$, since the resulting object arises in the 
four--point open string amplitude \cite{Motivic}. This will be discussed in subsection 5.1 and further discussions will follow there. 

\subsection{Third order recurrence relation}

The recurrence relation for the coefficients $v_k(z)$ of the series \req{33f2}, which has been derived in   \cite{Boels}, reads
\begin{align}\label{33rdrecrel}
v_k(z)&=c_1 v_{k-1}(z)+c_2 v_{k-2}(z)+c_3 v_{k-3}(z)\ ,
\end{align}
with
\begin{align}\label{3coeff}
\begin{split}
c_1&=\Delta_1 I(0,0,1,\theta,\theta)-Q_1I(0)\ ,\\
c_2&=\Delta_2 I(0,0,1,\theta)-Q_2I(0,0)\ ,\\
c_3&=\Delta_3 I(0,0,1)\ ,
\end{split}
\end{align}
and $v_0=1;~v_1,v_2=0$ as initial values.
Furthermore, we define:
\begin{align}\label{3deltaq}
\begin{split}
\Delta_1&=a_1+a_2+a_3-b_1-b_2\ ,\\
\Delta_2&=a_1 a_2+a_2 a_3+a_3 a_1-b_1 b_2\ ,\\
\Delta_3&=a_1 a_2 a_3\ ,\\
Q_1&=b_1+b_2\ ,\\
Q_2&=b_1 b_2\ .
\end{split}
\end{align}
According to \req{2solution} the solution of \req{33rdrecrel} is:
\begin{align}\label{3sol}
 v_k(z)=\sum\limits_{j_1+2j_2+3j_3=k-3}\left\{c_1^{j_1},c_2^{j_2},c_3^{j_3}\right\}c_3\ .
\end{align}
Applying the definitions \req{3coeff} as well as the identities \req{2commute}, \req{2multigop}, \req{3inverse} and 
\req{3rmvtheta} gives:
\begin{multline}\label{3result}
 v_k(z)=\sum\limits_{m_1+l_1+2(l_2+m_2)+3m_3=k-3}(-1)^{l_1+l_2}\Delta_1^{m_1}\Delta_2^{m_2}\Delta_3^{m_3+1}Q_1^{l_1}Q_2^{l_2}\\
 \times I(0,0)\left\{I(0)^{l_1},I(0,0)^{l_2},I(1)^{m_1},I(1,0)^{m_2},I(1,0,0)^{m_3}\right\} I(1)\ .
\end{multline}
This compact expression allows to extract any order of the expansion of the hypergeometric function \req{33f2}
without having to calculate lower orders. For Mathematica implementations the routine 'DistinctPermutations'
is useful to evaluate the generalized operator products. With this the iterated integrals can directly be written in terms of  MPLs or MZVs in the  case $z=1$. All functions are finite. As a consequence no singularities occur.

As already mentioned, expansions of other hypergeometric functions ${}_3F_2(\vec{m}+\vec{a};\vec{n}+\vec{b};z)$, can be 
obtained from the result \req{3result} with the relations \req{3hgrel}. Two such functions, which enter
the five--point open superstring amplitude, are the topic of subsection 5.2.

As in the second order case there is an identity, which allows to remove the generalized operator product yielding a
representation in terms of MPLs. How to get there will be discussed in section 6.

\subsection[Recurrence relation at $p$--th order]{Recurrence relation at $\bm{p}$--th order}

Finally, let us discuss the case of general $p$, which assumes the expansion:
\be\label{32fp}
{}_pF_{p-1}\left[
\begin{matrix} 
\alpha'a_{1},\ldots\alpha'a_{p} \\ 
1+\alpha'b_{1},\ldots,1+\alpha'b_{p-1} 
\end{matrix};z\right]=\sum\limits_{k=0}^{\infty}w^p_k(z)\ (\alpha')^k\ .
\ee
From the differential equation \req{3hgdgl} for the coefficients in \req{32fp} we obtain the following recurrence relation  \cite{Kalmykov}:
\begin{align}
 w_k^p(z)=\sum\limits_{\alpha=1}^pc_\alpha^p\ w_{k-\alpha}^p(z)\ \ \ ,\ \ \ k\geq p\ . \label{REKn}
\end{align}
The initial conditions are $w_0^p(z)=1$ and $w_k^p(z)=0$ for $0<k<p$. 
In \req{REKn} the coefficients are
\begin{align}
 c_\alpha^p=\Delta_\alpha^p\ I(0)^{p-1}I(1)\theta^{p-\alpha}
 -Q_\alpha^p\ I(0)^\alpha\ ,
\end{align}
with
\be 
\Delta_\al^p=P^p_\al-Q_\al^p\ \ \ ,\ \ \ \al=1,\ldots,p-1\ \ \ ,\ \ \ \Delta_p^p=P^p_p\ ,\label{SYMMD}
\ee 
and $P_\al^p$ the $\al$--th symmetric product (elementary symmetric function) of the parameters $a_1,\ldots,a_p$
and $Q_\bet^p$ the $\bet$--th symmetric product of the parameters $b_1,\ldots,b_{p-1}$, i.e.:
\bea
P_\al^p&=&\sum\limits_{i_1,\ldots,i_\al=1\atop i_1<i_2<\ldots<i_\al}^p a_{i_1}\cdot\ldots\cdot a_{i_\al}\ \ \ ,\ \ \ \al=1,\ldots,p\ ,\\[2mm]
Q_\bet^p&=&\sum\limits_{i_1,\ldots,i_\bet=1\atop i_1<i_2<\ldots<i_\bet}^{p-1} b_{i_1}\cdot\ldots\cdot b_{i_\bet}\ \ \ ,\ \ \ \bet=1,\ldots,p-1\ \ \ ,\ \ \ Q_p^p=0\ .\label{SYMMF}
\eea
According to \req{2solution} the solution follows from:
\begin{align}
 w_k^p(z)=\sum\limits_{j_1+2j_2+\ldots+pj_p=k-p}\{(c_1^p)^{j_1},(c_2^p)^{j_2},\ldots,(c_p^p)^{j_p}\}\ c_p^p\ .
\end{align}
Performing the same steps as for $p=2$ and $p=3$ leads to the following result:
\begin{align}
 w_k^p(z)&=\sum\limits_{\vec{l},\vec{m}}(-1)^{|\vec{l}|}(\Delta_1^p)^{m_1}(\Delta_2^p)^{m_2}\ldots(\Delta_{p-1}^p)^{m_{p-1}}
 (\Delta_p^p)^{m_p+1}(Q_1^p)^{l_1}(Q_2^p)^{l_2}\ldots (Q_{p-1}^p)^{l_{p-1}}\nonumber\\
 &\times I(0)^{p-1}\{I(0)^{l_1},\ldots,I(\underbrace{0,\ldots,0}_{p-1})^{l_{p-1}},I(1)^{m_1},\ldots,
 I(1,\underbrace{0,\ldots,0}_{p-1})^{m_p}\}I(1)\ .\label{qresult}
\end{align}
The sum is over the multi--indices $\vec{l}=(l_1,l_2,\ldots,l_{p-1})$ and 
$\vec{m}=(m_1,m_2,\ldots,m_{p})$, which solve the equation:
\begin{align}
 \sum\limits_{\alpha=1}^{p-1}\alpha(l_\alpha+m_{\alpha})+pm_p=k-p\ .
\end{align}
Note, that in contrast to the findings in \cite{Kalmykov} the result \req{qresult} allows to express  any order $w^p_k(z)$ directly without using lower orders.
A representation in terms of MPLs will be given in section~6.

As a final comment we note that in the series \req{32f1}, \req{33f2}  and \req{32fp}
the powers of  $\ap$ are always accompanied
by MPLs of uniform degree of transcendentality (maximal transcendental of weight $k$).
The transcendental weight $w$ is given by the degree of the MPL: $w(\ln x)=1, w({\cal L}i_a)=a$
and $w( {\cal L}i_{n_1,\ldots,n_d})=\sum_{i=1}^d n_i$. 
The degree of transcendentality for a product is defined to be the sum of the degrees of each factor. 
The maximal transcendentality in the power series expansions \req{32f1}, \req{33f2} and \req{32fp} 
w.r.t. $\ap$
simply follows from the underlying recursion relations \req{32ndstep1}, \req{3result}
and \req{qresult} for their expansion coefficients $w_k(z)$.

\section{Fuchsian equations and explicit solutions by iterations and recurrences}

\subsection{Generalized hypergeometric functions and Fuchsian equations}

For $y=\F(x)$, with $p=n=q+1$  the differential equation \req{3hgdgl} becomes
\be\label{DGLq}
x^{n-1}\ (1-x)\ \fc{d^ny}{dx^{n}}+\fa_0\ y+\sum_{\nu=1}^{n-1} x^{\nu-1}\ (\fa_\nu x-\fb_\nu)\ \fc{d^\nu y}{dx^\nu}=0\ ,
\ee
with the parameters $\fa_i,\fb_j$ given as polynomials in $a_r,b_s$.
This is a Fuchsian equation with regular singularities at $x=0,\ x=1$ and $x=\infty$
and $n$ linearly independent solutions for $|x|<1$ and $n-1$ independent solutions 
at $x=1$ to be specified below. Furthermore, with 
\bea
f_1&=&y\ ,\\
f_2&=&y'\ ,\\
&\vdots&\\
f_n&=&y^{(n-1)}\ ,\label{reduce}
\eea
eq. \req{DGLq} can be brought into a system of first order linear differential equations
\be\label{system}
\p_x\bm{f}=A(x)\ \bm{f}\ ,
\ee
with some quadratic matrix $A$ given by the parameters $a_i,b_j$.
The resulting system \req{system} implies non--simple or spurious poles, which can be 
transformed away by a suitable transformation $T$:
\bea
B&=&T^{-1}AT-T^{-1}\p_x T\\
\bm{f}&=&T\ \bm{g}\ .\label{transf}
\eea
Eventually, eq. \req{DGLq} can be cast into a Fuchsian system of first order
\be\label{Fuchs}
\fc{d\bm{g}}{dx}=\lf(\fc{B_0}{x}+\fc{B_1}{1-x}\ri) \bm{g}\ ,
\ee
with regular singularities at $x=0,\ x=1$ and $x=\infty$.
For $q=1$ we have
$$\fa_0=-a_1a_2,\ \fa_1=-(1+a_1+a_2),\ \fb_1=-b_1\ ,$$
$$A=\begin{pmatrix}
0&1  \\  
\fc{a_1a_2}{x}+\fc{a_1a_2}{1-x}&\fc{1+a_1+a_2-b_1}{1-x}-\fc{b_1}{x}
\end{pmatrix}\ ,$$
and with $T=\begin{pmatrix}
1&0  \\  
0&x^{-1}
\end{pmatrix}$ we obtain:
$$B_0=\begin{pmatrix}
0&1  \\  
0&\bet_1
\end{pmatrix}\ \ \ ,\ \ \ B_1=\begin{pmatrix}
0&0  \\  
\al_0&\al_1+\bet_1
\end{pmatrix}\ ,$$
with $\al_i$ being the $2-i$--th elementary symmetric function of $a_1,a_2$ and 
$\bet_i$ the $2-i$--th elementary symmetric function of $1-b_1$, i.e.
\be\begin{array}{lcl}
\al_0=a_1a_2\ &,& \bet_0=0\ ,\\
\al_1=a_1+a_2\ &,& \bet_1=1-b_1\ .
\end{array}
\ee
On the other hand, for $q=2$ we determine
\bea
\fa_0&=&-a_1a_2a_3\ ,\\
\fa_1&=&-(1+a_1+a_2+a_3+a_1a_2+a_1a_3+a_2a_3)\ ,\\
\fa_2&=&-(3+a_1+a_2+a_3)\ ,\\
\fb_1&=&-b_1b_2\ ,\\
\fb_2&=&-(1+b_1+b_2)\ ,\\
\eea
$$A=\begin{pmatrix}
0&1&0  \\  
0&0&1\\
\fc{a_1a_2a_3}{1-x}&\fc{b_1+b_2+a_1a_2+a_1a_3+a_2a_3-b_1b_2-1}{1-x}-\fc{1-b_1-b_2+b_1b_2}{x}&\fc{2-b_1-b_2}{x}-\fc{2+a_1+a_2+a_3-b_1-b_2}{1-x}\end{pmatrix}\ ,$$
and with
$T=\begin{pmatrix}
1&0&0  \\  
0&x^{-1}&0  \\
0&-x^{-2}&x^{-2}
\end{pmatrix}$ we arrive at:
\be
B_0=\begin{pmatrix}
0& 1&0 \\  
0& 0&1 \\
0&-\bet_1&\bet_2
\end{pmatrix}\ \ \ ,\ \ \ 
B_1=\begin{pmatrix}
0&0&0  \\  
0&0&0  \\
\al_0&\al_1-\bet_1&\al_2+\bet_2
\end{pmatrix}\ ,
\ee
with $\al_i$ being the $3-i$--th elementary symmetric function of $a_1,a_2,a_3$ and 
$\bet_i$ the $3-i$--th elementary symmetric function of $1-b_1,1-b_2$, i.e.
\be\begin{array}{lcl}
\al_0=a_1a_2a_3\ &,&\bet_0=0\ ,\\
\al_1=a_1a_2+a_2a_3+a_1a_3\ &,&\bet_1=(1-b_1)\ (1-b_2)\ ,\\
\al_2=a_1+a_2+a_3\ &,&\bet_2=2-b_1-b_2\ .
\end{array}
\ee
For the generic case $n=q+1$ one can recursively define the transformation \req{transf} 
\be\label{Tmatrix}
T_n=\lf(\begin{array}{c|c}
\raisebox{-10pt}{{\Large\mbox{{$T_{n-1}$}}}}& \ \raisebox{-10pt}{{\Large\mbox{{${\bm{0}}_{n-1}$}}}} 
\\[6mm] \hline \\
x^{-1}\ \omega_1\;T_{n-1}\;\omega_2&x^{-n}\\[2mm]  
\end{array}\ri)\ ,
\ee
with
$$\omega_1=(0^{n-2},1)\ \ \ ,\ \ \ \omega_2=
\begin{pmatrix}
0&1&0&0&\ldots&0&0 \\[2mm]  
0&-(n-2)!&1&0&\ldots &0&0\\[2mm]
\vdots& && \ddots&\ddots& & \vdots\\[2mm]
0&\ldots&& 0& & 1 &  0\\
0&\ldots&& 0& & -(n-2)! &  1\\
0&\ldots& &0 & &0&-(n-2)!
\end{pmatrix}\ ,$$
e.g.:
$$T_4=\begin{pmatrix}
1&0&0&0\\[2mm]  
0&\fc{1}{x}&0&0\\[2mm]
0&-\fc{1}{x^2}&\fc{1}{x^2}&0\\[2mm]
0&\fc{2}{x^3}&-\fc{3}{x^3}&\fc{1}{x^3}
\end{pmatrix},\ 
T_5=\begin{pmatrix}
1&0&0&0 &0 \\[2mm]  
0&\fc{1}{x}&0&0&0\\[2mm]
0&-\fc{1}{x^2}&\fc{1}{x^2}&0&0\\[2mm]
0&\fc{2}{x^3}&-\fc{3}{x^3}&\fc{1}{x^3}&0\\[2mm]
0&-\fc{6}{x^4}&\fc{11}{x^4}&-\fc{6}{x^4}&\fc{1}{x^4}\\[2mm]
\end{pmatrix}\ .$$
For \req{Fuchs} this transformation $T$ yields the matrices
\be\label{matB0}
B_0=\begin{pmatrix}
0 &1 &0 &\ldots &0 &0  \\  
0 &0 &1 &\ldots &0 &0  \\  
\vdots&\vdots&\vdots&&\vdots&\vdots&\\
0 &0 &0 &\ldots &1 &0  \\
0 &0 &0 &\ldots &0 &1  \\
0&(-1)^{n}\bet_1&(-1)^{n+1}\bet_2&\ldots&-\bet_{n-2}&\bet_{n-1}
\end{pmatrix}\ ,
\ee
and
\be\label{matB1}
B_1=\begin{pmatrix}
0 &0 &0 &\ldots &0 &0  \\  
\vdots&\vdots&\vdots&&\vdots&\vdots&\\
0 &0 &0 &\ldots &0 &0  \\
\al_0&\al_1+(-1)^{n}\bet_1&\al_2-(-1)^n\bet_2&\ldots&\al_{n-2}-\bet_{n-2}&\al_{n-1}+\bet_{n-1}
\end{pmatrix}\ ,
\ee
with $\al_i$ being the $n-i$--th elementary symmetric function of $a_1,\ldots,a_{n}$ and 
$\bet_i$ the $n-i$--th elementary symmetric function of $1-b_1,\ldots,1-b_{n-1}$.

There is a whole family of transformations $T$ yielding the form \req{Fuchs}.
E.g. for $q=1$ the two transformations
\be
T=\begin{pmatrix}
1&0  \\  
0&\fc{\lambda}{1-x}
\end{pmatrix}\ \ \ ,\ \ \ T=\begin{pmatrix}
1&0  \\  
0&\fc{\lambda}{x}
\end{pmatrix}\ \ \ ,\ \ \ \lambda\in{\bf R}-\{0\}
\ee
yield
\bea
B_0&=&\begin{pmatrix}
0&0  \\  
\fc{a_1a_2}{\lambda}&-b_1
\end{pmatrix}\ \ \ ,\ \ \ B_1=\begin{pmatrix}
0&\lambda  \\  
0&a_1+a_2-b_1
\end{pmatrix}\ ,\\[5mm]
B_0&=&\begin{pmatrix}
0&\lambda  \\  
0&-b_1
\end{pmatrix}\ \ \ ,\ \ \ B_1=\begin{pmatrix}
0&0  \\  
\fc{a_1a_2}{c}&a_1+a_2-b_1
\end{pmatrix}\ ,
\eea
respectively.

The entries of the matrices \req{matB0} and \req{matB1} are given by the elementary symmetric functions $\al_r$ and $\bet_s$ of the two sets of parameters $a_1,\ldots,a_n$ and 
$1-b_1,\ldots,1-b_{n-1}$, respectively. The latter naturally arise after expanding the differential equation \req{3hgdgl}, which yields:
\be\label{dgl} 
-(1-x)\ \fc{d}{dx}(\theta^{n-1}y)+\sum_{i=1}^{n-1}\lf(\al_i-\fc{\bet_i}{x}\ri)
\ \theta^iy+\al_0\ y=0\ \ \ ,\ \ \ \theta=x\fc{d}{dx}\ .
\ee
Furthermore, with
\be
f_k=\theta^{(k-1)}\ y\ \ \ ,\ \ \ k=1,\ldots,n
\label{reduce1}
\ee
eq. \req{dgl} can be brought into the system \req{Fuchs} of first order linear differential equations with the matrices \req{matB0} and \req{matB1} and
\be\label{gVec}
\bm{g}=\begin{pmatrix}
y\\
\theta y\\
\vdots\\
\theta^{(n-1)}y
\end{pmatrix}\ ,
\ee
i.e. $\bm{f}=T\bm{g}$ with $T\equiv T_n$ given in \req{Tmatrix}.

For a specific initial condition the system \req{Fuchs} can be solved
by Picard's iterative methods.
A notorious example is the series expansion of the KZ equation to be discussed in the next subsection.

\subsection{Drinfeld associator for expansions of hypergeometric functions}

In this subsection we establish the connection between generalized hypergeometric functions $\F$ and the Drinfeld associator.
The KZ equation of one variable based on the free Lie algebra with generators $e_0,e_1$ 
\be\label{KZ}
\fc{d}{dx}\Phi=\lf(\fc{e_0}{x}+\fc{e_1}{1-x}\ri) \Phi
\ee
is the universal Fuchsian equation for the case of three regular singular points at $0,1$ and $\infty$ on ${\bf P}^1$. 
The unique solution $\Phi(x)\in {\bf C}\vev{\{e_0,e_1\}}$ to \req{KZ} is known as the generating series of multiple polylogarithms (MPLs) in one variable \cite{BrownPoly}
\be\label{MPL}
\Phi(x)=\sum_{w\in \{e_0,e_1\}^\ast}L_w(x)\ w\ ,
\ee
with the coefficients $L_w(x)$ and the symbol $\{e_0,e_1\}^\ast$ denoting a non--commutative word $w=w_1w_2\ldots$ in the letters $w_i\in \{e_0,e_1\}$.
This alphabet specifies the underlying MPLs:
\bea\label{MPLs}
\ds{L_{e_0^n}(x)}&:=&\ds{\fc{1}{n!}\ \ln^n x\ ,\ n\in {\bf N}\ ,}\\[5mm] 
\ds{L_{e_1w}(x)}&:=&\ds{\int_0^x\fc{dt}{1-t}\ L_w(t)\ ,}\\[5mm]
\ds{L_{e_0w}(x)}&:=&\ds{\int_0^x\fc{dt}{t}\ L_w(t)\ \ \ ,\ \ \ L_1(x)=1\ .}
\eea
In particular, we have $L_{e_1}=-\ln(1-x)$ and $L_{e_0^{m-1}e_1}(x)={\cal L}i_m(x)$, with the classical polylogarithm \req{3MPL}. The solution \req{MPL} can be found recursively and built by Picard's iterative methods. 
It is not possible to find power series solutions in $x$ expanded at $x=0$ or $x=1$, because 
they have essential singularities at these points.
However, one can construct a unique analytic solution $\Phi_0$ normalized at $x=0$ with the asymptotic behaviour 
$\Phi_0(x)\ra x^{e_0}$.
Another fundamental solution $\Phi_1$ normalized at $x=1$ with $\Phi_1(x)\ra (1-x)^{-e_1}$ can be considered.
By analytic continuation the connection matrix between the solutions $\Phi_0$ and $\Phi_1$ is independent of $x$ and gives rise to the Drinfeld associator:
\be\label{Drinfeld}
Z(e_0,e_1) =\Phi_1(x)^{-1}\ \Phi_0(x)\ .
\ee
It is the regularized value of $\Phi$ at $x=1$ 
and given by the non--commutative generating series of (shuffle--regularized) MZVs \cite{LeMurakami}
\begin{align}
Z(e_0,e_1) &=\sum_{w\in\{e_0,e_1\}^\ast}\z(w)\ w=1+\z_2\ [e_0,e_1]+\z_3\ (\ [e_0,[e_0,e_1]]-[e_1,[e_0,e_1]\ )\nonumber\\
&+\z_4\ \Big(\ [e_0,[e_0,[e_0,e_1]]]-\fc{1}{4}\ [e_1,[e_0,[e_0,e_1]]]+[e_1,[e_1,[e_0,e_1]]]
+\fc{5}{4}\ [e_0,e_1]^2\ \Big)\nonumber\\
&+\z_2\ \z_3\ \Big(\ ( [e_0,[e_0,e_1]]-[e_1,[e_0,e_1] )\ [e_0,e_1]+[e_0,[e_1,[e_0,[e_0,e_1]]]]\cr
&-[e_0,[e_1,[e_1,[e_0,e_1]]]]\ \Big)
+\z_5\ \Big(\ [e_0,[e_0,[e_0,[e_0,e_1]]]]-\h\ [e_0,[e_0,[e_1,[e_0,e_1]]]]\nonumber\\
&-\fc{3}{2}\ [e_1,[e_0,[e_0,[e_0,e_1]]]]
+(e_0\leftrightarrow e_1)\ \Big)+\ldots\ ,\label{PhiKZ}
\end{align}
with $\z(e_0^{n_1-1}e_1\ldots e_0^{n_r-1}e_1)=\zeta(n_1,\ldots,n_r)$, the shuffle relation 
$\z(w_1\shuffle w_2)=\zeta(w_1)\ \zeta(w_2)=\zeta(w_1w_2)+\zeta(w_2w_1)$ and $\z(e_0)=0=\z(e_1)$.

By taking proper representations for $e_0$ and $e_1$ the KZ equation \req{KZ} gives rise to specific Fuchsian equations  \req{Fuchs}  describing generalized hypergeometric functions \req{Hyper}. 
Their local exponents at the regular singularities $0,1$ and $\infty$ are encoded in the Riemann
scheme
\be\label{scheme}
\begin{pmatrix}
\underline{0} &\underline{1} &\underline{\infty}   \\  
0&0&a_1\\
1-b_1&1&a_2\\
1-b_2&2&a_3\\
\vdots&\vdots&\vdots\\
1-b_{q-1}&q-1&a_q\\
1-b_q&d&a_{q+1}
\end{pmatrix}\ ,
\ee
with $d=-a_{q+1}+\sum_{j=1}^qb_j-a_j>0$. The inequality condition guarantees convergence of \req{Hyper} at the unit circle $|x|=1$. For non--integer parameters $b_j$, with $b_i\neq b_j$ a $q+1$--dimensional basis of solutions of \req{dgl} 
is given by \cite{Slater}
\begin{align}
&\F\lf[{a_1,\ldots,a_{q+1}\atop b_1,\ldots,b_q};x\ri]\ ,\label{fundamental}\\
x^{1-b_i}\ &\F\lf[{1+a_1-b_i,1+a_2-b_i,\ldots,1+a_{q+1}-b_i
\atop 1+b_1-b_i,1+b_2-b_i,\ldots,(\ast),\ldots,1+b_q-b_i,2-b_i};x\ri]\ ,\ \ \ i=1,\ldots,q\nonumber\ ,
\end{align}
with $(\ast)$ denoting omission of the expression $1+b_i-b_i$.

For \req{32fp} at $z=1$ let us now derive an alternative expression, which is described by the Drinfeld associator.
For this case the symmetric functions \req{SYMMF} can be used to parameterize the matrices \req{matB0} and \req{matB1} provided we change their form into:
\be\label{MatB}
B^p_0=\begin{pmatrix}
0 &1 &0 &\ldots &0 &0  \\  
0 &0 &1 &\ldots &0 &0  \\  
\vdots&\vdots&\vdots&&\vdots&\vdots&\\
0 &0 &0 &\ldots &1 &0  \\
0 &0 &0 &\ldots &0 &1  \\
0&-Q^p_{n-1}&-Q^p_{n-2}&\ldots&-Q^p_2&-Q^p_1
\end{pmatrix},\ 
B^p_1=\begin{pmatrix}
0 &0  &\ldots &0 &0  \\  
\vdots&\vdots&&\vdots&\vdots&\\
0 &0 &\ldots &0 &0  \\
\Delta^p_n&\Delta^p_{n-1}&\ldots&\Delta^p_{2}&\Delta^p_{1}
\end{pmatrix}\ .
\ee
Now, $Q_\bet^p$ is the $\bet$--th elementary symmetric function of $b_1,\ldots,b_{p-1}$ defined
in \req{SYMMF} and $\Delta_\bet^p$ is given in \req{SYMMD}, with $\bet=1,\ldots,n$ and $n=p$.

At a neighbourhood of $x=0$ and $x=1$ for \req{dgl} one can construct two sets of fundamental solutions $u_i$ and $v_i$ ($i=1,\ldots,n$), respectively.
According to \req{gVec} for \req{Fuchs} these two sets give rise to the following solution matrices
\be\label{solmatrix}
\bm{g}_0=\begin{pmatrix}
u_1&\ldots&u_n\\
\th u_1&\ldots&\th u_n\\
\vdots&\vdots&\vdots\\
\th^{(n-1)}u_1&\ldots&\th^{(n-1)}u_n
\end{pmatrix}\ \ \ ,\ \ \ \bm{g}_1=\begin{pmatrix}
 v_1&\ldots&v_n\\
\th v_1&\ldots&\th v_n\\
\vdots&\vdots&\vdots\\
\th^{(n-1)}v_1&\ldots&\th^{(n-1)}v_n
\end{pmatrix}\ ,
\ee
respectively. 
The first set of fundamental solutions $u_i$ is given in \req{fundamental} subject to the change $b_l\ra b_l+1$.
Hence, for $x\ra0$ we have:
\be\label{solmatrix1}
\bm{g}_0\stackrel{x \ra 0}{\lra} \begin{pmatrix}
1&x^{-b_1}&\ldots&x^{-b_q}\\
0&-b_1x^{-b_1}  & \ldots&-b_qx^{-b_q}\\
\vdots&\vdots& \vdots&\vdots\\
0&(-b_1)^{n-1}x^{-b_1}  &\ldots&(-b_q)^{n-1}x^{-b_q}
\end{pmatrix}=:\gamma_0,
\ee
On the other hand, with \req{scheme} for $x\ra1$ we have
\be\label{solmatrix2}
\bm{g}_1\stackrel{x \ra 1}{\lra}  \begin{pmatrix}
1&0&\ldots&0&0\\
\lf.\th v_1\ri|_{x=1}& \lf.\th v_2\ri|_{x=1} & \ldots &\lf.\th v_{n-1}\ri|_{x=1}&-d(1-x)^{d-1}\\
\vdots&\vdots& \ddots&\vdots&\vdots\\
\lf.\th^{(n-1)}v_1\ri|_{x=1}&\lf.\th^{(n-1)}v_2\ri|_{x=1}& \ldots &\lf.\th^{(n-1)}v_{n-1}\ri|_{x=1}
&(-1)^{n-1}(d)_{n-1}(1-x)^{d-n+1}
\end{pmatrix}=:\gamma_1 ,
\ee
with $(d)_{n-1}$  representing the falling factorial $(d)_{n-1}=d(d-1)\cdot\ldots\cdot (d-n+2)$.
We want to relate the solutions \req{solmatrix} to the normalized solutions $\Phi_i$ for the KZ equation \req{KZ} with $e_0$ and $e_1$ replaced by the representations $B_0$ and $B_1$, respectively.  
By comparing the limits \req{solmatrix1} and \req{solmatrix2} with the behaviour of 
the normalized solutions $\Phi_i$ of \req{KZ}
\be
\Phi_0 \stackrel{x \ra 0}{\lra} x^{B_0}\ \ \ ,\ \ \ \Phi_1 \stackrel{x \ra 1}{\lra} 
(1-x)^{-B_1}
\ee
one can define the following connection matrices $C_0,C_1$
\be\label{Connection}
\gamma_0=:x^{B_0}\ C_0\ \ \ ,\ \ \ \gamma_1=:(1-x)^{-B_1}\ C_1\ ,
\ee
which  allow to express the solutions \req{solmatrix} in terms of the normalized solutions $\Phi_0$ and $\Phi_1$, respectively:
\be\label{Connect}
{\bf g}_i=\Phi_i\ C_i\ \ \ ,\ \ \ i=0,1\ .
\ee
On the other hand, from the definition of the Drinfeld associator \req{Drinfeld}
it follows
\be\label{NICE}
Z(B_0,B_1)=C_1\ {\bf g}_1^{-1}\ {\bf g}_0\ C_0^{-1}\ ,
\ee
which is valid for any $x$.
By considering in \req{NICE} the limit $x\ra 1$ we are able to extract a relation between the first matrix element
of the Drinfeld associator and the value of $u_1$ at $x=1$, i.e.
\be\label{NICE1}
{}_pF_{p-1}\lf[{a_1,\ldots,a_{p}\atop 1+b_1,\ldots,1+b_{p-1}};1\ri]=
\lf.Z(B^p_0,B^p_1)\ \ri|_{1,1}\ ,
\ee
with $e_0,e_1$ replaced by the representations $B_0$ and $B_1$, respectively. The latter can be found in \req{MatB}. Eq. \req{NICE} is one of  the main results of this subsection. E.g. for $p=2$ and
$p=3$ we obtain
\begin{align}
{}_2F_1\lf[{a,b\atop 1+c};1\ri]&=
\lf.Z\lf[\begin{pmatrix} 0&1\\
0&-c\end{pmatrix},\begin{pmatrix} 0&0\\
ab&a+b-c\end{pmatrix}\ri]\ \ri|_{1,1},\label{2F13F2}\\[5mm]
{}_3F_2\lf[{a_1,a_2,a_3\atop 1+b_1,1+b_2};1\ri]&=
\lf.Z
\begingroup
\everymath{\scriptstyle}
\scriptsize
\lf[\begin{pmatrix} 0&1&0\\
0&0&1\\
0&-b_1b_2&-b_1-b_2\end{pmatrix},\begin{pmatrix} 0&0&0\\
0&0&0\\
\footnotesize a_1a_2a_3&a_1a_2+a_1a_3+a_2a_3-b_1b_2&a_1+a_2+a_3-b_1-b_2\end{pmatrix}\ri]
\endgroup\ri|_{1,1}\hskip-0.25cm,\nonumber
\end{align}
respectively. For what follows it is important, that the transformations $T$ given in \req{Tmatrix} have a block structure w.r.t. the first entry.
In order to prove \req{NICE} we derive the following block--structure of the matrices involved
$$\gamma_0=
\left(
\begin{array}{c|c}
  1 & x^{-b_1} \cdots x^{-b_q} \\ \hline
  0 & \Oc(x^{-b_1})\cdots \Oc(x^{-b_q})\\
  \vdots &  \vdots\hskip2cm          \vdots\\
  0 & \Oc(x^{-b_1})\cdots \Oc(x^{-b_q})
\end{array}
\right),\ x^{B_0}=
\left(
\begin{array}{c|c}
  1 &  \ast \cdots \ast\\ \hline
  0 & \raisebox{-15pt}{{\huge\mbox{{$\ast$}}}} \\[-3ex]
  \vdots & \\[-0.5ex]
  0 &
\end{array}
\right),\ \ \mbox{i.e.:}\ \ C_0=
\left(
\begin{array}{c|c}
  1 &  \ast \cdots \ast\\ \hline
  0 & \raisebox{-15pt}{{\huge\mbox{{$\ast$}}}} \\[-3ex]
  \vdots & \\[-0.5ex]
  0 &
\end{array}
\right)\ ,$$
with irrelevant contributions of the order $x^{-b_i}$ in the second block of $\gamma_0$, and:
$$\gamma_1=
\left(
\begin{array}{c|c}
  1 & 0 \cdots0 \\ \hline
  \ast & \raisebox{-15pt}{{\huge\mbox{{$\ast$}}}} \\[-3ex]
  \vdots & \\[-0.5ex]
  \ast &
\end{array}
\right),\ (1-x)^{-B_1}=
\left(
\begin{array}{c|c}
  1 & 0 \cdots0\\ \hline
  0 & \raisebox{-15pt}{{\mbox{{$\bm{1}_{n-2}$}}}} \\[-3ex]
  \vdots & \\
  0&  \\ \hline
  \ast &\ast\cdots\ast
\end{array}
\right),\ \ \mbox{i.e.:}\ \ C_1=
\left(
\begin{array}{c|c}
  1 &  0 \cdots 0\\ \hline
  \ast & \raisebox{-15pt}{{\huge\mbox{{$\ast$}}}} \\[-3ex]
  \vdots & \\[-0.5ex]
  \ast &
\end{array}
\right).$$
Eventually, with
$$
\bm{g}_0(x=1)=
\left(
\begin{array}{c|c}
  u_1(x=1) &  \ast \cdots \ast\\ \hline
  \ast & \raisebox{-15pt}{{\huge\mbox{{$\ast$}}}} \\[-3ex]
  \vdots & \\[-0.5ex]
  \ast &
\end{array}
\right)\ \ \ ,\ \ \ \bm{g}_1(x=1)\equiv\gamma_1=
\left(
\begin{array}{c|c}
  1 &  0 \cdots 0\\ \hline
  \ast & \raisebox{-15pt}{{\huge\mbox{{$\ast$}}}} \\[-3ex]
  \vdots & \\[-0.5ex]
  \ast &
\end{array}
\right)$$
we are able to verify \req{NICE}.

\def\Li{{\cal L}i}

Eq. \req{Connect} yields a connection between the hypergeometric solutions \req{solmatrix} and the fundamental solution $\Phi_0$ of the KZ equation \req{KZ}, which for the first matrix element reads:
\be
\lf.\bm{g}_0\ri|_{1,1}=
\lf.\Phi_0\ C_0\ri|_{1,1}=\lf.\Phi_0[B^p_0,B^p_1](x)\ri|_{1,1}\ \ .
\ee
The last relation follows from the block--structure of the connection matrix $C_0$ given above.
Furthermore, an explicit expression for the fundamental solution $\Phi_0$ is given in \cite{Furusho}
\begin{align}
\Phi_0[e_0,e_1](x)&=1+e_0\ \ln x-e_1\ \ln(1-x)+\h\ \ln^2x\ e_0^2+\Li_2(x)\ e_0e_1\nonumber\\
&-\lf[\ \Li_2(x)+\ln x\ \ln(1-x)\ \ri]e_1e_0+\h\ \ln^2(1-x)\ e_1^2+\fc{1}{6}\ \ln^3x\ e_0^3\nonumber\\
&+\Li_3(x)\ e_0^2e_1-\lf[\ 2\ \Li_3(x)-\ln x\ \Li_2(x)\ \ri] e_0e_1e_0+\Li_{2,1}(x)\ e_0e_1^2\nonumber\\[2mm]
&+\lf[\ \Li_3(x)-\ln x\ \Li_2(x)-\h\ln^2 x\ \ln(1-x)\ \ri] e_1e_0^2+\Li_{1,2}(x)\ e_1e_0e_1\nonumber\\[2mm]
&-\lf[\ \Li_{1,2}(x)+\Li_{2,1}(x)-\h\ \ln x\ \ln^2(1-x)\ \ri] e_1^2e_0-
\fc{1}{6}\ \ln^3(1-x)\ e_1^3+\ldots\ ,\label{Phi0}
\end{align}
with the polylogarithms \req{3MPL}. This series follows by successively evaluating the sum
\req{MPL}. For generic $p$ the first matrix element of $\bm{g}_0$ introduced in \req{solmatrix}
represents the generalized hypergeometric function \req{32fp}.
With \req{Phi0} we are able to generalize the relation \req{NICE1} to arbitrary positions $x$
\be\label{NICE2}
{}_pF_{p-1}\lf[{a_1,\ldots,a_{p}\atop 1+b_1,\ldots,1+b_{p-1}};x\ri]=
\lf.\Phi_0[B^p_0,B^p_1](x)\ \ri|_{1,1}\ ,
\ee
with $e_0,e_1$ replaced by the representations $B^p_0$ and $B^p_1$, respectively. The latter can be found in \req{MatB}. As a consequence of their block--structure eventually only a few terms of \req{Phi0} contribute in \req{NICE2}, cf. eq. \req{Fundamental} as an example.

The series \req{32fp} is described for $z=1$ by the relation \req{NICE1} and for generic $z$
by \req{NICE2}.
It is worth pointing out the compactness and simplicity of the results 
\req{NICE1} and \req{NICE2}. For a given 
power $\ap^k$ in \req{32fp} the coefficients $w_k^p(1)$ are given by the terms of degree $k$ of the Drinfeld associator \req{PhiKZ} described by MZVs of weight $k$  supplemented 
by matrix products of $k$ matrices $B^p_0$ and $B^p_1$ given in \req{MatB}.
Furthermore, for a given 
power $\ap^k$ in \req{32fp} the coefficients $w_k^p(z)$ are given by the terms of degree $k$ of the fundamental solution \req{Phi0} of the KZ equation.
As a consequence, computing higher orders in the expansion \req{32fp} is reduced to simple matrix multiplications.

The entries of the matrices \req{MatB} are homogeneous polynomials of a given degree. 
This property makes sure, that the matrix elements
$\lf.Z(B^p_0,B^p_1)\ri|_{1,1}$ in eq. \req{NICE1} and 
$\lf.\Phi_0[B^p_0,B^p_1](x)\ri|_{1,1}$ in eq. \req{NICE2} are also homogeneous in agreement with the expansion coefficients $w_k^p$ in \req{32fp}.
The degree of the polynomials $Q_\bet^p$ and $\Delta_\bet^p$ entering in the matrices \req{MatB} 
is in one--to--one correspondence with the corresponding number of derivatives of the corresponding 
entry of the vector \req{gVec}. This fact is 
crucial to perform iterative methods at each order in the parameter~$\eps$.

Besides, for $z=1$ in the series \req{32fp}  the powers of  $\ap$ are always accompanied
by MZVs with a fixed ``degree of transcendentality'' (maximal transcendental).
The degree of transcendentality (transcendentality level \cite{Fleischer:1998nc}) of $\pi$ is defined to be 1, while
that of $\zeta(n)$ is $n$, and for multiple zeta values $\zeta(n_1, . . . , n_r)$ it is 
$\sum\limits_{i=1}^r n_i$.
The degree of transcendentality for a product is defined to be the sum of the degrees of each factor. 
General graphical criteria for maximal transcendentality of multiple Gaussian  
hypergeometric functions have been given in \cite{Mafra:2011nw,Stieberger:2012rq}.
Similar transcendentality properties hold for the generic case $z\neq 1$.

\subsection{Generic first--order Fuchsian equations, iterations and recurrences}

A generic system $\bm{g}'=A(x)\bm{g}$ of $n$ equations of (first--order) Fuchsian class has the form
\be\label{genFuchs}
\fc{d\bm{g}}{dx}=\sum_{i=0}^l\fc{A_i}{x-x_i}\ \bm{g}\ ,
\ee
with the $l+1$ distinct points $x_0,\ldots,x_l$ and constant quadratic 
non--commutative matrices $A_i\in M(n)$. If $\sum\limits_{i=0}^l A_i\neq 0$ the system of equations \req{genFuchs} has $l+2$ regular singular points at $x=x_i$ and $x=\infty$ and is known
as Schlesinger system. 

At a regular singular point any solution can be expressed explicitly by the combination of elementary functions and power series convergent within a circle around the singular point. 
A solution to \req{genFuchs}  taking values in ${\bf C}\vev{A}$  with the alphabet $A=\{A_0,\ldots,A_l\}$
 can be given as formal weighted sum over iterated integrals  (with the weight given by the number of iterated integrations)
\be\label{genSOL}
\bm{g}(x)=\sum_{w\in A^\ast}L_w(x)\ w\ ,
\ee
generalizing the case \req{MPL} and leading to hyperlogarithms~\cite{Goncharov}.
The latter are defined recursively from words $w$ built from an alphabet $\{w_0,w_1,\ldots\}$
 (with $w_i\simeq A_i$) with $l+1$ letters:
\bea\label{hyperlog}
\ds{L_{w_0^n}(x)}&:=&\ds{\fc{1}{n!}\ \ln^n (x-x_0)\ ,\ n\in {\bf N}\ ,}\\[5mm]
\ds{L_{w_i^n}(x)}&:=&\ds{\fc{1}{n!}\ \ln^n\lf(\fc{x-x_i}{x_0-x_i}\ri)\ ,\ 1\leq i\leq l\ ,}\\[5mm] 
\ds{L_{w_iw}(x)}&:=&\ds{\int_0^x\fc{dt}{t-x_i}\ L_w(t)\ \ \ ,\ \ \ L_1(x)=1\ .}
\eea
The alphabet $A$ is directly related to the differential forms 
$$\fc{dx}{x-x_0},\ldots,\fc{dx}{x-x_l}$$ 
appearing in \req{genFuchs}. 
Generically the functions \req{MPLs} and \req{hyperlog} may also be 
written as Goncharov  polylogarithms \cite{goncharov}
\be
L_{w_{\sigma_1}\ldots w_{\sigma_l}}(x)=G(x_{\sigma_1},\ldots,x_{\sigma_l};x)\ ,
\ee
given by
\be\label{MPOLY}
G(a_1,\ldots,a_n;z)=\int_0^z\fc{dt}{t-a_1}\ G(a_2,\ldots,a_n;t)\ ,
\ee
with $G(z):=G(;z)=1$ except $G(;0)=0$ and $a_i,z\in\bm{C}$.
Typically, for a given class of amplitudes one only needs a certain special subset of allowed indices $a_i$ referring to a specific alphabet. E.g.  for the evaluation of loop integrals arising in massless quantum field theories one has $a_i\in\{0,1\}$. However, the inclusion of particle masses in loop integrals  may give rise to  $a_i\in\{0,1,-1\}$.
The objects \req{MPOLY} are related to the MPLs (cf. also eq. \req{3MPL})
\be
{\cal L}i_{n_1,\ldots,n_l}(z_1,\ldots,z_l)=\sum_{0<k_l<\ldots <k_1}
\fc{z_1^{k_1}\cdot\ldots\cdot z_l^{k_l}}{k_1^{n_1}\cdot\ldots\cdot k_l^{n_l}}
\ee
as follows:
\be
{\cal L}i_{n_1,\ldots,n_l}(z_1,\ldots,z_l)=(-1)^l\ G\lf(\underbrace{0,\ldots,0}_{n_1-1},\fc{1}{z_1},
\ldots,\underbrace{0,\ldots,0}_{n_l-1},\fc{1}{z_1z_2\cdot\ldots\cdot z_l};1\ri)\ .
\ee
Furthermore we have:
\be
L_{w_0^{n_l-1}w_{\sigma_l}\ldots w_0^{n_2-1}w_{\sigma_2}w_0^{n_1-1}w_{\sigma_1}}(x)=
(-1)^l\ {\cal L}i_{n_l,\ldots,n_1}\lf(\fc{x-x_0}{x_{\sigma_l}-x_0},
\fc{x_{\sigma_l}-x_0}{x_{\sigma_{l-1}}-x_0},\ldots
\fc{x_{\sigma_3}-x_0}{x_{\sigma_2}-x_0},
\fc{x_{\sigma_2}-x_0}{x_{\sigma_1}-x_0}\ri)\ .
\ee
For any $0\leq i\leq l$ there exists a unique solution $\bm{g}_i(x)$ to \req{genFuchs} 
with the leading behaviour $x\ra x_i$ of $\bm{g}_i$ given by (cf. eq. \req{Connection})
\be\label{behav}
\bm{g}_i(x)= P_i(x)\ (x-x_i)^{ A_i} \ \bm{C}_i\ ,
\ee
with the normalization vector $\bm{C}_i$ and some holomorphic power series 
\be
P_i=1+\sum\limits_{n\geq 1}P_{in}\ (x-x_i)^n\ ,
\ee
with complex coefficients $P_{in}$. 
Based on  the building blocks \req{hyperlog} and by applying Picard's iterative methods
a solution at $x=x_i$ can be constructed in a similar way as \req{genSOL}. One can compare two solutions  referring to the  two points $x=x_0$ and $x=x_i$.
The quotient of any two such solutions is a  non--commutative 
series with constant coefficients known as regularized zeta series  giving rise to 
an associator: 
\be
Z^{(x_i)}(A_0,\ldots,A_l)\ \ \ ,\ \ \ i=1,\ldots,l\ . 
\ee
The latter determines the monodromy of the hyperlogarithms \cite{BrownFuchs}.
We refer to \cite{GPSS} for an explicit treatment of the case $l=2$.

In the following let us assume, that in \req{genFuchs} the matrices $A_i$ have some polynomial dependence on $\eps$ with integer powers as:
\be
A_i=\sum_{n=1}^{n_0}a_{in}\ \eps^n\ .
\ee
In this case in \req{behav} the factor $\bm{C}_i$ and $P_i$ contain subleading contributions in $\eps$. We are looking for a power series solution in $\epsilon$:
\be\label{gA}
\bm{g}(x)=\sum_{k\geq 0}\bm{u}_k(x)\ \eps^k\ .
\ee
Eventually, each order $\eps^k$ of the  power series  is supplemented by a 
${\bf Q}$--linear combination of iterated integrals of weight $k$.
After inserting the Ansatz \req{gA} into \req{genFuchs} we obtain a recursive differential equation for the functions $\bm{u}_k(x)$, which can be integrated to the following recursive relation
\be\label{setupF}
\bm{u}_k(x)=\bm{u}_k(0)+\sum_{i=0}^l\sum_{n=1}^{\min\{n_0,k\}}a_{in}\ \int_0^x\fc{\bm{u}_{k-n}(t)}{t-x_i}\ dt\ ,
\ee
which translates into the following operator equation:
\be\label{recFuchs}
\bm{u}_k(x)=\bm{u}_k(0)+\sum\limits_{i=0}^l\sum_{n=1}^{\min\{n_0,k\}}a_{in}\ 
I(x_i)\ \bm{u}_{k-n}(x)\ .
\ee
Above, $\bm{u}_k(0)$ represents a possible inhomogeneity accounting for an integration constant, which is determined by the initial value problem. Evidently, we have\footnote{Already at $k\!=\!1$ the equation \req{setupF} translates into the non--trivial recursion
$\bm{u}_1(x)=\bm{u}_1(0)+\sum\limits_{i=0}^la_{i1}\ \int_0^x\fc{\bm{u}_0(t)}{t-x_i}  dt$.}
 $\bm{u}_0(x)= \bm{u}_0(0)=const$.
We may find a general solution to \req{recFuchs} by considering the following recurrence relation
\begin{align}
 \bm{u}_k(x)=\bm{u}_k(0)+\sum_{n=1}^{\min\{n_0,k\}}c_n\ \bm{u}_{k-n}(x)\ ,\label{recFuchs2}
\end{align}
with  the coefficients:
\begin{align}
 c_n=\sum\limits_{i=0}^la_{in}\ I(x_i)\ .
\end{align}
For \req{recFuchs2} we can directly apply our general solution for inhomogeneous recurrence relations \req{2inhsol} to obtain
\begin{align}\label{genSol}
\bm{u}_k(x)&=\sum\limits_{\alpha=0}^{n_0-1}\sum\limits_{\substack{|\vec{j}_1|+2|\vec{j}_2|+\ldots+n_0|\vec{j}_{n_0}|\\=k-n_0-\alpha}}
 \{\ldots\}\sum\limits_{\beta=\alpha+1}^{n_0}\sum\limits_{\gamma=0}^la_{\gamma\beta}\ I(x_\gamma)\ \bar{\bm{u}}_{n_0-\beta-\alpha}\\
 &+\sum\limits_{\alpha=n_0}^k\sum\limits_{\substack{|\vec{j}_1|+2|\vec{j}_2|+\ldots+n_0|\vec{j}_{n_0}|\\=k-\alpha}}\{\ldots\}\ \bm{u}_{\alpha}(0)\ ,
\end{align}
with the generalized operator product
\begin{align}\label{Fuchsiangop}
\{\ldots\}=\{(\hat{I}_{1,0})^{j_{1,0}},\ldots,(\hat{I}_{1,l})^{j_{1,l}},\ldots,
(\hat{I}_{n_0,0})^{j_{n_0,0}}, \ldots,(\hat{I}_{n_0,l})^{j_{n_0,l}}\}\ ,
\end{align}
and:
\begin{align}
 \hat{I}_{i,j}:=a_{ji}\ I(x_j)\ .
\end{align}
Eq. \req{genSol} is valid for $k\geq n_0$ and it uses the initial values $\bm{u}_k(x)=\bar{\bm{u}}_k$,\ $k=0,1,\ldots,n_0-1$.
The vectors $\vec{j}_i$,\ $i=1,\ldots,n_0$ consist of $l+1$ non--negative integers: $\vec{j_i}=(j_{i,0},\ldots,j_{i,l})$. Each of these elements $j_{a,b}$ is an index of a corresponding argument $\hat{I}_{a,b}$ in the generalized operator product \req{Fuchsiangop}. 

Since the matrices $a_{ji}$ are non--commutative they cannot be factorized from the generalized 
operator product. Therefore if possible it is better to avoid matrix notations. 
If the matrices $a_{ji}$ commute the generalized operators product \req{Fuchsiangop} simplifies to
\begin{align}
 \{\ldots\}&=\left(\prod\limits_{\substack{0\leq\beta\leq l\\1\leq\gamma\leq n_0}}(a_{\beta,\gamma})^{j_{\gamma,\beta}}\right)
\binom{j_{1,0}+\ldots+j_{n_0,0}}{j_{1,0},~\ldots,j_{n_0,0}}\cdots\binom{j_{1,l}+\ldots+j_{n_0,l}}{j_{1,l},~\ldots,j_{n_0,l}}\\
 &\times \{I(x_0)^{j_{1,0}+\ldots+j_{n_0,0}},\ldots,I(x_l)^{j_{1,l}+\ldots+j_{n_0,l}}\}\ ,
\end{align}
where every integral operator $I(x_i)$ appears only once.

The system \req{genFuchs} generically appears after expressing  Feynman integrals
as first--order coupled systems of differential equations, cf. e.g. Ref. \cite{Henn:2014qga}
for some recent report on differential equations for Feynman integrals. 
In this context alternatively we could consider higher order differential equations instead of coupled
first order equations, treat equations that can be decoupled as inhomogeneities
and  use our recurrence methods from section 2 and 3.

\section{Low--energy expansion of superstring amplitudes}

The full $\ap$--dependence of tree--level string amplitudes is encoded in 
generalized Euler integrals, which integrate to multiple Gaussian hypergeometric functions 
\cite{Oprisa:2005wu}. Extracting from the latter the power series expansion in $\ap$ is of both phenomenological
\cite{Lust:2008qc} and mathematical interest \cite{Motivic}. 
The computation of $\ap$--expansions of generalized Euler integrals, which leads to MPLs
integrating to MZVs, has been initiated in \cite{Oprisa:2005wu,Stieberger:2006te}, while a more systematic way by making profit of the underlying algebra of MPLs has been presented in \cite{Broedel:2013tta}. 
Further attempts can be found in \cite{Broedel:2013aza,Boels,Barreiro:2013dpa}.
However, obtaining in a fully systematic way a closed, compact and analytic expression for a given order in $\ap$, which does not rely on its lower orders to be computed in advance and which is straightforwardly 
applicable, is desirable.
We have found two methods, which exactly meet these requirements, 
one way by solving the underlying recurrence relations (presented in section 3) 
and the other way by matching a given $\ap$--order in
the power series expansion  with the corresponding coefficient of some fundamental and universal solution  of the KZ equation (presented in section 4).

In this section we want to apply these two new methods to the four--point \cite{Green:1981xx}
and five--point \cite{Barreiro:2005hv,Oprisa:2005wu,Stieberger:2006te} superstring 
disk amplitudes. An elegant and   unifying picture of both amplitudes 
in terms of super Yang--Mills building blocks and generic string form factors 
has been elaborated in \cite{Mafra:2011nv,Mafra:2011nw,Stieberger:2012rq}.
The latter are given by generalized hypergeometric functions \req{Hyper} with $p=2$ for the four--point case and $p=3$ for the five--point case, respectively. 
The all--order expressions \req{3result}, \req{2F13F2} and \req{NICE2} 
of these functions will now be used
to obtain all--order expansions for these superstring amplitudes.

\subsection{Four--point superstring amplitude}

The four--point amplitude is written in terms of the hypergeometric function
\begin{align}\label{44pt}
 F(\alpha's,\alpha'u)={}_2F_1\left[
\begin{matrix} 
-\alpha's,~\alpha'u\\ 
1+\alpha's 
\end{matrix}
;1\right],
\end{align}
with the Mandelstam variables $s=(k_1+k_2)^2$ and $u=(k_1+k_4)^2$. According to \req{3mzv} and \req{32ndresult} the $\alpha'$--expansion takes the following
form:
\begin{align}\label{44ptexp}
F(\alpha's,\alpha'u)=1-\sum\limits_{k=2}^{\infty}(-\alpha')^k\sum\limits_{\alpha=1}^{k-1}s^{k-\alpha}u^{\alpha}
\zeta(\alpha+1,\{1\}^{k-\alpha-1})\ .
\end{align}
By this result the string corrections to the four--point Yang--Mills amplitudes  can easily be 
calculated for any order in $\alpha'$. The well known duality--symmetry of $F(s,u)$ w.r.t.  the exchange $s\leftrightarrow u$ is not automatically 
fulfilled in \req{44ptexp}. Instead, this leads to the relation
\begin{align}\label{4mzvid}
 \zeta(\alpha_1+1,\{1\}^{\alpha_2-1})=\zeta(\alpha_2+1,\{1\}^{\alpha_1-1})\ , \ \ \alpha_1,\alpha_2\geq1\ ,
\end{align}
which is a special case of the well--known duality formula for MZVs. Not only the identity \req{4mzvid}
but also the representation \req{44ptexp} for $F(\alpha's,\alpha'u)$ is already known. 
However, it is interesting
to compare these results with those for the five--point amplitude in the next sections.

\subsection{Five--point superstring amplitude}

For five--point amplitudes the basis of generalized Euler integrals is two--dimensional. A possible choice are the two functions \cite{Mafra:2011nv,Mafra:2011nw}
\begin{align}\label{45pt1}
F_1&=F(\ap s_1,\ap s_2)\ F(\ap s_3,\ap s_4)\;{}_3F_2\left[
\begin{matrix} 
\ap s_1,~1+\ap s_4,~-\ap s_{24}\\ 
1+\ap s_1+\ap s_2,~1+\ap s_3+\ap s_4 
\end{matrix};1\right]\ ,\\[3mm]
F_2&=\ap^2\ s_{13}s_{24}\ \frac{F(\ap s_1,\ap s_2)\ F(\ap s_3,\ap s_4)}{(1+\ap s_1+\ap s_2)(1+\ap s_3+\ap s_4)}\;{}_3F_2\left[
\begin{matrix} 
1+\ap s_1,~1+\ap s_4,~1-\ap s_{24}\\ 
2+\ap s_1+\ap s_2,~2+\ap s_3+\ap s_4 
\end{matrix};1\right]\ ,\label{45pt2}
\end{align}
which depend on the kinematic invariants $s_{ij}=(k_i+k_{j})^2$ and $s_i=s_{i\;i+1}$. 
Both hypergeometric functions can be related to \req{33f2} and derivatives thereof by using eqs. \req{3hgrel}:
\begin{align}\label{43f21}
 _3F_2\left[
\begin{matrix} \scriptstyle
\alpha'a_{1},~1+\alpha'a_{2},~\alpha'a_{3} \\ \scriptstyle
1+\alpha'b_{1},~1+\alpha'b_{2} \end{matrix}
;1\right]&=\left(\frac{\theta}{\alpha'a_2}+1\right)\left. {}_3F_2\left[
\begin{matrix} \scriptstyle
\alpha'a_{1},~\alpha'a_{2},~\alpha'a_{3} \\ \scriptstyle
1+\alpha'b_{1},~1+\alpha'b_{2} \end{matrix}
;z\right]\right|_{z=1}\ ,\\[3mm]
 {}_3F_2\left[
\begin{matrix} \scriptstyle
1+\alpha'a_{1},~1+\alpha'a_{2},~1+\alpha'a_{3} \\ \scriptstyle
2+\alpha'b_{1},~2+\alpha'b_{2} \end{matrix}
;1\right]&=\frac{(1+\alpha'b_1)(1+\alpha'b_2)}{a_1a_2a_3(\alpha')^3z}\ \theta\left. {}_3F_2\left[
\begin{matrix} \scriptstyle
\alpha'a_{1},~\alpha'a_{2},~\alpha'a_{3} \\ \scriptstyle
1+\alpha'b_{1},~1+\alpha'b_{2} \end{matrix}
;z\right]\right|_{z=1}\ ,\label{43f22}
\end{align}
with
\begin{align}\label{4inv1}
\begin{split}
a_1&=s_1\ \ \ ,\ \ \ a_2=s_4\ \ \ ,\ \ \ a_3=s_2+s_3-s_5\ ,\\
b_1&=s_1+s_2\ \ \ ,\ \ \ b_2=s_3+s_4\ .
\end{split}
\end{align}
Eqs. \req{43f21} and \req{43f22} lead to:
\begin{align}\label{4f1}
 F_1&=F(\ap s_1,\ap s_2)\ F(\ap s_3,\ap s_4)\;{}_3F_2\left[
\begin{matrix} \scriptstyle
\ap s_1,~\ap s_4,~-\ap s_{24}\\ \scriptstyle
1+\ap s_1+\ap s_2,~1+\ap s_3+\ap s_4 
\end{matrix};1\right]\\
&-\ap^2\ s_1s_{24}\ F(\ap s_1,\ap s_2)\ F(\ap s_3,\ap s_4)\ \frac{\theta}{(\alpha')^3\Delta_3}\left.{}_3F_2\left[
\begin{matrix} \scriptstyle
\ap s_1,~\ap s_4,~-\ap s_{24}\\ \scriptstyle
1+\ap s_1+\ap s_2,~1+\ap s_3+\ap s_4 
\end{matrix};z\right]\right|_{z=1},\nonumber\\[3mm]
F_2&=\ap^2\ s_{13}s_{24}\ F(\ap s_1,\ap s_2)\ F(\ap s_3,\ap s_4)\ \frac{\theta}{(\alpha')^3\Delta_3}\left.{}_3F_2\left[
\begin{matrix} \scriptstyle
\ap s_1,~\ap s_4,~-\ap s_{24}\\ \scriptstyle
1+\ap s_1+\ap s_2,~1+\ap s_3+\ap s_4 
\end{matrix};z\right]\right|_{z=1}.\label{4f2}
\end{align}
In these two expressions the $\alpha'$--expansions of all factors  are known. As a consequence  we are able to derive the expansions of  $F_1$ and $F_2$. In the following this is accomplished in two different ways.
While in subsection 5.2.1 in the representation of \req{3result} the expressions for the parameters \req{3deltaq} are left 
 as symbols,  in  subsection 5.2.2  we take the explicit values for the parameters \req{3deltaq} subject to  \req{4inv1} and use the corresponding elementary symmetric functions:
\begin{align}\label{4inv2}
\begin{split}
\Delta_1&=-s_5\ ,\\
\Delta_2&=s_1s_2-s_2s_3+s_3s_4-s_4s_5-s_5s_1\ ,\\
\Delta_3&=s_1s_2s_4+s_1s_3s_4-s_1s_4s_5\ ,\\
Q_1&=s_1+s_2+s_3+s_4\ ,\\
Q_2&=s_1s_3+s_2s_3+s_1s_4+s_2s_4\ .
\end{split}
\end{align}
With the parameters \req{4inv2} we will derive an explicit expression for $F_2$. Although this representation is not as compact as the one in subsection 5.2.1, it is
more suitable for further applications to be discussed in section 6.

\subsubsection{Representation in terms of elementary symmetric functions $\bm{\Delta_\al}$ and $\bm{Q_\al}$}

Applying the results of section 3 for \req{4f1} and \req{4f2} yields
\begin{align}\label{4f11}
F_1&=\sum\limits_{k=0}^\infty(\alpha')^{k}
 \sum\limits_{\substack{k_1,k_2\geq 0\\k_1+k_2\leq k}}u_{k_1}(s_1,s_2)u_{k_2}(s_3,s_4)v_{k-k_1-k_2}(1)
 \nonumber\\
 &-\sum\limits_{k=2}^\infty(\alpha')^{k}\ s_{1}s_{24}
 \sum\limits_{\substack{k_1,k_2\geq 0\\k_1+k_2\leq k-2}}u_{k_1}(s_1,s_2)u_{k_2}(s_3,s_4)\left.\frac{\theta}{\Delta_3}
 v_{k-k_1-k_2+1}(z)\right|_{z=1}\ ,\\
F_2&=\sum\limits_{k=2}^\infty(\alpha')^{k}\ s_{13}s_{24}
 \sum\limits_{\substack{k_1,k_2\geq 0\\k_1+k_2\leq k-2}}u_{k_1}(s_1,s_2)u_{k_2}(s_3,s_4)\left.\frac{\theta}{\Delta_3}
 v_{k-k_1-k_2+1}(z)\right|_{z=1}\ ,\label{4f22}
\end{align}
with $u_k(a,b)$ defined in accordance with $u_k(z)$ in \req{32ndstep2}
\begin{align}\label{42f1con}
 u_k(a,b)=\begin{cases}
     \ds{1} & \text{for } k=0, \\
     \ds{\sum\limits_{\alpha=1}^{k-1}(-1)^{k+1}a^{k-\alpha}b^{\alpha}\zeta(\alpha+1,\{1\}^{k-\alpha-1})} & \text{otherwise,}
   \end{cases}
\end{align}
and with
\begin{multline}\label{43f2con}
 \left.\frac{\theta}{\Delta_3} v_k(z)\right|_{z=1}=\sum\limits_{l_1+m_1+2(l_2+m_2)+3m_3
 =k-3}(-1)^{l_1+l_2}\Delta_1^{m_1}\Delta_2^{m_2}\Delta_3^{m_3}Q_1^{l_1}Q_2^{l_2}\\
 \times \left. I(0)\left\{I(0)^{l_1},I(0,0)^{l_2},I(1)^{m_1},I(1,0)^{m_2},I(1,0,0)^{m_3}\right\} I(1)\right|_{z=1},
\end{multline}
representing the expansion of the $p=3$ hypergeometric functions, which follows easily from \req{3result}.
Eqs. \req{4f11} and \req{4f22} give all orders of the $\alpha'$--expansions of the five--point open  superstring amplitude.
Besides the kinematic variables at each order only products of at most three MZVs are produced.
This is to be contrasted with the procedure in \cite{Broedel:2013tta} where also higher products
of MZVs appear.

As already discussed at the end of subsection 3.3 there is a way to remove the generalized operator product to obtain
a representation in terms of MZVs.

\subsubsection{Representation in terms of kinematic invariants $\bm{s_{i}}$}

The polynomial parts of $u_{k_1}(s_1,s_2)$, $u_{k_2}(s_3,s_4)$ and $v_k(z)$ can be combined in \req{4f22}
to obtain
\begin{align}\label{4f2alt}
 F_2=s_{13}s_{24}\sum\limits_{k=2}^\infty(\ap)^k\sum\limits_{j_1+j_2+j_3+j_4+j_5=k-2}s_1^{j_1}
 s_2^{j_2}s_3^{j_3}s_4^{j_4}s_5^{j_5}f(j_1,j_2,j_3,j_4,j_5)\ ,
\end{align}
where all MZVs and integral operators are contained in:
\begin{align}\label{4mzvpart}
 f(j_1,j_2,j_3,j_4,j_5)=\sum\limits_{\vec{l}=(l_1,l_2,l_3,l_4)}(-1)^{|\vec{l}|}\;\zeta_{l_1,l_2}'\;
 \zeta_{l_3,l_4}'v(j_1-l_1,j_2-l_2,j_3-l_3,j_4-l_4,j_5)\ .
\end{align}
The function \req{4mzvpart} involves the MZVs of \req{42f1con}
\begin{align}\label{42f1part}
   \zeta_{i_1,i_2}'=
   \begin{cases}
    \zeta(i_1+1,\{1\}^{i_2-1})  & \text{for } i_1,i_2\geq1,\\
    -1 & \text{for } i_1,i_2=0,\\
    0 & \text{else,} 
   \end{cases}
\end{align}
and the integral operators of \req{43f2con}:
\begin{align}\label{43f2part}
 \lefteqn{v(j_1,j_2,j_3,j_4,j_5)}\nonumber\\
 &= \sum\limits_{\vec{\alpha},\vec{\beta},\gamma,\vec{\delta},\vec{\epsilon}\in L}
 \left.I(0) \left\{I(0)^{|\vec{\alpha}|},I(0,0)^{|\vec{\beta}|},I(1)^{\gamma},I(1,0)^{|\vec{\delta}|},
 I(1,0,0)^{|\vec{\epsilon}|}\right\}I(1)\right|_{z=1}\\
 &\times(-1)^{|\vec{\alpha}|+|\vec{\beta}|+\delta_3+j_5} \binom{|\vec{\alpha}|}{\alpha_1,\alpha_2,\alpha_3,\alpha_4}
 \binom{|\vec{\beta}|}{\beta_1,\beta_2,\beta_3,\beta_4}
 \binom{|\vec{\delta}|}{\delta_1,\delta_2,\delta_3,\delta_4,\delta_5}
 \binom{|\vec{\epsilon}|}{\epsilon_1,\epsilon_2,\epsilon_3}\ .\nonumber
\end{align}
The summation is over non--negative integers $\gamma$ and the multiple indices:
\begin{align}\label{4indices}
\begin{split}
 \vec{\alpha}&=(\alpha_1,\alpha_2,\alpha_3,\alpha_4)\ ,\\
 \vec{\beta}&=(\beta_1,\beta_2,\beta_3,\beta_4)\ ,\\
 \vec{\delta}&=(\delta_1,\delta_2,\delta_3,\delta_4,\delta_5)\ ,\\
 \vec{\epsilon}&=(\epsilon_1,\epsilon_2,\epsilon_3)\ .
\end{split}
\end{align}
The summation region of $\vec{\alpha}$, $\vec{\beta}$, $\gamma$, $\vec{\delta}$ and $\vec{\epsilon}$ 
is the solution set $L$ of the five equations:
\begin{align}\label{4sumreg}
\begin{split}
j_1&= \alpha_1 + \beta_1 + \beta_2 + \delta_1 + \delta_2 + |\vec{\epsilon}|\ ,\\
j_2&= \alpha_2 + \beta_3 + \beta_4 + \delta_1 + \delta_3 + \epsilon_1\ ,\\
j_3&= \alpha_3 + \beta_1 + \beta_3 + \delta_3 + \delta_4 + \epsilon_2\ ,\\
j_4&= \alpha_4 + \beta_2 + \beta_4 + \delta_4 + \delta_5 + |\vec{\epsilon}|\ .\\
j_5&= \gamma + \delta_2 + \delta_5 + \epsilon_3.
\end{split}
\end{align}
The function $v(j_1,j_2,j_3,j_4,j_5)$ is related to $v_k(z)$ through:
\begin{align}\label{4compare}
\left.\frac{\theta}{\Delta_3} v_k(z)\right|_{z=1}=
\sum\limits_{j_1+j_2+j_3+j_4+j_5=k-3} s_1^{j_1}s_2^{j_2}s_3^{j_3}s_4^{j_4}s_5^{j_5}\ 
v(j_1,j_2,j_3,j_4,j_5)\ .
\end{align}
Obviously, the r.h.s. of \req{4compare} together with \req{43f2part} is less 
compact than the r.h.s. of
\req{43f2con}, which uses the quantities $\Delta_1$, $\Delta_2$, $\Delta_3$, $Q_1$ and $Q_2$. But the advantage of the former is,
that the symmetry of $F_2\;(s_{13}s_{24})^{-1}$ can directly be analyzed  in \req{4mzvpart}. This function is invariant w.r.t.
cyclic permutation of the kinematic invariants ($s_1,s_2,s_3,s_4,s_5$). Therefore $f(j_1,j_2,j_3,j_4,j_5)$ is invariant
w.r.t. cyclic permutations of $(j_1,j_2,j_3,j_4,j_5)$. Just like for the four--point amplitude (cf. subsec. 5.1) this
symmetry is non--trivially fulfilled and various MZV identities are generated. To obtain them, the generalized operator product
in \req{43f2part} has to be written in terms of MZVs. This is of course the same operator product as the one in \req{3result}
and \req{43f2con}. In addition, there are summations over a total of 17 indices,  of which five can be evaluated with 
eqs. \req{4sumreg}. 
Thus besides the issue of converting the generalized product of integral operators into MZVs it is
interesting to see, whether some of the remaining twelve sums can be evaluated. 
For the $\alpha'$--expansion of the 4--point amplitude \req{44ptexp}, this is performed in section 3.2. The identity,
which leads from eq. \req{32ndstep1} to \req{32ndstep2} eliminates both the generalized operator product and the inner sum.
This identity and similar ones, which apply to the 5--point case, are discussed in section 6.

\subsection{Open superstring amplitudes and the Drinfeld associator}

World--sheet disk integrals describing $N$--point open string tree--level  amplitudes 
integrate to multiple Gaussian hypergeometric functions \cite{Oprisa:2005wu} whose singularity structure is generically given by divisors whose monodromy is described by  generalized KZ equations. 
Prime examples are the cases of $N=4$ and $N=5$ whose singularity structure is  described by the 
three regular singular points $0,1$ and $\infty$ leading to the Fuchsian differential equations of first order \req{Fuchs}. The latter is related to the KZ equation \req{KZ}, whose basic solutions can be adjusted to specific  
solutions of \req{Fuchs}, cf. section 4 for more details.
Hence, the solutions of the KZ equation are {\it directly}  related to the  world--sheet disk integrals
and their singularity structure. As a consequence the monodromies on the string world--sheet are {\it explicitly} furnished by the Drinfeld associator.
By contrast in \cite{Broedel:2013aza} a relation between the Drinfeld associator \req{PhiKZ}
and $N$--point world--sheet disk integrals has been given by connecting the boundary values of the 
$N$--point and $N-1$--point integrals thereby giving a relation between 
their underlying world--sheet disk integral. 

The four--point superstring amplitude is written in terms of the hypergeometric function
\req{44pt}. With the choice
\be\label{choi}
a\equiv a_1=-\ap\ s\ \ \ ,\ \ \ b\equiv a_2 =\ap u\ \ \ ,\ \ \ c\equiv b_1=\ap s
\ee
the matrices \req{MatB} become:
\be
B_0=\begin{pmatrix}
0&1\\
0&-\ap s
\end{pmatrix}\ \ \ ,\ \ \ B_1=\begin{pmatrix}
0&0\\
-(\ap)^2\ su&\ap (u-2 s)
\end{pmatrix}\ .
\ee
With this matrix representation eq. \req{2F13F2} expresses the hypergeometric function \req{44pt} in terms of the Drinfeld associator $Z$ as:
\be
{}_2F_1\left[
\begin{matrix} 
-\alpha's,~\alpha'u\\ 
1+\alpha's 
\end{matrix}
;1\right]=Z\lf(B_0,B_1)\ri|_{1,1}\ .
\ee
This gives a {\it direct} relation between the four--point superstring amplitude and 
the Drinfeld associator by {\it exactly} matching the monodromy of the string world--sheet 
to the corresponding monodromy of the underlying hypergeometric function ${}_2F_1$.
We refer the reader to \cite{Drummond:2013vz} for a different
relation with a matrix representation which is linear in the kinematic invariants.

The five--point superstring amplitude uses the two functions \req{45pt1} and \req{45pt2}.
According to \req{43f21} and \req{43f22} the latter can be generated from the single function
\be\label{fundamentall}
{}_3F_2\left[
\begin{matrix} \scriptstyle
\alpha'a_{1},~\alpha'a_{2},~\alpha'a_{3} \\ \scriptstyle
1+\alpha'b_{1},~1+\alpha'b_{2} \end{matrix}
;x\right]\ ,
\ee
with the parameters $a_i$ and $b_j$ defined in \req{4inv1}. 
According to the relation \req{NICE2} the function \req{fundamentall} can be related to the fundamental solution $\Phi_0$ of the KZ equation \req{KZ}.
For $p=3$ we obtain
\bea
{}_3F_2\lf[{\ap a_1,\ap a_2,\ap a_3\atop 1+\ap b_1,1+\ap b_2};x\ri]&=&\ds{\lf.\Phi_0[B_0,B_1](x)\ri|_{1,1}}\\[3mm]
&=&\ds{1+\Delta_3\ \Li_3(x)+\Delta_1\ \Delta_3\ \Li_{2,2}(x)-Q_1\ \Delta_3\ \Li_4(x)+\ldots\ ,}\label{Fundamental}
\eea
with the matrix representations \req{MatB}
\be\label{REP5}
B_0=\begin{pmatrix}
0&1&0\\
0&0&1\\
0&-\ap^2 Q_2&-\ap Q_1
\end{pmatrix}\ \ \ ,\ \ \ B_1=\begin{pmatrix}
0&0&0\\
0&0&0\\
\ap^3\Delta_3&\ap^2\Delta_2&\ap\Delta_1
\end{pmatrix}\ ,
\ee
and the parameters $Q_\al$ and $\Delta_\bet$  corresponding to the elementary symmetric functions and given in \req{4inv2}.
Hence, all details for the two functions \req{45pt1} and \req{45pt2} 
can be derived from the fundamental solution \req{Fundamental} of the KZ equation \req{Fuchs}.
To summarize, the whole $\ap$--dependence of the five--point superstring amplitude is
described by the KZ equation \req{Fuchs}  and its fundamental solution \req{Phi0} with the representations \req{REP5}.
Again, this gives a {\it direct} relation between the five--point superstring amplitude and 
the Drinfeld associator by {\it exactly} matching the monodromy of the string world--sheet 
to the corresponding monodromy of the underlying hypergeometric function ${}_3F_2$.

Similar relations can be found for any  $N$--point superstring amplitude (with $N\geq6$).
This will be exhibited in a future publication \cite{work}.

\section{From generalized operator products to MZVs}

The results of the sections 3 and 5 left some open questions: how to obtain the all--order expression \req{32ndstep2}
for the hypergeometric function ${}_2F_1$  from its representation \req{32ndstep1} with the latter  involving generalized operator
products. Likewise, how to achieve similar transformations on the operator products arising for the hypergeometric functions
${}_3F_2$ and ${}_pF_{p-1}$ in general. These questions will be discussed in section 6.1. 
The relations to be derived there will be applied to the results from sections 3 and 5 in section 6.2. The MZV identities, which
follow from cyclic symmetry of the function $f(j_1,j_2,j_3,j_4,j_5)$ will be discussed in section 6.3.

\subsection{Identities for generalized operator products}

In this subsection three types of operator products will be discussed. Starting from simple cases 
involving  independent arguments the complexity increases step by step to finally obtain identities, which can be applied
to \req{3result}, \req{qresult} and \req{43f2part}. Therefore, not all identities will be needed, at least not for the results of
this paper. However, simpler identities provide a consistency check for the most complicated ones.

All identities for generalized operator products presented in this section contain MZVs.
Identical relations hold for MPLs as well. Before and after every generalized operator product there is an $I(0)$ and an $I(1)$ 
operator, respectively, to ensure finiteness of the corresponding MZV.
The following equations make extensive use of two notations, which we introduced at the end of section 3.1
in eqs. \req{3summzv} - \req{notation}: sums over all sets of
indices $\vec{n}$ of MZVs $\zeta(\vec{n})$ and multiple index sums.

\subsubsection{Independent arguments}

We start with generalized operator products, which include only independent arguments.

\paragraph{Example 1.1:}

The simplest example is:
\begin{align}\label{5ind1}
 \left.I(0)\{I(0)^{j_1},I(1)^{j_2}\}I(1)\right|_{z=1}=\sum\limits_{\substack{w=j_1+j_2+2\\d=j_2+1}}
 \zeta(\vec{n})\ .
\end{align}
It is clear, that the sum is over the given weight and depth, since these quantities correlate 
directly to the number of integral operators. It needs to be proven, that \textit{all} MZVs
of given weight and depth are generated by the generalized operator product on the l.h.s. To do this, it is sufficient to show, that
\begin{enumerate}
 \item both sides contain the same number of terms
 \item and that there are no identical terms on the l.h.s.
\end{enumerate}
The second point is clear due to the definition of the generalized operator product as the sum of \textit{distinct} permutations
and the fact, that both arguments $I(0)$ 
and $I(1)$ are independent. 
The first property is also true.
The number of different MZVs of weight $w$ and depth $d$ is $\binom{w-2}{d-1}$, which equals
$\binom{j_1+j_2}{j_2}$ in this case. According to eq. \req{2multi} this quantity is identical to the number of terms on the l.h.s. of \req{5ind1}.

\paragraph{Example 1.2:}

With $I(1,0)$ instead of $I(1)$ the identity \req{5ind1} becomes:
\begin{align}\label{5ind2}
 \left.I(0)\{I(0)^{j_1},I(1,0)^{j_2}\}I(1)\right|_{z=1}=\sum
 \limits_{\substack{w=j_1+2j_2+2\\d=j_2+1;~n_i\geq2}} \zeta(\vec{n})\ .
\end{align}
The additional condition $n_i\geq2$ is self--explanatory. For the first integer $n_1$ it is obvious that $n_1\geq 2$,
because the l.h.s. starts with the operator $I(0)$. For any other integers $n_i$ to be one, any sequence
of integral operators would have to include the product $I(1,1)$. With the arguments $I(0)$ and $I(1,0)$ this is obviously not
possible. 
To prove that the l.h.s. of \req{5ind2} produces \textit{all} MZVs of given weight
and depth, which do not include an index $n_i=1$, similar arguments as for \req{5ind1} hold here. 
The generalization to cases with $I(1,0)$ replaced by other arguments of the type $I(1,0,\ldots,0)$ is  straightforward.

\paragraph{Examples 1.3 and 1.4:}

Other generalized operator products with independent arguments are
\begin{align}\label{5ind3}
\left.I(0)\{I(1)^{j_1},I(1,0)^{j_2}\}I(1)\right|_{z=1}&=\left.I(0,1)\{I(1)^{j_1},I(0,1)^{j_2}\}\right|_{z=1}\nonumber\\
&=\sum\limits_{\substack{w=j_1+2j_2+2;~d=j_1+j_2+1\\
n_1=2;~d_1=j_1;~d_2=j_2+1}}\zeta(\vec{n})\ ,
\end{align}
and
\begin{align}\label{5ind4}
\left.I(0)\{I(1)^{j_1},I(1,0)^{j_2},I(1,0,0)^{j_3}\}I(1)\right|_{z=1}&=\left.I(0,1)\{I(1)^{j_1},I(0,1)^{j_2},I(0,0,1)^{j_3}\}
\right|_{z=1}\nonumber\\
&=\sum\limits_{\substack{w=j_1+2j_2+3j_3+2;~d=j_1+j_2+j_3+1\\
n_1=2;~d_1=j_1;~d_2=j_2+1;~d_3=j_3}}\zeta(\vec{n})\ ,
\end{align}
with two and three arguments, respectively. In both first equations we used 
\begin{align}\label{5ind5}
\{I(1)^{j_1},I(1,0)^{j_2},\ldots,I(1,\underbrace{0,\ldots,0}_{j_n-1})^{j_n}\}I(1)= I(1)\{I(1)^{j_1},
I(0,1)^{j_2},\ldots,I(\underbrace{0,\ldots,0}_{j_n-1},1)^{j_n}\}\ ,
\end{align}
to make evident the conditions on $d_i$. To check 
consistency, note that $d=\sum_id_i$ and that relation \req{5ind4} becomes \req{5ind3} for $j_3=0$. For a 
strict proof, the same strategy as for \req{5ind1} should work. Even though more combinatorics is needed
here, to determine for example the number of MZVs of given $w$, $d$, $d_1$, $d_2$ and $d_3$.
The generalization to more arguments of the kind $I(1,0,\ldots,0)$ is straightforward. 

\subsubsection{Dependent arguments}

The following identities involve generalized products of dependent operators. Distinct permutations of
dependent factors can be identical. Consider for 
instance the generalized operator product
\begin{align}\label{61stex}
\{I(0),I(1),I(1,0)\}=I(0,1,1,0)+I(0,1,0,1)+I(1,0,0,1)+I(1,1,0,0)+2I(1,0,1,0)\ .
\end{align}
Clearly, the third arguments $I(1,0)$ can be written as a product of the first two $I(1)$ and $I(0)$. As a consequence
the two distinct permuations
\begin{align}\label{6dperms}
 I(1) I(0) I(1,0)\ \text{ and }\ I(1,0) I(1) I(0)
\end{align}
are identical and the corresponding product $I(1,0,1,0)$ appears twice in \req{61stex}. No other products appears
more than once, since $I(1,0,1,0)$ is the only one to contain twice the sequence $I(1)I(0)$. For the more complicated
identities in the following, the task is to count identical permutations such as \req{6dperms}. Since these identities
translate generalized operator products into sums of MZVs, identical permutations correspond to MZVs, which appear more than once.
Therefore a weighting is required in the sum of MZVs, i.e. a function, which can be evaluated for every single MZV in that sum
and thereby describes how often every single MZV appears. On one hand weightings can depend on quantities, which are identical for all MZVs of the corresponding sum. These
quantities are determined by the relevant generalized operator product. Examples are the indices of the generalized operator
product or the weight $w$ and depth $d$, which are given by the number of integral operators. On the other
hand, in order to give different factors for different MZVs, the weighting has to depend on the indices
$\vec{n}=(n_1,\ldots,n_d)$ of MZVs $\zeta(\vec{n})$. This dependence can be explicit or in terms of related quantities such as
the number $d_j$ of indices, which equal $j$.

We start with the discussion on the generalized operator product of the most general case \req{qresult}. 
There are two types of arguments in the generalized operator products we encountered in our results of sections 3 and 5.
Those, that consist only of integral operators $I(0)$ and those, which have an operator $I(1)$ 
to the left of all $I(0)$ operators. In the following relation the former have an index $j_\mu$, $\mu=1,\ldots,a$ and the latter have $j_\nu'$, $\nu=1,\ldots,b$:
\begin{multline}\label{5depgen}
  I(0)\{I(0)^{j_1},I(0,0)^{j_2},\ldots,I(\underbrace{0,\ldots,0}_{a})^{j_a},I(1)^{j_1'},I(1,0)^{j_2'},\ldots,
  I(1,\underbrace{0,\ldots,0}_{b-1})^{j_b'}\}I(1)\big|_{z=1}\\
  =\sum\limits_{\substack{w=j_1+2j_2+\ldots+aj_a\\
  +j_1'+2j_2'+\ldots+bj_b'+2\\d=j_1'+j_2'+\ldots+j_b'+1}}\zeta(\vec{n})\;
  \omega_{a,b}(\vec{n}'-\vec{1};j_1,j_2,\ldots,j_a;j_1',j_2',\ldots,j_b')\ ,
\end{multline} 
with the constant vector $\vec{1}=(\underbrace{1,\ldots,1}_d)$ and the elements
\begin{align}\label{5n1}
n_i'=
 \begin{cases} n_1-1 &\mbox{for } i=1\ , \\
n_i & \mbox{for }i=2,\ldots,d\ , 
\end{cases} 
\end{align}
of the vector $\vec{n}'$.
The conditions for the weight $w$ and the depth $d$ in the sum of MZVs are self--explanatory. They are related to the number of integral operators, which can easily be 
read off from the first line. Of greater interest is the weighting
\begin{multline}\label{5outer}
 \omega_{a,b}(\vec{n};j_1,j_2,\ldots,j_a;j_1',j_2',\ldots,j_b')\\
 =\sum\limits_{\substack{\vec{\beta}_1+\vec{\beta}_2+\ldots+\vec{\beta}_b=\vec{1}\\|\vec{\beta}_\nu|=j_\nu';~\beta_{\nu,1}=0}}
  \omega_a(\vec{n}-\vec{\beta}_2-2\vec{\beta}_3-\ldots-(b-1)\vec{\beta}_b;j_1,j_2,\ldots,j_a)\ ,
\end{multline}
with
\begin{align}\label{5inner}
\omega_{a}(\vec{n};j_1,j_2,\ldots,j_a)
 =\sum\limits_{\substack{\vec{\alpha}_1+2\vec{\alpha}_2+\ldots+a\vec{\alpha}_a=\vec{n}\\
 |\vec{\alpha}_\mu|=j_\mu}}\binom{\vec{\alpha}_1+\vec{\alpha}_2
 +\ldots+\vec{\alpha}_a}{\vec{\alpha}_1,~\vec{\alpha}_2,~\ldots,~\vec{\alpha}_a}\ .
\end{align}
The weighting $\omega_{a,b}$ depends on the indices $\vec{n}$ of the MZVs and the indices $j_\mu$ and $j_\nu'$  of the generalized operator product. 
Besides the multinomial coefficient, essentially there are two summations in the definitions \req{5outer} and \req{5inner}.
The one in \req{5outer} over the indices
\begin{align}\label{6ib}
\vec{\beta}_\nu=(\beta_{\nu,1},\ldots,\beta_{\nu,d})\ ,\ \  \nu=1,\ldots,b
\end{align}
take permutations into account, which involve the operators $I(1,0,\ldots,0)$, while the sums in \req{5inner} over the indices
\begin{align}\label{6ia}
\vec{\alpha}_\mu=(\alpha_{\mu,1},\ldots,\alpha_{\mu,d})\ ,\ \  \mu=1,\ldots,a
\end{align}
refer to the operators of the type $I(0,\ldots,0)$. To explain the expressions \req{5outer} and \req{5inner} in detail, we consider permutations of the arguments of the type $I(0,\ldots,0)$ before discussing those of the arguments of the type $I(1,0,\ldots,0)$. Finally we explain how both types are related.

Both sums in \req{5outer} and \req{5inner} use several multi--indices \req{6ib} and \req{6ia} with $d$ elements each. This is motivated by the following idea. First we count the number of identical permutations of operators, 
which are part of the integral operator representation
\begin{align}\label{5intrep}
I(\underbrace{0,\ldots,0}_{n_i-1},1)\ 
\end{align}
of a single MZV index $n_i$. Then we use the multi--index notation to combine the results of all indices $\vec{n}$, which yields the total number of identical permutations, i.e. the weighting.

Let us start with identical permutations of arguments of the type $I(0,\ldots,0)$, which are part of one index \req{5intrep}. Assuming there are $a$ different arguments of the type $I(\underbrace{0,\ldots,0}_{\mu})\equiv I(0)^\mu$, $\mu=1,\ldots,a$, which appear $\alpha_\mu$ times, then for fixed $\alpha_\mu$ there are
\begin{align}
 \binom{\alpha_1+\alpha_2+\ldots+\alpha_a}{\alpha_1,~\alpha_2,~\ldots,~\alpha_a}
\end{align}
distinct permutations of these arguments. Every permutation is a sequence of $\alpha_1+2\alpha_2+\ldots+a\alpha_a$ operators 
$I(0)$. Now we assume the $\alpha_\mu$ are not fixed, but the total number of operators $I(0)$ has to add up to $n_i-1$. To count the
permutations we have to sum over all sets $(\alpha_1,\alpha_2,\ldots,\alpha_a)$ and take the fixed
total number of $n_i-1$ operators $I(0)$ into account:
\begin{align}\label{6om}
\sum\limits_{\alpha_1+2\alpha_2+\ldots+a\alpha_a=n_i-1} 
\binom{\alpha_1+\alpha_2+\ldots+\alpha_a}{\alpha_1,~\alpha_2,~\ldots,~\alpha_a}\ .
\end{align}
These are the number the possibilities to permute the arguments of the type $I(0,\ldots,0)$, which contribute to one index
\req{5intrep}. Multiplying the possibilities of all indices $\vec{n}$ of a MZV yields the
multi--index notation presented in \req{5inner} and \req{6ia}, with $\alpha_{\mu,i}$ being the number of operators $I(0)^{\mu}$, which contribute to the index $n_i$. The elements of a multi--index are not independent, since their sum $\alpha_{\mu,1}+\ldots+\alpha_{\mu,d}=|\vec{\alpha}_\mu|$,
i.e. the total number of operators $I(0)^\mu$, is fixed by the index $j_\mu$ of the generalized operator
product. As a consequence we have the additional conditions $|\vec{\alpha}_\mu|=j_\mu$, $\mu=1,\ldots,a$ in \req{5inner}. 
The first index $n_1$ is an exceptional case. The first $I(0)$ in the integral operator representation is the one to the left
of the generalized operator product in \req{5depgen}. Thus only $n_1-2$ operators are relevant for the permutations. 
This explains the use of \req{5n1} for the arguments of the weighting $\omega_{a,b}$ in the identity \req{5depgen}.

Now that we have discussed the arrangements of the arguments $I(0,\ldots,0)$, described by $\omega_a$, let us consider the arguments $I(1,0,\ldots,0)$.
All identical permutations of the $b$ arguments $I(1,\underbrace{0,\ldots,0}_{\nu-1})\equiv I(1)I(0)^{\nu-1}$, $\nu=1,\ldots,b$ are taken into account by the multi--index sums in \req{5outer}.
The summation indices \req{6ib} take only two different values:
\begin{align}\label{6li}
\beta_{\nu,i}\in\{0,1\}\ , \ \ \nu=1,\ldots,b\ ,\ \ i=2,\ldots,d\ .
\end{align}
The case $\beta_{\nu,i}=1$ corresponds to $I(1)I(0)^{\nu-1}$ contributing\footnote{By saying ``$n_i$ originates from $c_j$'' or 
``$c_j$ contributes to $n_i$'' it is meant, that $I(1)$ in the integral operator representation
\req{5intrep} of the index $n_i$ is part of the operator $c_j$.} to the index $n_{i-1}$.
On the other hand $\beta_{\nu,i}=0$ means, that $I(1)I(0)^{\nu-1}$ does not contribute to $n_{i-1}$.
Of course no more (and no less) than one argument of the type $I(1,0,\ldots,0)$ can contribute to a single index $n_i$, therefore \req{5outer} uses the conditions $\beta_{1,i}+\ldots+\beta_{b,i}=1$ for all $i=1,\ldots,d$ or in the multi--index notation: $\vec{\beta}_{1}+\ldots+\vec{\beta}_{b}=\vec{1}\ $.
Similar to the conditions for $|\vec{\alpha}_{\mu}|$ in \req{5inner}, the conditions for $|\vec{\beta}_\nu|$ are necessary due to the fixed total number $j_\nu'$ of arguments $I(1)I(0)^{\nu-1}$ in the generalized operator product.
The first index $n_{1}$ provides again an exceptional case. Since there is no previous index to which any argument could contribute, we set $\beta_{\nu,1}=0$, $\nu=1,\ldots,b$. 

Finally we can analyse how permutations between arguments of the type $I(0,\ldots,0)$ and those between arguments $I(1,0,\ldots,0)$ affect each other.  For every configuration of the arguments of the type $I(1,0,\ldots,0)$, i.e. for every set of indices $\beta_{\nu,i}$, we need to count the identical permutations of the arguments of the type $I(0,\ldots,0)$ via $\omega_a$. That is why $\omega_a$ appears in \req{5outer}. However, this combination of the two types of permutations is not just a product, since they are not independent. In other words, it is not possible to obtain identical products, out of permutations of two operators $I(1)I(0)^{\nu_1}$ and $I(1)I(0)^{\nu_2}$ with $\nu_1\neq \nu_2$, without changing the positions of operators of the type $I(0,\ldots,0)$ as well.
This effect of the sums over \req{6ib} on the sums over \req{6ia} is described by the first argument of $\omega_a$ in \req{5outer}. It contains the contributions $-\vec{\beta}_2-2\vec{\beta}_3-\ldots-(b-1)\vec{\beta}_b$. As a result, after inserting \req{5outer} and \req{5inner} in \req{5depgen}, we get
the conditions
\begin{align}\label{6cond}
\alpha_{1,i}+2 \alpha_{2,i}+\ldots+a\alpha_{a,i}=n_i'-1-\beta_{2,i}-2\beta_{3,i}-\ldots-(b-1)\beta_{b,i}\ , \ \ i=1,\ldots,d
\end{align}
for the sums over the indices \req{6ia}. This can be explained as follows. Recall, that the l.h.s. represents the number of operators $I(0)$, which are relevant for permutations of arguments $I(0,\ldots,0)$. In general this does not equal $n_i'-1$ as suggested by \req{6om} and the discussions in that paragraph. It depends on which of the arguments $I(1,0,\ldots,0)$ contributes to the previous index $n_{i-1}$. E.g. there are only $n_i-2$ relevant operators in case $I(1,0)$ contributes to $n_{i-1}$ ($i>1$), since the first $I(0)$ in \req{5intrep} comes from the argument $I(1,0)$ and is therefore fixed. This case is represented by $\beta_{2,i}=1$, which indeed gives $n_i-2$ on the r.h.s. of \req{6cond}. In general the number of relevant operators $I(0)$ is $n_i-\nu$ with $I(1)I(0)^{\nu-1}$ contributing to $n_{i-1}$. In accordance with eq. \req{6cond} this is represented by $\beta_{\nu,i}=1$ for $i>1$. 

The function \req{5inner} has a remarkable property. Dropping the conditions for $|\vec{\alpha}_\mu|$, which is equivalent to
summing over all sets $\vec{j}=(j_1,j_2,\ldots,j_a)$, gives
\begin{align}\label{5fibok}
 \sum\limits_{\vec{j}}\omega_{a}(\vec{n};j_1,j_2,\ldots,j_a)=F^{(a)}_{n_1+1}F^{(a)}_{n_2+1}\ldots F^{(a)}_{n_d+1}\ ,
\end{align}
with the generalized Fibonacci numbers:
\begin{align}
F^{(k)}_n=\sum_{\alpha=1}^{k}F^{(k)}_{n-\alpha}\ , \ \ F^{(k)}_1= F^{(k)}_2=1\ , \ \ F^{(k)}_{n\leq0}=0\ .
\end{align}
 The weighting \req{5outer} depends on the indices of the MZVs but not on their order, except for $n_1$.
Furthermore $\omega_{a,1}(\vec{n};j_1,j_2,\ldots,j_a;j_1')$ is equivalent to $\omega_{a}(\vec{n};j_1,j_2,\ldots,j_a)$ as long as both functions are used as weightings in identical sums of MZVs.
A few special cases of the generalized operator product in identity \req{5depgen} are discussed in the following.  

\paragraph{Example 2.1:}
Identity \req{5depgen} with $(a,b)=(2,3)$ is relevant for the generalized hypergeometric function ${}_3F_2$:
\begin{multline}\label{5dep8}
\lefteqn{\left.I(0)\{I(0)^{j_1},I(0,0)^{j_2},I(1)^{j_3},I(1,0)^{j_4},I(1,0,0)^{j_5}\}I(1)\right|_{z=1}}\\
 =\sum\limits_{\substack{
  w=j_1+2j_2+j_3+2j_4+3j_5+2\\ d=j_3+j_4+j_5+1}}\zeta(\vec{n})\; \omega_{2,3}(\vec{n}'-\vec{1};j_2,j_4,j_5)\ ,
\end{multline}
with
\begin{align}\label{5om1c}
 \omega_{2,3}(\vec{n};j_x,j_y,j_z)=\sum\limits_{\substack{\vec{\beta}+\vec{\gamma}\leq \vec{1};~\beta_1=\gamma_1=0\\|\vec{\beta}|=j_y;
 ~|\vec{\gamma}|=j_z}}
 \omega_{2}(\vec{n}-\vec{\beta}-2\vec{\gamma};j_x)\ ,
\end{align}
and
\begin{align}\label{5om1}
 \omega_2(\vec{n};j)=\sum\limits_{\substack{\vec{\alpha}\leq \lfloor\vec{n}/2\rfloor
 \\|\vec{\alpha}|=j}}\binom{\vec{n}-\vec{\alpha}}{\vec{\alpha}}\ .
\end{align}
Eq. \req{5fibok} gives the relation to the Fibonacci
numbers $F_n\equiv F^{(2)}_n$:
\begin{align}\label{5fibo}
 \sum\limits_{j}\omega_{2}(\vec{n};j)=F_{n_1+1}F_{n_2+1}\ldots F_{n_d+1}\ .
\end{align}

Alternatives to the multi--index sum representation \req{5depgen} are possible for generalized operator products with $a=1$, i.e.
the cases where the weighting needs only to consider the permutations of the $I(1,0,\ldots,0)$ operators. As a result the weighting depends on $d_j$ rather than on $n_i$. 

\paragraph{Example 2.2:}
The simplest nontrivial case is:
\begin{align}\label{5dep1}
\left. I(0)\{I(0)^{j_1},I(1)^{j_2},I(1,0)^{j_3}\}I(1)\right|_{z=1}=\sum\limits_{\substack{w=j_1+j_2+2j_3+2\\
 d=j_2+j_3+1}}\zeta(\vec{n})\binom{d-1-d_1}{j_3}\ .
\end{align}
In contrast to $\omega_{1,2}$ the weighting is written as a binomial coefficient without any sums. It can be understood as the number of ways how the operators $I(1,0)$ are distributed among the indices $\vec{n}$. This explains the lower line of the binomial coefficient,
since the number of operators $I(1,0)$ is $j_3$. The upper line of the binomial coefficient represents the number of integers
$n_1,\ldots,n_d$, to which the third argument $I(1,0)$ can contribute. This is the depth $d$ minus one, 
because the $I(1)$ of the MZV index $n_d$ is fixed. Furthermore, one has to subtract $d_1$, since the third argument $I(1,0)$ cannot contribute to $n_{i}$ when $n_{i+1}=1$.

For example $(j_1,j_2,j_3)=(1,1,1)$ gives $w=6$ and $d=3$. One MZV with these properties is $\zeta(3,2,1)$.
Since the $I(1)$ of $n_3$ is fixed, the third argument $I(1,0)$ can only come with
$n_1$ and $n_2$. The latter is not possible, since otherwise the sequence for $n_3$ would start with an $I(0)$ 
and therefore $n_3\geq 2$, which contradicts $n_3=1$. So there is only one way to obtain this MZV with the given
arguments: $I(0,0,(1,0),1,1)$. Parentheses are included to indicate the position of the third argument. This is in 
accordance with the weighting in \req{5dep1}: $\binom{3-1-1}{1}=1$. The same holds for all other MZVs of weight
$w=6$ and depth $d=3$, which have $d_1=1$: $\zeta(3,1,2)$, $\zeta(2,3,1)$ and $\zeta(2,1,3)$.
There is one MZV with $d_1=2$, namely $\zeta(4,1,1)$. The corresponding weighting is zero, since the third argument cannot
contribute to any $n_i$. The last MZV to consider for the given weight and depth is $\zeta(2,2,2)$ with $d_1=0$. 
This one appears twice, since $I(1,0)$ can contribute both to $n_1$ and $n_2$: $I(0,1,0,(1,0),1)+I(0,(1,0),1,0,1)$.
It is easy to check, that this example gives indeed
\begin{multline}\label{5depexp}
 I(0)\{I(0),I(1),I(1,0)\}I(1)\big|_{z=1}\\=\zeta(3,2,1)+\zeta(3,1,2)+\zeta(2,3,1)+\zeta(2,1,3)+2\zeta(2,2,2)\ ,
\end{multline}
in agreement with \req{5dep1}.
\paragraph{Example 2.3:}
A similar relation applies to the case with $I(1,0,0)$ instead of $I(1,0)$:
\begin{align}\label{5dep2}
 \left.I(0)\{I(0)^{j_1},I(1)^{j_2},I(1,0,0)^{j_3}\}I(1)\right|_{z=1}=\sum\limits_{\substack{w=j_1+j_2+3j_3+2\\
 d=j_2+j_3+1}}\zeta(\vec{n})\binom{d-1-d_1-\bar{d}_2}{j_3}\ .
\end{align}
The third argument $I(1,0,0)$ contributing to $n_i$ implies $n_{i+1}\geq 3$. Therefore, the number of indices $n_i$ which can
originate from $I(1,0,0)$ (corresponding to the upper line of the binomial coefficient) is  the total
number $d$ minus one due to $n_d$. In addition the term $(d_1+\bar{d}_2)$, representing the number of integers with
$n_{i+1}<3$, has to be subtracted. The first integer $n_1$ has to be excluded from these considerations simply because there is no preceding 
integer to which $I(1,0,0)$ could contribute. So $\bar{d}_2$, which is the number of indices $n_i$ which equal 2, does not count
$n_1=2$: $\bar{d}_2=d_2-\delta_{2,n_1}$. This distinction is not necessary for $d_1$, as $n_1\neq 1$.

\paragraph{Example 2.4:}
The following relation includes three arguments of the kind $I(1,0,\ldots,0)$:
\begin{multline}\label{5dep3}
 \lefteqn{ \left.I(0)\{I(0)^{j_1},I(1)^{j_2},I(1,0)^{j_3},I(1,0,0)^{j_4}\}I(1)\right|_{z=1}}\\
  =\sum\limits_{\substack{w=j_1+j_2+2j_3+3j_4+2\\
 d=j_2+j_3+j_4+1}}\zeta(\vec{n})\binom{d-1-d_1-j_4}{j_3}
 \binom{d-1-d_1-\bar{d}_2}{j_4}\ .
\end{multline} 
The weighting has a similar explanation as the ones in \req{5dep1} and \req{5dep2}. The first binomial coefficient counts all 
identical terms, which follow from the distribution of the third argument. The second binomial coefficient represents
the same for the fourth argument. It is easy to check that the cases $j_3=0$ and $j_4=0$ reproduce \req{5dep2} and \req{5dep1},
respectively. 

Identities \req{5dep1}, \req{5dep2} and \req{5dep3} give more compact weightings than the ones, which follow from \req{5depgen}.
It is possible to derive these binomial coefficients from the multi--index sums but it is not obvious how to achieve this. Also, note that $j_1=0$ in \req{5dep1} and \req{5dep3} is in accordance with eqs. \req{5ind3} and \req{5ind4}, respectively.

\subsubsection{Identities with sums}

All identities discussed so far are sufficient to write all generalized operator products, which appear in section 3 and 5, in terms of MPLs or MZVs. 
\paragraph{Example 3.1:}
For instance, by using eq. \req{5dep1} it is possible to close the gap in the calculation of the hypergeometric function ${}_2F_1$ in section 3.2:
\begin{align}
\sum\limits_{\alpha}(-1)^\alpha  \left.I(0)\{I(0)^{j_1-\alpha},I(1)^{j_2-\alpha},I(1,0)^{\alpha}\}I(1)\right|_{z=1}
&=\sum\limits_{\alpha}(-1)^\alpha \sum\limits_{\substack{w=j_1+j_2+2\\
 d=j_2+1}}\zeta(\vec{n})\binom{j_2-d_1}{\alpha}\nonumber\\
 &=\sum\limits_{\substack{w=j_1+j_2+2\\
 d=j_2+1}}\zeta(\vec{n})\sum\limits_{\alpha}(-1)^\alpha\binom{j_2-d_1}{\alpha}\nonumber\\
 &=\sum\limits_{\substack{w=j_1+j_2+2\\
 d=j_2+1}}\zeta(\vec{n})\ \delta_{j_2,d_1}\nonumber\\
 &=\zeta(j_1+2,\{1\}^{j_2})\ .\label{5sumex1}
\end{align}
The sum in the last step disappears, since there is only one set of indices $\vec{n}$ with weight
$w=j_1+j_2+2$, depth $d=j_2+1$ and $j_2$ times the index 1. In this simple case it is possible to combine the outer
sum over $\alpha$ with the sum of MZVs to obtain a simple expression. 
However, in \req{43f2part} there are summations over 17 indices \req{4indices}  and performing the evaluation in the same way gives a rather complicated weighting.

\paragraph{Example 3.2:}
Summations over two indices are already problematic:
\begin{align}
 \sum\limits_{\alpha_1,\alpha_2}(-1)^{\alpha_1+\alpha_2}I(0)&\{I(0)^{j_1-\alpha_1-\alpha_2},I(0)^{j_2-\alpha_1},
 I(0,0)^{\alpha_1},\left.I(1)^{j_3-\alpha_2},I(1,0)^{\alpha_2}\}I(1)\right|_{z=1}\nonumber\\[3mm]
 =&\sum\limits_{\alpha_1,\alpha_2}(-1)^{\alpha_1+\alpha_2} \binom{j_1+j_2-2\alpha_1-\alpha_2}{j_2-\alpha_1}\nonumber\\[3mm]
 \times&\;I(0)\{I(0)^{j_1+j_2-2\alpha_1-\alpha_2},I(0,0)^{\alpha_1},\left. I(1)^{j_3-\alpha_2},I(1,0)^{\alpha_2}\}I(1)\right|_{z=1}
 \nonumber\\[3mm]
=&\sum\limits_{\substack{w=j_1+j_2+j_3+2\\d=j_3+1}}\zeta(\vec{n})
 \sum\limits_{\alpha_1,\alpha_2}(-1)^{\alpha_1+\alpha_2}\binom{j_1+j_2-2\alpha_1-\alpha_2}{j_2-\alpha_1}\nonumber\\
 \times&\;\omega_{2,2}(\vec{n}'-\vec{1};j_1+j_2-2\alpha_1-\alpha_2,\alpha_1;j_3-\alpha_2,\alpha_2)\ .\label{5sumex2}
 \end{align}
In the first step \req{2identical} is used to combine the identical arguments.  Applying identity \req{5depgen} in the next step
leads to the given weighting. There is no obvious way how to simplify this expression. Both summations in the first line have
a form similar to the one on the l.h.s. of the first line of \req{5sumex1}. Hence, the question arises, whether  they can be evaluated
as well. The formalism discussed in the following yields indeed a much simpler expression for \req{5sumex2}. 

The general form of the expressions, which are discussed in this section, is
\begin{align}
\sum\limits_{\alpha} (-1)^{\alpha} \{c_1^{j_1-\alpha},c_2^{j_2-\alpha},(c_2 c_1)^{\alpha},\ldots\}\ ,\label{5sumgen}
\end{align}
i.e. one argument is a product of two others and their indices share the same summation index $\alpha$.
The dots represent additional arguments. Their indices may depend on other summation indices but not on $\alpha$. 
Obviously all operator products in \req{5sumgen} contain the same number of $c_1$'s and $c_2$'s independent of $\alpha$.
Thus identical products arise, not only from single generalized operator products due to dependent arguments, but also from 
products with different $\alpha$. The goal is to combine all the identical terms. Due to the factor $(-1)^{\alpha}$ many of these
terms cancel, which leads to simplifications in both the generalized operator product and the summation regions. Eventually, it is
possible to obtain for \req{5sumgen} a compact representation in terms of MZVs, not only for the generalized operator product,
but for the complete expression including the sum.

Obviously, products arising from the generalized product of \req{5sumgen} can only be identical, if they contain the same number 
$m$ of sequences $c_2c_1$. Denoting the sum of all terms, which include $m$ times the sequence $c_2c_1$, by $s_m$, the
$\alpha=0$ operator product of \req{5sumgen} can be written as:
\begin{align}\label{5sum0}
 \{c_1^{j_1},c_2^{j_2},\dots\}=\sum\limits_{m\geq 0}s_m\ .
\end{align}
The sum is over all possible numbers of sequences $c_2c_1$. The next term ($\alpha=1$) gives:
\begin{align}\label{5sum1}
 \{c_1^{j_1-1},c_2^{j_2-1},c_2 c_1,\ldots\}=\sum\limits_{m\geq1}m\ s_m\ .
\end{align}
Here the sum starts with $m=1$, since there is at least one sequence $c_2c_1$ coming from the third argument.
In addition, the summands are weighted by $m$. The reason is, that some terms appear more than once: the sequences
$c_2c_1$, which come from the first two arguments, can be exchanged with the ones coming from the
third argument without changing the product. E.g. for $m=2$ there are terms of the form ($\ldots c_2c_1\ldots (c_2c_1)\ldots$),
where the inner brackets indicate, that the second sequence comes from the third argument. To all of these products, there
is one identical term: ($\ldots (c_2c_1)\ldots c_2c_1\ldots$). This explains the weight 2 for the case $m=2$ in \req{5sum1}. 
For general $\alpha$ and $m$, there are $\alpha$ sequences $c_2c_1$ coming from the third argument, while the
remaining $m-\alpha$ sequences $c_2c_1$ originate from the first two arguments.
This explains the binomial coefficient in the general relation
\begin{align}\label{5sumalpha}
 \{c_1^{j_1-\alpha},c_2^{j_2-\alpha},(c_2 c_1)^\alpha,\ldots\}=\sum\limits_{m\geq\alpha}s_m\binom{m}{\alpha}\ .
\end{align}
Inserting this expression in \req{5sumgen} yields:
\begin{align}
 \sum\limits_{\alpha} (-1)^{\alpha} \{c_1^{j_1-\alpha},c_2^{j_2-\alpha},(c_2 c_1)^{\alpha},\ldots\}
 &=\sum\limits_{m\geq0}s_m \sum\limits_{\alpha=0}^{m} (-1)^{\alpha} \binom{m}{\alpha}\nonumber\\
 &=\sum\limits_{m\geq0}s_m  \delta_{m,0}=s_0\ .\label{5sums0}
\end{align}
Only $s_0$ is left. This is the sum of all products, which do not include the sequence $c_2c_1$.
Thus, summations of the form \req{5sumgen} can be interpreted as restrictions for the sequences of operators, which appear in the 
non--commutative products.

For expressions with more sums of the form \req{5sumgen}, eq. \req{5sums0} has to be applied to each of these individually. It appears,
that the indices of some arguments include more than one summation index. A compact representation is possible, when all
summation indices appear in the first entry of the generalized operator product and therefore all composed arguments contain $c_1$:
\begin{multline}\label{5id}
\sum\limits_{\vec{\alpha}}(-1)^{|\vec{\alpha}|}\{c_1^{j_1-|\vec{\alpha}|},c_2^{j_2-\alpha_1},(c_2c_1)^{\alpha_1},
c_3^{j_3-\alpha_2},(c_3c_1)^{\alpha_2},\ldots,c_n^{j_n-\alpha_{n-1}},(c_nc_1)^{\alpha_{n-1}}\}\\
=c_1^{j_1}\{c_2^{j_2},c_3^{j_3},\ldots,c_n^{j_n}\}\ .
\end{multline}
Applying \req{5sums0} to each summation allows to identify the forbidden sequences $c_2c_1$, $c_3c_1$, ..., $c_nc_1$, i.e.
only those products remain, in which all $c_1$'s appear to the left of all other operators $c_2$, $c_3$, ..., $c_n$.
These are the terms on the r.h.s. of \req{5id}. 
All dependent arguments and the sums over the corresponding indices are
removed. The number of arguments is reduced from $2n-1$ to $n-1$.

The identity \req{5id} provides an alternative to determine example 3.1.
Setting $n=2$, $c_1=I(0)$ and $c_2=I(1)$ gives the relation:
\begin{align}
 \sum\limits_{\alpha}(-1)^\alpha  \left.I(0)\{I(0)^{j_1-\alpha},I(1)^{j_2-\alpha},I(1,0)^{\alpha}\}I(1)\right|_{z=1}=
 I(0)^{j_1+1}I(1)^{j_2+1}\ .
\end{align}
This matches eq. \req{5sumex1}. Also the $n=3$ version of \req{5id} with $c_1=I(0)$, $c_2=I(0)$ and $c_3=I(1)$ can be used to 
evaluate the sums over $\alpha_1$ and $\alpha_2$ in example 3.2. What remains, is:
\begin{align}
 I(0)^{j_1+1}\{I(0)^{j_2},\left.I(1)^{j_3}\}I(1)\right|_{z=1}=\sum\limits_{\substack{w=j_1+j_2+j_3+2\\d=j_3+1;~n_1\geq j_1+2}}
 \zeta(\vec{n})\ .
\end{align}
After the sums are removed, the generalized operator product can be written easily in terms of MZVs by using \req{5ind1}. 
Instead of the complicated weighting in \req{5sumex2}, there is only the additional condition for $n_1$ in the sum of MZVs.

The formalism used to obtain the important relations \req{5sums0} and \req{5id} can be generalized to cases involving general functions
$f(\alpha)$ instead of $(-1)^\alpha$:
 \begin{align}\label{5gen}
\sum\limits_{\alpha\geq \alpha_0}f(\alpha)
\{c_1^{j_1-\alpha},c_2^{j_2-\alpha},(c_2c_1)^{\alpha},\ldots \}=\sum\limits_{m\geq\alpha_0}s_{m}
\sum\limits_{\alpha=\alpha_0}^{m}f(\alpha)\binom{m}{\alpha}\ .
\end{align}
Furthermore, the lower bound $\vec{\alpha}_0$ of the summation is kept general. This allows to handle expressions, where the
indices of composed arguments are not simply summation indices, but depend on other quantities as well. 
Eq. \req{43f2part} has indeed the more general form \req{5gen}, since it contains multinomial coefficients. However, all of them can be 
removed by using eq. \req{2identical}. On the other hand, the resulting expression involves an increased number of arguments. Hence, 
in general one has to decide, whether to handle more complicated functions $f(\alpha)$ or to deal with a larger number of
arguments in the generalized operator product. The latter turned out to be more appropriate for the relations considered in this paper, because in this case
only the factors $(-1)^{\alpha}$ remain, which allows to apply the advantageous relation \req{5sums0}. 

The strategy to simplify  
\req{43f2part} after all multinomial coefficients are removed, is to use shifts in the summation indices to bring as many of the twelve sums
(after application of eqs. \req{4sumreg}) as possible into the form \req{5sums0}.
This allows us to identify all forbidden sequences. However, the configuration for the generalized operator product in
\req{43f2part} is not as convenient as the one in identity \req{5id}. In contrast to $c_1$ in \req{5id}, there is no argument 
of the generalized operator product in \req{43f2part} with an index, that includes all indices of summations of the form 
\req{5sums0}. Furthermore, there are arguments in \req{43f2part}, whose indices do not depend on indices of summations of the 
form \req{5sums0} at all. As a consequence of these two issue, we are not able to present \req{43f2part} in terms of a simplified generalized operator product, as it happens in identity \req{5id}.
Instead the corresponding forbidden sequences are used to write generalized operator products directly in terms of MZVs. This is achieved in a similar manner as for the relations in section 6.1.2. First all
permutations of integral operators, which consist only of $I(0)$, are counted. Then it is analysed how the contribution of operators, which include $I(1)$, affects the weighting.
Thus inner multiple index sums are related to the first step and outer ones to the second step.
This is demonstrated on some examples in the following. We start with generalized operator products with only a few arguments and
minor deviations from the form in \req{5id}, to ultimately present an identity, which can be applied to \req{43f2part}.

\paragraph{Example 3.3:}
A simple case to start with is
\begin{multline}\label{5sumex3}
 \sum\limits_{\alpha}(-1)^{\alpha}\left.I(0)\{I(0)^{j_1-\alpha},I(0)^{j_2-\alpha},I(0,0)^{\alpha},I(1,0)^{j_3}\}I(1)
 \right|_{z=1}\\
 =\sum\limits_{\substack{w=2+j_1+j_2+2j_3\\d=j_3+1}}\zeta(\vec{n})\;\omega_2'(\vec{n}-\vec{2};j_1)\ ,
\end{multline}
with the weighting:
\begin{align}\label{5om2}
 \omega_2'(\vec{n};j)=\sum\limits_{\substack{\vec{\alpha}\leq\vec{n}\\|\vec{\alpha}|=j}}1\ .
\end{align}
The arguments\footnote{Some of the arguments
in the generalized operator products of this and the following examples are identical. Hence, in order to avoid confusion
the argument with the  index $j_i$ is referred to as $c_i$.}, which can contribute to a sequence of operators $I(0)$, are $c_1$ and $c_2$. 
The sum over $\alpha$ removes all products with the sequence $c_2c_1$. Therefore, there is only one way to arrange them:
all $c_1$ to the left of all $c_2$ resulting in the factor of $1$ in the multiple index sum \req{5om2}. 
The sequence of $n_i-1$ operators $I(0)$ related to $n_i$ starts with one $I(0)$ stemming either from $c_3$
for $i>1$ or from the $I(0)$ to the left of the generalized operator product for $i=1$. As a consequence there can be up to $n_i-2$ arguments $c_1$ contributing
to $n_i$. This explains the range of the sum in \req{5om2}, when the arguments of $\omega_2'$ given in identity \req{5sumex3} are inserted. The additional condition for $|\vec{\alpha}|$ takes the fixed number of factors $c_1$ into account. Since the argument $c_3$ is independent of all others, it is irrelevant for the weighting.

\paragraph{Example 3.4:}
The following relation includes two arguments, which are not related to summations:
\begin{multline}\label{5sumex4}
 \sum\limits_{\alpha}(-1)^{\alpha}\left.I(0)\{I(0)^{j_1-\alpha},I(0)^{j_2-\alpha},I(0,0)^{\alpha},I(1)^{j_3},I(1,0)^{j_4}\}
 I(1)\right|_{z=1}\\
 =\sum\limits_{\substack{w=2+j_1+j_2+j_3+2j_4\\d=j_3+j_4+1}}\zeta(\vec{n})\;\omega_{2a}'(\vec{n}-\vec{2};j_1,j_3)\ ,
\end{multline}
with:
\begin{align}\label{5om2a}
 \omega_{2a}'(\vec{n};j_x,j_y)=\sum\limits_{\substack{\vec{\beta}\leq\vec{1}\\|\vec{\beta}|=j_y;~\beta_1=0}}
 \omega_2'(\vec{n}+\vec{\beta};j_x)\ .
\end{align}
The generalized operator product contains the same operators of the type $I(0,\ldots,0)$ as example 3.3. This is why the inner sum uses $\omega_2'$.
The only difference is the range. It can be either $n_i-1$ or $n_i-2$. This depends on wether $c_3$ or $c_4$ contribute to
$n_{i-1}$ ($i>1$). Similar to the identity \req{5depgen}, this is taken into account by the outer sum over the multi--index $\vec{\beta}=(\beta_1,\ldots,\beta_d)$.
Identical terms, which follow from the exchange of arguments $c_3$ and $c_4$ are counted this way. For $\beta_i=1$ the
argument $c_3$ contributes to $n_{i-1}$, while for $\beta_i=0$ $c_4$ does. The condition for $|\vec{\beta}|$ is due to the
fixed number $j_3$ of arguments $c_3$. Again $n_1$ is not affected by these discussions, therefore $\beta_1=0$.

\paragraph{Example 3.5:}
Next, there are two sums of the form \req{5sums0}:
\begin{multline}\label{5sumex5}
 \sum\limits_{\alpha_1,\alpha_2}(-1)^{\alpha_1+\alpha_2}I(0)\{I(0)^{j_1-\alpha_1-\alpha_2},I(0)^{j_2-\alpha_1},
 I(0,0)^{\alpha_1},I(0)^{j_3-\alpha_2},I(0,0)^{\alpha_2},\\
\left. I(1,0)^{j_4}\}I(1)\right|_{z=1}
 =\sum\limits_{\substack{w=2+j_1+j_2+j_3+2j_4\\d=j_4+1;~n_i\geq 2}}\zeta(\vec{n})\;\omega_3'(\vec{n}-\vec{2};j_2,j_3)\ ,
\end{multline}
with:
\begin{align}\label{5om3}
 \omega_3'(\vec{n};j_x,j_y)=\sum\limits_{\substack{\vec{\alpha}+\vec{\beta}\leq\vec{n}\\
 |\vec{\alpha}|=j_x;~|\vec{\beta}|=j_y}}\binom{\vec{\alpha}+\vec{\beta}}{\vec{\beta}}\ .
\end{align}
The relevant operators of the type $I(0,\ldots,0)$ are $c_1$, $c_2$ and $c_3$. The forbidden sequences are $c_2c_1$ and $c_3c_1$. 
So all $c_1$'s have to appear to the left of all the $c_2$'s and $c_3$'s contributing to the same $n_i$. 
The positions of $c_2$ and $c_3$ are not completely fixed, since they may be permuted. With $\alpha_i$ and $\beta_i$ being the
number of factors $c_2$ and $c_3$, respectively, the number of permutations are of course $\binom{\alpha_i+\beta_i}{\alpha_i}$.
The range of the sum is the same as in example 3.3, since the operators of the type $I(1,0,\ldots,0)$ are the same.
The additional conditions in the sum in \req{5om3} are self--explanatory.

\paragraph{Example 3.6:}
This example includes three sums:
\begin{multline}\label{5sumex7}
 \sum\limits_{\alpha_1,\alpha_2,\alpha_3}(-1)^{\alpha_1+\alpha_2+\alpha_3}I(0)\{I(0)^{j_1-\alpha_1-\alpha_2},
 I(0)^{j_2-\alpha_1-\alpha_3},I(0,0)^{\alpha_1},I(0)^{j_3-\alpha_2},I(0,0)^{\alpha_2},\\
 I(1)^{j_4-\alpha_3},
 I(1,0)^{\alpha_3},\left.I(1,0)^{j_5}\}I(1)\right|_{z=1}=\sum\limits_{\substack{w=j_1+j_2+j_3+j_4+2j_5+2\\d=j_4+j_5+1}}\zeta(\vec{n})\;
 \omega_4'\left(\vec{n}-\vec{2};
\begin{matrix} 
j_4\\ 
j_2,~j_3
\end{matrix}
\right)\ ,
\end{multline}
with
\begin{align}\label{5sumom4}
 \omega_4'\left(\vec{n};
\begin{matrix} 
j_x\\ 
j_a,~j_b
\end{matrix}
\right)=\sum\limits_{\substack{\vec{\mu}\leq\vec{1}\\|\vec{\mu}|=j_x;~\mu_1=0}}
 \sum\limits_{\substack{\vec{\alpha}+\vec{\beta}\leq\vec{n}+\vec{\mu}\\|\vec{\alpha}|=j_a;~|\vec{\beta}|=j_b}}
 \binom{\vec{\alpha}+\vec{\beta}+\vec{\mu}(\delta_{\vec{\alpha},0}-1)}{\vec{\alpha}}\ .
\end{align}
Forbidden products are $c_1c_2$, $c_1c_3$ and $c_4c_2$. Therefore, from all arguments contributing to the same sequence
of integral operators $I(0)$, the $c_1$'s appear to the right of all $c_2$'s and $c_3$'s. With $\alpha_i$ and $\beta_i$
being the numbers of $c_2$'s and $c_3$'s, respectively, there are
\begin{align}\label{5bin1}
 \binom{\alpha_i+\beta_i}{\alpha_i}
\end{align}
possibilities to arrange given numbers of $c_1$'s, $c_2$'s and $c_3$'s without the sequences $c_1c_2$ or $c_1c_3$. In case $c_5$ appears
to the left of these operators, the coefficient \req{5bin1} stays the same. But for $c_4$ the third forbidden product $c_4c_2$ has to be
respected. For $\alpha_i>0$ there are
\begin{align}\label{5bin2}
 \binom{\alpha_i+\beta_i-1}{\alpha_i}
\end{align}
possibilities, while for $\alpha_1=0$ the coefficient remains as in \req{5bin1}. Thus for all $\alpha_i$ there are
\begin{align}\label{5bin3}
 \binom{\alpha_i+\beta_i+\delta_{\alpha_i,0}-1}{\alpha_i}
\end{align}
permutations of given numbers of $c_1$'s, $c_2$'s and $c_3$'s with $c_4$ appearing to the left and without the sequences $c_1c_2$, $c_1c_3$ and $c_4c_2$. Through the sums over the multi--index $\vec{\mu}=(\mu_1,\ldots,\mu_d)$ both cases are taken into account: $\mu_i=1$ represents $c_4$ contributing to $n_{i-1}$ $(i>1)$, which yields the coefficient \req{5bin3} in \req{5sumom4}. On the other hand $\mu_i=0$ represents $c_5$ contributing to $n_{i-1}$ $(i>1)$, which gives the coefficient \req{5bin1}.
Also the range of the inner sums depends on wether $c_4$ or $c_5$ contributes to $n_{i-1}$. It is $n_i-1$ for the former and $n_i-2$ for the latter ($i>1$). The additional conditions for the sums in \req{5sumom4} are there for the same reasons as the ones in the previous examples.

\paragraph{Example 3.7:}
The following expression involves sums over six indices $\vec{\alpha}=(\alpha_1,\ldots,\alpha_6)$:
\begin{align}
 &\sum\limits_{\vec{\alpha}}(-1)^{|\vec{\alpha}|}I(0)\{I(0)^{j_1-\alpha_1-\alpha_2-\alpha_5},I(0)^{j_2-\alpha_3-\alpha_4},
 I(0)^{j_3-\alpha_1-\alpha_3},I(0)^{j_4-\alpha_2-\alpha_4-\alpha_6},\nonumber\\
 &I(0,0)^{\alpha_1},I(0,0)^{\alpha_2},I(0,0)^{\alpha_3},I(0,0)^{\alpha_4},I(1,0)^{j_5-\alpha_5},I(1,0,0)^{\alpha_5},
 I(1,0)^{j_6-\alpha_6},\nonumber\\
 &I(1,0,0)^{\alpha_6},\left.I(1,0)^{j_7}\}I(1)\right|_{z=1}\nonumber\\
 =&\sum\limits_{\substack{w=2+j_1+j_2+j_3+j_4+2j_5+2j_6+2j_7\\ d=1+j_5+j_6+j_7}}
 \zeta(\vec{n})\;\omega_5'\left(\vec{n}-\vec{2};
\begin{matrix} 
j_5,~j_6\\ 
j_1,~j_2,~j_3,~j_4
\end{matrix}
\right)\ ,\label{5sumex8}
\end{align}
with the weighting
\begin{multline}\label{5om5}
 \omega_5'\left(\vec{n};
\begin{matrix} 
j_x,~j_y\\ 
j_a,~j_b,~j_c,~j_d
\end{matrix}
\right)\\
 =\sum\limits_{\substack{\vec{\mu}+\vec{\nu}\leq \vec{1}\\|\vec{\mu}|=j_x;~|\vec{\nu}|=j_y\\
 \alpha_1,\beta_1=0}}\sum\limits_{\substack{\vec{\alpha}+\vec{\beta}+\vec{\gamma}+\vec{\delta}=\vec{n}\\
 |\vec{\alpha}|=j_a;~|\vec{\beta}|=j_b\\|\vec{\gamma}|=j_c;~|\vec{\delta}|=j_d}}
 \binom{\vec{\alpha}+\vec{\beta}+\vec{\mu}(\delta_{\vec{\alpha},0}-1)}{\vec{\alpha}}
 \binom{\vec{\gamma}+\vec{\delta}+\vec{\nu}(\delta_{\vec{\delta},0}-1)}{\vec{\delta}}\ .
\end{multline}
The forbidden sequences related to the six summations are $c_3c_1$, $c_4c_1$, $c_3c_2$, $c_4c_2$, $c_5c_4$ and 
$c_6c_1$. The first four sequences exclusively affect the operators $I(0)$ and are therefore relevant for the inner sums. 
A sequence of operators $I(0)$ consisting of the arguments $c_1$, $c_2$, $c_3$ and $c_4$ without the forbidden sequences 
has to start with all $c_1$ and all $c_2$. All permutations of these two factors are allowed. Also all permutations
of $c_3$ and $c_4$ are allowed. Hence, there are
\begin{align}\label{5bin4}
 \binom{\alpha_i+\beta_i}{\alpha_i}\ \binom{\gamma_i+\delta_i}{\delta_i}
\end{align}
possibilities to build this sequence with $\alpha_i$, $\beta_i$, $\gamma_i$ and $\delta_i$ being the numbers
of factors $c_1$, $c_2$, $c_3$ and $c_4$, respectively. 
Some of the forbidden sequences interfere with each other: $c_5c_4$ with $c_4c_{1}$ and $c_4c_{2}$. This means, that products, in which these
sequences are combined ($c_5c_4c_1$ and $c_5c_4c_2$), do not just vanish but they appear with a negative sign. In other words, 
an expression, which simply ignores the sequences $c_5c_4$,  $c_4c_{1}$ and $c_4c_{2}$ involves too many products. An
elegant way to solve this problem, is to introduce the additional forbidden sequences $c_{1}c_4$ and $c_{2}c_4$, which have
to be considered, if and only if $c_5$ contributes to $n_i$. This is possible because $c_4=c_1=c_2$.

The freedom of choosing the forbidden sequences $c_2 c_1$ or $c_1c_2$ for  $c_2=c_1$ is used in the previous examples to avoid interfering forbidden sequences. 
However, this manipulation is not possible for the 
example under consideration.

The coefficient within the inner sums of \req{5om5} depends on which argument of the type $I(1,0,\ldots,0)$ contributes to $n_{i-1}$: $c_5$, $c_6$ or $c_7$. The upper bound of the inner sums is $n_i-2$ in all three cases. For $c_7$ the coefficient is identical
to \req{5bin4}. For $c_6$ those contributions have to be subtracted, which start with $c_1$. For $\alpha_i>0$ the first binomial
coefficient in \req{5bin4} changes to \req{5bin2}. Using Kronecker deltas this can  be written for all $\alpha_i$ as \req{5bin3}.
For $c_5$ the second binomial coefficient in \req{5bin4} has to be modified. In case $\delta_i>0$ there are
\begin{align}\label{5bin5}
 \binom{\gamma_i-1+\delta_i}{\delta_i}
\end{align}
possible permutations of $c_3$ and $c_4$. Therefore, for all $\delta_i$ the second binomial coefficient becomes:
\begin{align}\label{5bin6}
 \binom{\gamma_i+\delta_{\delta_i,0}-1+\delta_i}{\delta_i}\ .
\end{align}
All these cases are respected by the summations over the $d$--dimensional multi--indices $\vec{\mu}$ and $\vec{\nu}$.
For $\mu_i= \nu_i=0$ the binomial 
coefficients remain as in \req{5bin4}, so this represents the contribution of $c_7$ to $n_{i-1}$. The first binomial coefficient changes to
\req{5bin3} for $\nu_i=1$, while the second becomes \req{5bin6} for $\mu_i=1$, thus representing the contributions of $c_6$ and
$c_5$, respectively.

Note, that the introduction of the additional forbidden sequences leads to the symmetric form of the binomial
coefficients in \req{5om5}, which is in agreement with the symmetry of the corresponding generalized operator product. 
Of course one of the multiple indices $\vec{\alpha}$, $\vec{\beta}$, $\vec{\gamma}$ and $\vec{\delta}$
can be removed, e.g. to obtain the summation region $\vec{\alpha}+\vec{\beta}+\vec{\gamma}\leq\vec{n}$ for the
inner sum. This would however destroy the symmetric form of the binomial coefficients.

\paragraph{Example 3.8:} This is the most general relation, which includes the previous examples 3.3--3.7 as special cases:
\begin{align}
 \sum\limits_{\vec{\alpha}=(\alpha_1,\alpha_2,\ldots,\alpha_9)}&(-1)^{|\vec{\alpha}|}I(0)
 \{I(0)^{j_1-\alpha_1-\alpha_2-\alpha_5-\alpha_7-\alpha_8}, I(0)^{j_2-\alpha_3-\alpha_4},\nonumber\\
 & I(0)^{j_3-\alpha_1-\alpha_3},I(0)^{j_4-\alpha_2-\alpha_4-\alpha_6-\alpha_9},I(0,0)^{\alpha_1},I(0,0)^{\alpha_2},
 I(0,0)^{\alpha_3},I(0,0)^{\alpha_4},\nonumber\\
 &I(1,0)^{j_5-\alpha_5},I(1,0,0)^{\alpha_5},I(1,0)^{j_6-\alpha_6},I(1,0,0)^{\alpha_6},I(1)^{j_7-\alpha_7-\alpha_9},
 I(1,0)^{\alpha_7},\nonumber\\
 &I(1,0)^{\alpha_9-\alpha_8},I(1,0,0)^{\alpha_8},\left.I(1,0)^{j_8}\}I(1)\right|_{z=1}\nonumber\\
 =&\sum\limits_{\substack{w=j_1+\ldots+j_5+2j_6+2j_7+2j_8+2\\d=j_5+j_6+j_7+j_8+1}}\omega_6'\left(\vec{n}-\vec{2};
\begin{matrix} 
j_5,~j_6,~j_7\\ 
j_1,~j_2,~j_3,~j_4
\end{matrix}
\right)\ ,\label{5sumex9}
\end{align}
with:
\begin{multline}\label{5om6}
 \omega_6'\left(\vec{n};
\begin{matrix} 
j_x,~j_y,~j_z\\ 
j_a,~j_b,~j_c,~j_d
\end{matrix}
\right)=
 \sum\limits_{\substack{\vec{\mu}+\vec{\nu}+\vec{\sigma}\leq\vec{1}\\|\vec{\mu}|=j_x;~ |\vec{\nu}|=j_y
;~ |\vec{\sigma}|=j_z\\
\mu_1,\nu_1,\sigma_1=0}}
\sum\limits_{\substack{\vec{\alpha}+\vec{\beta}+\vec{\gamma}+\vec{\delta}=\vec{n}+\vec{\sigma}\\
|\vec{\alpha}|=j_a;~ |\vec{\beta}|=j_b\\ |\vec{\gamma}|=j_c;~|\vec{\delta}|=j_d}}\\
\times\binom{\vec{\alpha}+\vec{\beta}+(\vec{\mu}+\vec{\sigma})(\delta_{\vec{\alpha},0}-1)}{ \vec{\alpha}}
\binom{\vec{\gamma}+\vec{\delta}+(\vec{\nu}+\vec{\sigma})(\delta_{\vec{\delta},0}-1)}{ \vec{\delta}}\ .
\end{multline}
Forbidden sequences are $c_3c_1$, $c_4c_1$, $c_3c_2$, $c_4c_2$, $c_5c_1$, $c_6c_4$, $c_7c_1$ and $c_7c_4$. 
The arguments of the type $I(0,\ldots,0)$ are the same as in example 3.7. Hence, the number of possibilities \req{5bin4} 
for a sequence of operators $I(0)$ apply here as well. There are four arguments of the type $I(1,0,\ldots,0)$: 
$c_5$, $c_6$, $c_7$ and $c_8$. The range of the inner sums is $n_i-1$ for $c_7$ and $n_i-2$ for all others. 
For $c_8$ the coefficient \req{5bin4} is unaffected. For $c_5$ the first binomial coefficient changes again to \req{5bin3} and
for $c_6$ the second one changes again to \req{5bin6}. For $c_7$ both forbidden sequences including $c_5$ and $c_6$ are combined,
so both binomial coefficients change to \req{5bin3} and \req{5bin6}, respectively. All these modifications are taken into account 
in \req{5om6}.

The following relations hold in case the functions on both sides are used as weightings within identical sums of MZVs.
These relations provide a consistency check, because they result both from the definitions of the weightings
and the corresponding generalized operator products: 
\begin{align}\label{5om6limit}
\begin{split}
 \omega_6' \left(\vec{n};
\begin{matrix} 
j_x,~j_y,~0\\ 
j_a,~j_b,~j_c,~j_d
\end{matrix}
\right) &= \omega_5'\left(\vec{n};
\begin{matrix} 
j_x,~j_y\\ 
j_a,~j_b,~j_c,~j_d
\end{matrix}
\right)\\
 \omega_6' \left(\vec{n};
\begin{matrix} 
0,~0,~j_z\\ 
j_a,~j_b,~j_c,~0
\end{matrix}
\right) &= \omega_4' \left(\vec{n};
\begin{matrix} 
j_z\\ 
j_a,~j_b
\end{matrix}
\right)\\
\omega_6' \left(\vec{n};
\begin{matrix} 
0,~0,~0\\ 
j_a,~j_b,~j_c,~0
\end{matrix}
\right) &= \omega_3'(\vec{n};j_a,j_b)\\
\omega_6' \left(\vec{n};
\begin{matrix} 
0,~0,~j_z\\ 
0,~j_b,~j_c,~0
\end{matrix}
\right) &= \omega_{2a}'(\vec{n};j_b,j_z)\\
\omega_6' \left(\vec{n};
\begin{matrix} 
0,~0,~0\\ 
0,~j_b,~j_c,~0
\end{matrix}
\right) &= \omega_2'(\vec{n};j_b)\ .
\end{split}
\end{align}
Furthermore the following symmetry holds:
\begin{align}\label{5om6sym}
 \omega_6' \left(\vec{n};
\begin{matrix} 
j_x,~j_y,~j_z\\ 
j_a,~j_b,~j_c,~j_d
\end{matrix}
\right) =\omega_6' \left(\vec{n};
\begin{matrix} 
j_y,~j_x,~j_z\\ 
j_d,~j_c,~j_b,~j_a
\end{matrix}
\right) \ .
\end{align}

\subsection{Applications}
Two methods to get from expression \req{32ndstep1} to \req{32ndstep2} for the hypergeometric function ${}_2F_1$
have been demonstrated in subsection 6.1.3 by using either 
identity \req{5dep1} or \req{5id}. 

Identity \req{5dep8} allows to express the all--order expansion \req{3result} of the $p=3$ 
hypergeometric function in terms of MPLs:
\begin{multline}\label{5fin1}
  v_k(z)=\sum\limits_{l_1+m_1+2(l_2+m_2)+3m_3=k-3}(-1)^{l_1+l_2}\Delta_1^{m_1}\Delta_2^{m_2}
  \Delta_3^{m_3+1}Q_1^{l_1}Q_2^{l_2}\\
\times\sum\limits_{\substack{
  w=k;~n_1\geq3\\ d=m_1+m_2+m_3+1}}{\cal L}i_{\vec{n}}(z)\;\omega_{2,3}(\vec{n}''-\vec{1};l_2,m_2,m_3)\ ,
\end{multline}
with $\omega_{2,3}$ defined in \req{5om1c}, $n_1''=n_1-2$ and $n_i''=n_i$ for $i=2,3,\ldots,d$.

Eq. \req{5depgen} leads to the following representation in terms of MPLs for the coefficient functions \req{qresult} of
${}_pF_{p-1}$:
\begin{multline}
 w_k^p(z)=\sum\limits_{\vec{l},\vec{m}}(-1)^{|\vec{l}|}(\Delta_1^p)^{m_1}(\Delta_2^p)^{m_2}\ldots(\Delta_{p-1}^p)^{m_{p-1}}
 (\Delta_p^p)^{m_p+1}(Q_1^p)^{l_1}(Q_2^p)^{l_2}\ldots (Q_{p-1}^p)^{l_{p-1}}\\
 \times\sum\limits_{\substack{w=k;~n_1\geq p\\d=m_1+m_2+\ldots+m_p+1}}{\cal L}i_{\vec{n}}(z)\;
  \omega_{p-1,p}(\vec{n}^*-\vec{1};l_1,l_2,\ldots,l_{p-1};m_1,m_2,\ldots,m_p)\ ,
\end{multline}
with the weighting $\omega_{p-1,p}$ defined in \req{5outer}, $n_1^*=n_1-(p-1)$ and $n_i^*=n_i$ for $i=2,3,\ldots,d$.

The expression \req{43f2con}, which enters the all--order expansions \req{4f11} and \req{4f22} of the 5--point open string
amplitude, can now be written in terms of MZVs as:
\begin{multline}\label{5fin2}
  \left.\frac{\theta}{\Delta_3}v_k(z)\right|_{z=1}=\sum\limits_{l_1+m_1+2(l_2+m_2)+3m_3
 =k-3}(-1)^{l_1+l_2}\Delta_1^{m_1}\Delta_2^{m_2}\Delta_3^{m_3}Q_1^{l_1}Q_2^{l_2}\\
\times \sum\limits_{\substack{ w=k-1\\ d=m_1+m_2+m_3+1}}\zeta(\vec{n})\;\omega_{2,3}(\vec{n}'-\vec{1};l_2,m_2,m_3)\ .
\end{multline}
For the alternative representation \req{4f2alt} of the 5--point string amplitude we use eqs. \req{4sumreg} in \req{43f2part} 
to evaluate five of the 17 sums. Some shifts in the remaining 12 summation indices allow to bring nine summations into the form 
\req{5sums0}. Then it is possible to apply the identity \req{5sumex9} to arrive at:
\begin{align}
 &v(j_1,j_2,j_3,j_4,j_5)\nonumber\\
 ={}&\sum\limits_{\vec{\beta},\vec{\delta},\vec{\epsilon}}(-1)^{|\vec{j}|+\delta_3+
 |\vec{\beta}|+\delta_2+\delta_5+|\vec{\epsilon}|}I(0)\{I(0)^{j_1-\delta_1-\beta_1-\beta_2-\delta_2-\epsilon_2-\epsilon_3},
 I(0)^{j_2-\delta_1-\delta_3-\beta_3-\beta_4},\label{5op}\\
 & I(0)^{j_3-\delta_3-\delta_4-\beta_1-\beta_3},
 I(0)^{j_4-\delta_4-\delta_5-\beta_2-\beta_4-\epsilon_1},I(0,0)^{\beta_1},I(0,0)^{\beta_2},I(0,0)^{\beta_3},I(0,0)^{\beta_4},\nonumber\\
 & I(1,0)^{\delta_4-\epsilon_2},I(1,0,0)^{\epsilon_2},I(1,0)^{\delta_1-\epsilon_1},I(1,0,0)^{\epsilon_1},I(1)^{j_5-\delta_5-\delta_2},I(1,0)^{\delta_2},
 I(1,0)^{\delta_5-\epsilon_3},\nonumber\\
 &\left.I(1,0,0)^{\epsilon_3},I(1,0)^{\delta_3}\}I(1)\right|_{z=1}\nonumber\\
 ={}&(-1)^{|\vec{j}|}\sum\limits_{\delta_1,\delta_3,\delta_4}(-1)^{\delta_3} \sum\limits_{\substack{w=j_1+j_2+j_3+j_4+j_5+2\\
 d=j_5+\delta_1+\delta_3+\delta_4+1}}\zeta(\vec{n})\nonumber\\
 \times&\;\omega_6'\left(\vec{n}-\vec{2};
\begin{matrix} 
\delta_4,~\delta_1,~j_5\\ 
j_1{-}\delta_1,~j_2{-}\delta_1{-}\delta_3,~j_3{-}\delta_3{-}\delta_4,~j_4{-}\delta_4,~j_5
\end{matrix}
\right)\nonumber\\
 ={}&(-1)^{|\vec{j}|}\sum\limits_{w=j_1+j_2+j_3+j_4+j_5+2}\zeta(\vec{n})\sum\limits_{\delta_1,\delta_4}
 (-1)^{d-1-j_5-\delta_1-\delta_4} \nonumber\\
 \times&\;\omega_6'\left(\vec{n}-\vec{2};
\begin{matrix} 
\delta_4,~\delta_1,~j_5\\ 
j_1{-}\delta_1,~j_2{+}j_5{+}\delta_4{-}d{+}1,~j_3{+}j_5{+}\delta_1{-}d{+}1,~j_4{-}\delta_4
\end{matrix}
\right)\ .\label{5g}
\end{align}
The weighting $\omega_6'$ is defined in \req{5om6}. In the second step the three sums over $\delta_1$, $\delta_3$ and $\delta_4$
are combined with the sum of MZVs. 
Apart from the multi--index sums in $\omega_6'$ only two of the sums over the indices \req{4indices} remain.

The transformations of the generalized operator products to sums of MPLs or MZVs in the all--order expressions for 
$v_{k}$, $w_{k}^p$ and $v(j_1,j_2,j_3,j_4,j_5)$ as presented in eqs. \req{5fin1} -- \req{5fin2} and \req{5g}
involve weightings, which are rather complicated, due to
to the variety of multiple index sums and the large number of conditions therein. 
On one hand the expressions presented in this section have the advantage, that they allow to pick a specific MPL or MZV, respectively, and directly determine the factor they appear with via the weighting.
On the other hand the corresponding representations \req{3result}, \req{qresult}, \req{43f2con} and \req{43f2part} in terms of generalized operator products provide more compact alternatives.

We showed in the previous section that there are identities for generalized operator products, which yield sums of MZV with less complicated weightings, e.g. binomial coefficients. Our objective in the following is to use such identities for limits of the function $v(j_1,j_2,j_3,j_4,j_5)$ with some of
its arguments $j_1,j_2,j_3,j_4,j_5$ set to zero. Not only is it interesting to see how certain generalized operator products simplify significantly this way, but also will we use these results in the next section to identify particular MZV identities. These limits do not follow directly from \req{5g}. Going instead one step backwards to the generalized operator product in \req{5op} allows to use different identities than \req{5sumex9} (in many cases eq. \req{5id}). For completeness and as a consistency check we give the MZV representation for all limits.
To reduce the number of limits to be calculated we use
\begin{align}\label{6sym1}
v(j_1,j_2,j_3,j_4,j_5)=v(j_4,j_3,j_2,j_1,j_5)\ .
\end{align}
This symmetry can easily be proven using for instance \req{43f2part} or the symmetry \req{5om6sym} of the weighting $\omega_6'$ in \req{5g}. Those cases, that involve weightings of similar complexity as $\omega_6'$, i.e. weighting with multiple index sums, are summarized in the appendix.

\paragraph[Four $j_i$ set to zero:]{Four $\bm{j_i}$ set to zero:} We start with the simplest limits, where four of the arguments of $v(j_1,j_2,j_3,j_4,j_5)$ are zero. These calculations are trivial and do not require any identities of section 6.1. No generalized operator products remain and therefore the integral operators can easily be written in terms of MZVs using \req{3mzv}. Setting $j_2=j_3=j_4=j_5=0$ in \req{5op} yields
\begin{align}\label{64e1}
 v(j_1,0,0,0,0)=(-1)^{j_1}\left.I(0)^{j_1+1}I(1)\right|_{z=1}=(-1)^{j_1}\zeta(j_1+2).
\end{align}
The case $j_1=j_3=j_4=j_5=0$ results in the same MZV and we can then use the symmetry \req{6sym1} to determine $v(0,0,j_3,0,0)$ and $v(0,0,0,j_4,0)$:
\begin{align}
v(j_1,0,0,0,0)=v(0,j_1,0,0,0)=v(0,0,j_1,0,0)=v(0,0,0,j_1,0)\ .
\end{align}
The limit $j_1=j_2=j_3=j_4=0$ in \req{5op} gives
\begin{align}\label{64e3}
 v(0,0,0,0,j_5)=(-1)^{j_5}\left.I(0)I(1)^{j_5+1}\right|_{z=1}=(-1)^{j_5}\zeta(2,\{1\}^{j_5})\ .
\end{align}

\paragraph[Three $j_i$ set to zero:]{Three $\bm{j_i}$ set to zero:} There are ten different cases with three arguments set to zero. For $j_3=j_4=j_5=0$ we get:
\begin{align}
 v(j_1,j_2,0,0,0)&=(-1)^{j_1+j_2}\sum\limits_{\delta_1}\binom{j_1+j_2-2\delta_1}{j_1-\delta_1}
 I(0)\{\left.I(0)^{j_1+j_2-2\delta_1},I(1,0)^{\delta_1}\}I(1)\right|_{z=1}\nonumber\\
 &=(-1)^{j_1+j_2}\sum\limits_{\delta_1}\binom{j_1+j_2-2\delta_1}{j_1-\delta_1}
 \sum\limits_{\substack{w=j_1+j_2+2\\d=\delta_1+1;~n_i\geq2}}\zeta(\vec{n})\nonumber\\
 &=(-1)^{j_1+j_2}\sum\limits_{\substack{w=j_1+j_2+2\\n_i\geq2}}\zeta(\vec{n})
 \binom{w-2d}{j_1+1-d}\ .\label{612}
\end{align}
We used identity \req{5ind2} in the first step and combined the sum over $\delta_1$ with the sum of MZVs in the last line. For $j_2=j_3=j_4=0$ the $n=2$ version of identity \req{5id} can be used to obtain:
\begin{align}
v(j_1,0,0,0,j_5)&=(-1)^{j_1+j_5}\sum\limits_{\delta_2}(-1)^{\delta_2}I(0)\{I(0)^{j_1-\delta_2},\left.
I(1)^{j_5-\delta_2},I(1,0)^{\delta_2}\}I(1)\right|_{z=1}\nonumber\\
&=(-1)^{j_1+j_5}\left.I(0)^{j_1+1}I(1)^{j_5+1}\right|_{z=1}=(-1)^{j_1+j_5}\zeta(j_1+2,\{1\}^{j_5})\ .\label{615}
\end{align}
The same identity applies to the limit $j_2=j_4=j_5=0$, thus we get:
\begin{align}
 v(j_1,0,j_3,0,0)&=(-1)^{j_1+j_3}\sum\limits_{\beta_1}(-1)^{\beta_1}I(0)\{I(0)^{j_1-\beta_1},\left.
 I(0)^{j_3-\beta_1},I(0,0)^{\beta_1}\}I(1)\right|_{z=1}\nonumber\\
 &=(-1)^{j_1+j_3}\left.I(0)^{j_1+j_3+1}I(1)\right|_{z=1}=(-1)^{j_1+j_3}\zeta(j_1+j_3+2)\ .\label{613}
\end{align}
The calculations for the case $j_2=j_3=j_5=0$ take the same steps as for $ v(j_1,0,j_3,0,0)$, so that:
\begin{align}
v(j_1,0,0,j_4,0)=v(j_1,0,j_4,0,0)\ .
\end{align}
Identity \req{5ind1} is useful for the limit $j_1=j_3=j_4=0$ : 
\begin{align}\label{5sub0}
  v(0,j_2,0,0,j_5)&=(-1)^{j_2+j_5} I(0)\{I(0)^{j_2}\left.,I(1)^{j_5}\}I(1)\right|_{z=1}\nonumber\\
  &=(-1)^{j_2+j_5}\sum\limits_{\substack{w=j_2+j_5+2\\d=j_5+1}}\zeta(\vec{n}).
\end{align}
Obtaining the MZV representation for $v(0,j_2,j_3,0,0)$ requires the identity \req{5sumex3} and therefore involves the multi--index sum $\omega_2'$. The result for $v(0,j_2,j_3,0,0)$ is given in \req{a23}. The limits $v(0,0,j_3,j_4,0)$, $v(0,0,0,j_4,j_5)$, $v(0,j_2,0,j_4,0)$ and $v(0,0,j_3,0,j_5)$ follow from eqs. \req{612}, \req{615}, \req{613} and \req{5sub0}, respectively, through the symmetry \req{6sym1}.

\paragraph[Two $j_i$ set to zero:]{Two $\bm{j_i}$ set to zero:} There are ten different cases with two $j_i$ set to zero. Setting $j_2=j_3=0$ in \req{5op} yields:  
\begin{align}\label{5sub3}
 v(j_1,0,0,j_4,j_5)
 ={}&(-1)^{j_1+j_4+j_5}\sum\limits_{\delta_5,\beta_2,\delta_2,\epsilon_3}(-1)^{\delta_5+\beta_2+\delta_2+\epsilon_3}
 I(0)\{I(0)^{j_1-\beta_2-\delta_2-\epsilon_2},I(0)^{j_4-\delta_5-\beta_2},\nonumber\\
 &I(0,0)^{\beta_2},
 I(1)^{j_5-\delta_5-\delta_2},I(1,0)^{\delta_2},\left.I(1,0)^{\delta_5-\epsilon_3},I(1,0,0)^{\epsilon_3}\}
 I(1)\right|_{z=1}\nonumber\\
 ={}&(-1)^{j_1+j_4+j_5}\sum\limits_{\delta_5}(-1)^{\delta_5}I(0)^{j_1+1}\{I(0)^{j_4-\delta_5},\left.I(1)^{j_5-\delta_5},
 I(1,0)^{\delta_5}\}I(1)\right|_{z=1}\nonumber\\
 ={}&(-1)^{j_1+j_4+j_5} \left.I(0)^{j_1+j_4+1}I(1)^{j_5+1}\right|_{z=1}\nonumber\\
 ={}&(-1)^{j_1+j_4+j_5} \zeta(j_1+j_4+2,\{1\}^{j_5})\ .
\end{align}
The identity \req{5id} had to be applied two times. In the first step with $n=4$ and in the second with $n=2$. 
This limit demonstrates how valuable the identities derived in section 6.1 can be. Starting with four sums of generalized 
operator products involving seven aguments only a single MZV remains in \req{5sub3}.
For the limit $j_3=j_4=0$ identity \req{5dep1} leads to
\begin{align}\label{5sub1}
 v(j_1,j_2,0,0,j_5)={}&(-1)^{j_1+j_2+j_5}\sum\limits_{\delta_1,\delta_2}(-1)^{\delta_1+\delta_2}
 \binom{j_1+j_2-\delta_1-\delta_2}{j_2-\delta_1}\binom{\delta_2}{\delta_1}I(0)\nonumber\\
 \times&\;\{I(0)^{j_1+j_2-\delta_1-\delta_2},\left.I(1)^{j_5+\delta_1-\delta_2},I(1,0)^{\delta_2}\}I(1)\right|_{z=1}\nonumber\\
 ={}&(-1)^{j_1+j_2+j_5}\sum\limits_{\delta_1=0}^{\min\{j_1,j_2\}}\sum\limits_{\substack{w=j_1+j_2+j_5+2\\d=j_5+\delta_1+1}}
 \zeta(\vec{n})\;\omega(j_1{+}1,j_2{+}1,d{-}d_1,\delta_1{+}1)\nonumber\\
 ={}&(-1)^{j_1+j_2+j_5}\sum\limits_{\substack{w=j_1+j_2+j_5+2\\j_5< d\leq j_5+1+\min\{j_1,j_2\}}}
 \zeta(\vec{n})\;\omega(j_1{+}1,j_2{+}1,d{-}d_1,d{-}j_5)\ ,
\end{align}
with the weighting\footnote{To write summation regions for sums of MZVs more compact, we set binomial coefficients to zero for negative arguments (cf. fn. 3 on p. 8). This is, however, not sufficient in \req{5sub1}, since the hypergeometric function in the weighting \req{5om0} has singularities in the region, where the binomial coefficients are set to zero. To ensure convergence we introduce explicit bounds of summation for the sum over $\delta_1$ in the third line of \req{5sub1}. This leads to the condition $j_5{+}1\leq d\leq j_5{+}1{+}\min\{j_1,j_2\}$ for the sum of MZVs in the last line of \req{5sub1}.}
\begin{align}\label{5om0}
 \omega(j_x,j_y,\delta_x,\delta_y)=\binom{\delta_x-1}{\delta_y-1}\binom{j_x+j_y-2\delta_y}{j_x-\delta_y}
 {}_2F_1\left[
\begin{matrix} 
\delta_y-j_x,~\delta_y-\delta_x\\ 
2\delta_y-j_x-j_y
\end{matrix}
;1\right].
\end{align}
For $j_3=j_5=0$ we obtain:
\begin{align}
 &v(j_1,j_2,0,j_4,0)\nonumber\\
 ={}&(-1)^{j_1+j_2+j_4}\sum\limits_{\delta_1,\beta_2,\beta_4,\epsilon_1}(-1)^{
 \beta_2+\beta_4+\epsilon_1}I(0)\{I(0)^{j_4-\beta_2-\beta_4-\epsilon_1},I(0)^{j_1-\delta_1-\beta_2},I(0,0)^{\beta_2},
 \nonumber\\&\left.I(0)^{j_2-\delta_1-\beta_4},
 I(0,0)^{\beta_4},I(1,0)^{\delta_1-\epsilon_1},I(1,0,0)^{\epsilon_1}\}I(1)\right|_{z=1}\nonumber\\
 ={}&(-1)^{j_1+j_2+j_4}\sum\limits_{\delta_1}\binom{j_1+j_2-2\delta_1}{j_1-\delta_1}I(0)^{j_4+1}
 \{\left.I(0)^{j_1+j_2-2\delta_1},I(1,0)^{\delta_1}\}I(1)\right|_{z=1}\nonumber\\
 ={}&(-1)^{j_1+j_2+j_4}\sum\limits_{\delta_1}\binom{j_1+j_2-2\delta_1}{j_1-\delta_1}
 \sum\limits_{\substack{w=j_1+j_2+j_4+2\\d=\delta_1+1;~n_1\geq j_4+2;~n_i\geq2}}\zeta(\vec{n})\nonumber\\
 ={}&(-1)^{j_1+j_2+j_4}\sum\limits_{\substack{w=j_1+j_2+j_4+2\\n_1\geq j_4+2;~n_i\geq2}}\zeta(\vec{n})
 \binom{w-j_4-2d}{j_1-d+1}\ .\label{6124}
\end{align}
The first step requires the $n=4$ version of \req{5id} and the second step eq. \req{5ind2}. Eventually the sum over $\delta_1$ is combined with the sum of MZVs.
With $j_2=j_4=0$ eq. \req{5op} becomes:
\begin{align}\label{5sub4}
  &v(j_1,0,j_3,0,j_5)\nonumber\\
  ={}&(-1)^{j_1+j_3+j_5}\sum\limits_{\beta_1,\delta_2}(-1)^{\beta_1+\delta_2} I(0)\{I(0)^{j_1-\beta_1-
  \delta_2},I(0)^{j_3-\beta_1},I(0,0)^{\beta_1},\left.I(1)^{j_5-\delta_2},I(1,0)^{\delta_2}\}I(1)\right|_{z=1}\nonumber\\
  ={}&(-1)^{j_1+j_3+j_5}I(0)^{j_1+1}\{I(0)^{j_3},\left.I(1)^{j_5}\}I(1)\right|_{z=1}\nonumber\\
  ={}&(-1)^{j_1+j_3+j_5}\sum\limits_{\substack{w=j_1+j_3+j_5+2\\d=j_5+1;~n_1\geq j_1+2}}\zeta(\vec{n})\ .
\end{align}
Here the $n=3$ version of identity \req{5id} is used in the first and \req{5ind1} in the second step.
The MZV representations for $v(j_1,j_2,j_3,0,0)$ and $v(0,j_2,j_3,0,j_5)$ include the weightings $\omega_3'$ and $\omega_{2a}'$, respectively, which involve multi--index sums. These cases can be found in eqs. \req{a123} and \req{a235}. The symmetry \req{6sym1} allows to straightforwardly determine $v(0,j_2,j_3,j_4,0)$, $v(0,0,j_3,j_4,j_5)$, $v(j_1,0,j_3,j_4,0)$ and $v(0,j_2,0,j_4,j_5)$ using the results in \req{a123}, \req{5sub1}, \req{6124} and \req{5sub4}, respectively.

\paragraph[One $j_i$ set to zero:]{One $\bm{j_i}$ set to zero:}
Setting $j_3=0$ in \req{5op} yields:
\begin{align}
 &v(j_1,j_2,0,j_4,j_5)\nonumber\\
 ={}&(-1)^{j_1+j_2+j_4+j_5}\sum\limits_{\delta_1,\delta_2,\beta_2,\beta_4,\delta_5,\epsilon_1,\epsilon_3}
 (-1)^{\delta_2+\beta_2+\beta_4+\delta_5+\epsilon_1+\epsilon_3}I(0)\{I(0)^{j_4-\delta_5-\beta_2-\beta_4-\epsilon_1-\epsilon_3},
 \nonumber\\
 &I(0)^{j_1-\delta_1-\delta_2-\beta_2},I(0,0)^{\beta_2},I(0)^{j_2-\delta_1-\beta_4},I(0,0)^{\beta_4},
 I(1)^{j_5-\delta_2-\delta_5},I(1,0)^{\delta_5},I(1,0)^{\delta_1-\epsilon_1},\nonumber\\
  &I(1,0,0)^{\epsilon_1},\left.I(1,0)^{\delta_2-\epsilon_3},I(1,0,0)^{\epsilon_3}\}I(1)\right|_{z=1}\nonumber\\
 ={}&(-1)^{j_1+j_2+j_4+j_5}\sum\limits_{\delta_1,\delta_2}(-1)^{\delta_1+\delta_2}
 \binom{j_1+j_2-\delta_1-\delta_2}{j_1-\delta_2}\binom{\delta_2}{\delta_1}\nonumber\\
 \times&\;I(0)^{j_4+1}
 \{I(0)^{j_1+j_2-\delta_1-\delta_2},I(1)^{j_5+\delta_1-\delta_2},\left.I(1,0)^{\delta_2}\}I(1)\right|_{z=1}\nonumber\\ 
 ={}&(-1)^{j_1+j_2+j_4+j_5}\sum\limits_{\delta_1=0}^{\min\{j_1,j_2\}}\sum\limits_{\substack{
 w=j_1+j_2+j_4+j_5+2\\d=j_5+\delta_1+1;~n_1\geq j_4+2}}\zeta(\vec{n})\;\omega(j_1{+}1,j_2{+}1,d{-}d_1,\delta_1{+}1)\nonumber\\
 ={}&(-1)^{j_1+j_2+j_4+j_5}\sum\limits_{\substack{
 w=j_1+j_2+j_4+j_5+2;~n_1\geq j_4+2\\j_5< d\leq j_5+1+\min\{j_1,j_2\}}}\zeta(\vec{n})\;\omega(j_1{+}1,j_2{+}1,d{-}d_1,d{-}j_5)\ .\label{61245}
\end{align}
The $n=5$ version of \req{5id} is applied in the first step and identity \req{5dep1} in step two.
The weighting $\omega$ is defined in \req{5om0}. The weightings in the MZV representations of the limits $j_5=0$ and $j_4=0$ involve multiple index sums. They are given in eqs. \req{a1234} and \req{a1235}. The limits $j_1=0$ and $j_2=0$ follow from eqs. \req{a1235} and \req{61245}, respectively, through the symmetry \req{6sym1}.

\subsection{Identities for MZVs}

In subsection 5.2.2 we presented the $\alpha'$--expansion \req{4f2alt} of the 5--point integral $F_2$, in which the kinematic part is separated from the MZVs. The latter are summarized in the function $f(j_1,j_2,j_3,j_4,j_5)$. 
Combining eqs. \req{45pt2} and \req{4f2alt} allows to write $f(j_1,j_2,j_3,j_4,j_5)$ as the coefficient function of the series:
\begin{multline}\label{mzvgen}
\frac{{}_2F_1\left[
{-s_1,~s_2\atop 
1+s_1}
;1\right]{}_2F_1\left[
{-s_3,~s_4\atop 
1+s_3}
;1\right]}{(1+s_1+s_2)(1+s_3+s_4)}\;  {}_3F_2\left[
{  1+ s_1,~1+ s_4,~1+s_2+s_3-s_5\atop  
2+ s_1+ s_2,~2+ s_3+ s_4};1\right]\\=\sum\limits_{j_1,j_2,j_3,j_4,j_5\geq0}s_1^{j_1}
 s_2^{j_2}s_3^{j_3}s_4^{j_4}s_5^{j_5}f(j_1,j_2,j_3,j_4,j_5)\ .
\end{multline}
The formula for $f(j_1,j_2,j_3,j_4,j_5)$ can be found in eq. \req{4mzvpart}, which essentially is a sum of products of three
types of MZVs representing the three generalized hypergeometric functions on the l.h.s. of \req{mzvgen}. Two of the factors in
\req{4mzvpart} are given as the object $\zeta_{i_1,i_2}'$, subject to the definition \req{42f1part}, which represent MZVs of
the kind $\zeta(i_1+1,\{1\}^{i_2-1})$ stemming from the hypergeometric functions ${}_2F_1$.
These MZVs can be written in terms of single zeta values \cite{D,C}.
The MZVs, which originate from
${}_3F_2$, are contained in the third factor $v(j_1,j_2,j_3,j_4,j_5)$. This function is given in \req{43f2part} in terms of generalized operator products and in \req{5g} in the MZV representation. Combining eqs. \req{4mzvpart} and \req{5g} we can write $f(j_1,j_2,j_3,j_4,j_5)$ in terms of MZVs as well:
\begin{multline}\label{5f}
 f(j_1,j_2,j_3,j_4,j_5)=(-1)^k\sum\limits_{\vec{l}=(l_1,l_2,l_3,l_4)}\zeta_{l_1,l_2}'\zeta_{l_3,l_4}'\sum\limits_{w=k-|\vec{l}|}\zeta(\vec{n})
 \sum\limits_{\delta_1,\delta_4}(-1)^{d-1-j_5-\delta_1-\delta_4}\\
 \times\omega_6'\left(\vec{n}-\vec{2};
\begin{matrix} 
\delta_4,~\delta_1,~j_5\\ 
j_1{-}l_1{-}\delta_1,~j_2{+}j_5{-}l_2{+}\delta_4{-}d{+}1,~j_3{+}j_5{-}l_3{+}\delta_1{-}d{+}1,~j_4{-}l_4{-}\delta_4
\end{matrix}
\right)\ ,
\end{multline}
with $k=j_1+j_2+j_3+j_4+j_5+2$.

Our motive in presenting the $\alpha'$--expansion of $F_2$ in the form \req{4f2alt} was to directly extract identities for MZVs. The object $F_2(s_{13}s_{24})^{-1}$, which equals the product of generalized hypergeometric functions on the l.h.s. of eq. \req{mzvgen}, is invariant w.r.t. cyclic permutations of the kinematic invariants $s_1,s_2,s_3,s_4,s_5$. For this symmetry to hold on the r.h.s. of eq. \req{mzvgen}, the function $f(j_1,j_2,j_3,j_4,j_5)$ has to be invariant w.r.t. cyclic permutations of its arguments:
\begin{multline}\label{cysy}
f(j_1,j_2,j_3,j_4,j_5)=f(j_5,j_1,j_2,j_3,j_4)=f(j_4,j_5,j_1,j_2,j_3)\\=f(j_3,j_4,j_5,j_1,j_2)=f(j_2,j_3,j_4,j_5,j_1)\ .
\end{multline}
This is not trivially fulfilled. Instead identities for MZVs are generated. Since we know the representation \req{5f} for $f(j_1,j_2,j_3,j_4,j_5)$ in terms of MZV, these identities can now be analysed. We mentioned a similar symmetry for the 4--point function \req{44ptexp}, which generates identities for MZVs as well.

The general identities, which follow from \req{5f} through eqs. \req{cysy}, are rather complicated due to the multi--index sums appearing in $\omega_6'$. However, some more interesting families of MZV identities are included. Instead of multi--index sums, they involve known functions such as binomial coefficients and hypergeometric functions ${}_2F_1$.  These identities appear for specific limits, where some of the arguments $j_1,j_2,j_3,j_4,j_5$ are set to zero. Below we present the MZV identities related to the limits we considered for $v(j_1,j_2,j_3,j_4,j_5)$ in section 6.2. Hence we ignore the cases, which yield multi--index sums, i.e. the ones, which satisfy $(j_2 \neq0)\land(j_3\neq0)$ (cf. appendix A).
Similar to the previous section, we start with limits, where up to four arguments of $f(j_1,j_2,j_3,j_4,j_5)$ equal zero and eventually discuss cases with only one $j_i$ set to zero. This way we see, which families of MZV identities are included in more general ones. Of course all identities follow from \req{5f}. We will, however, not use eq. \req{5f} to compute the limits for $f(j_1,j_2,j_3,j_4,j_5)$, since it is not obvious how the weighting $\omega_6'$ simplifies in many cases. Instead we work with eq. \req{4mzvpart} and insert the corresponding limits for $v(j_1,j_2,j_3,j_4,j_5)$, which are computed in the previous section. With those expressions for the MZVs, that originate from ${}_3F_2$, already given, it is straightforward to combine them in eq. \req{4mzvpart} with the contributions of the hypergeometric functions ${}_2F_1$. This step is particular trivial in case the condition $[(j_1=0)\lor(j_2=0)]\land[(j_3=0)\lor(j_4=0)]$ holds for $f(j_1,j_2,j_3,j_4,j_5)$.  Eq. \req{4mzvpart} then becomes $f(j_1,
j_2,j_3,j_4,j_5)=v(j_1,j_2,j_3,j_4,j_5)$. Thus, instead of giving explicit expressions for limits of $f(j_1,j_2,j_3,j_4,j_5)$, we will directly present the corresponding MZV identities. 
 
Setting $s_i=0$ on the r.h.s. of eq. \req{mzvgen} gives non--zero contributions for $j_i=0$ only. Therefore the generating function of MZV identities appearing for $j_i=0$ can be obtained by setting $s_i=0$ on the l.h.s. of \req{mzvgen}. 

Similar to the symmetry \req{6sym1} we have
\begin{align}\label{6sym2}
f(j_1,j_2,j_3,j_4,j_5)=f(j_4,j_3,j_2,j_1,j_5)\ .
\end{align}
It can be seen on the l.h.s. of eq. \req{mzvgen}, that the corresponding replacements $$(s_1,s_2,s_3,s_4,s_5)\rightarrow(s_4, s_3, s_2, s_1,s_5)$$ vary only the hypergeometric functions ${}_2F_1$. As a consequence the symmetry \req{6sym2} generates solely the MZV identity \req{4mzvid}. In contrast to that the cyclic symmetry \req{cysy} generates more interesting identities, as can be seen in the following.

\paragraph[Four $j_i$ set to zero:]{Four $\bm{j_i}$ set to zero:} There are five different limits for the simplest case:
\begin{align}\label{64}
f(j_1,0,0,0,0)=f(0,j_1,0,0,0)=f(0,0,j_1,0,0)=f(0,0,0,j_1,0)=f(0,0,0,0,j_1)\ .
\end{align}
According to eqs. \req{64e1} -- \req{64e3} the first 4 terms in \req{64} are identical, while $f(0,0,0,0,j_1)$ contains a different MZV. Hence, we obtain the relation
\begin{align}\label{6id1}
\zeta(\alpha_1+1)=\zeta(2, \{1\}^{\alpha_1-1})\ , \ \ \alpha_1\geq1 \ .
\end{align}
This is an instance of the more general relation \req{4mzvid}.

\paragraph[Three $j_i$ set to zero:]{Three $\bm{j_i}$ set to zero:} The limits with three arguments $j_i$ set to zero can be divided into two types:
\begin{align}\label{63a}
f(j_1,j_2,0,0,0)=f(0,0,j_1,j_2,0)=f(0,0,0,j_1,j_2)=f(j_2,0,0,0,j_1)
\end{align}
and
\begin{multline}\label{63b}
f(j_1,0,j_3,0,0)=f(0,j_1,0,j_3,0)=f(0,0,j_1,0,j_3)\\=f(j_3,0,0,j_1,0)=f(0,j_3,0,0,j_1)\ .
\end{multline}
We ignore $f(0,j_1,j_2,0,0)$, since it involves multi--index sums (cf. \req{a23}). While \req{63a} includes three independent relations, only one independent family of MZV identities is generated. Comparing eqs. \req{612} and \req{615} allows to write:
\begin{multline}\label{idq}
\sum\limits_{l_1,l_2\geq1}\zeta(l_1+1,\{1\}^{l_2-1})\sum\limits_{\substack{w=\alpha_1+\alpha_2-l_1-l_2\\n_i\geq2}}\zeta(\vec{n})\binom{w-2d}{\alpha_1-l_1-d}\\=\sum\limits_{\substack{w=\alpha_1+\alpha_2\\n_i\geq2}}\zeta(\vec{n})\binom{w-2d}{\alpha_1-d}-\zeta(\alpha_1+1,\{1\}^{\alpha_2-1})
\end{multline}
with $\alpha_1,\alpha_2\geq1$. Up to the relation \req{4mzvid} this identity is invariant w.r.t. $\alpha_1\leftrightarrow \alpha_2$, which allows us to give the additional restriction $\alpha_1\geq \alpha_2$ in order to generate less linear dependent identities. For examples at weight $w=5$ with $\alpha_1=3$, $\alpha_2=2$ identity \req{idq} yields
\begin{align}
2 \zeta(2) \zeta(3) = 3 \zeta(5) + \zeta(2, 3) + \zeta(3, 2) - \zeta(4, 1)
\end{align}
and at weight $w=6$ with $\alpha_1=4$, $\alpha_2=2$ we get
\begin{align}
\zeta(3)^2 + 2 \zeta(2) \zeta(4) = 4 \zeta(6) + \zeta(2, 4) + \zeta(3, 3) + \zeta(4, 2) - \zeta(5, 1)\ .
\end{align}
The second type of limits \req{63b} yields another class of identities. From eqs. \req{613} and \req{5sub0} follows the well known
sum theorem \cite{Granville} :
\begin{align}\label{6sth}
\zeta(\alpha_1)=\sum\limits_{\substack{w=\alpha_1\\d=\alpha_2}}\zeta({\vec{n}})\ , \ \ \ \ \alpha_1>\alpha_2\geq 1 \ .
\end{align}
At weights $w=5$ and $w=6$ it includes the following identities:
\begin{align}
\begin{split}\label{6ex2}
\zeta(5)&=\zeta(4,1)+\zeta(3,2)+\zeta(2,3)\\
&=\zeta(3,1,1)+\zeta(2,2,1)+\zeta(2,1,2)\\
&=\zeta(2,1,1,1) \ ,\\
\zeta(6)&=\zeta(5,1)+\zeta(4,2)+\zeta(3,3)+\zeta(2,4)\\
&=\zeta(4,1,1)+\zeta(3,2,1)+\zeta(3,1,2)+\zeta(2,2,2)+\zeta(2,3,1)+\zeta(2,1,3)\\
&=\zeta(3,1,1,1)+\zeta(2,2,1,1)+\zeta(2,1,2,1)+\zeta(2,1,1,2)\\
&=\zeta(2,1,1,1,1) \ .
\end{split}
\end{align}
The sum theorem is a particular beautiful identity, because it can be described in one sentence: The sum of all MZVs of given weight $w$ and depth $d$ is independent of $d$. 

\paragraph[Two $j_i$ set to zero:]{Two $\bm{j_i}$ set to zero:} There are two classes of limits:
\begin{align}\label{62a}
f(0,0,j_3,j_4,j_5)=f(j_5,0,0,j_3,j_4)=f(j_4,j_5,0,0,j_3)
\end{align}
and
\begin{align}\label{62b}
f(j_1,j_2,0,j_4,0)=f(j_4,0,j_1,j_2,0)=f(0,j_4,0,j_1,j_2)=f(j_2,0,j_4,0,j_1)\ .
\end{align}
We obtain one interesting relation via \req{62a} with eqs.
\req{5sub3} and \req{5sub1}:
\begin{multline}
\sum\limits_{l_1,l_2\geq1}\zeta(l_1+1,\{1\}^{l_2-1})\sum\limits_{\substack{w=\alpha_1+\alpha_2+\alpha_3-l_1-l_2\\ 
\alpha_3< d\leq \alpha_3+\min\{\alpha_1{-}l_1,\alpha_2{-}l_2\}}}
\zeta(\vec{n})\;\omega(\alpha_1{-}l_1,\alpha_2{-}l_2,d{-}d_1,d{-}\alpha_3)\\
=\sum\limits_{\substack{w=\alpha_1+\alpha_2+\alpha_3\\ \alpha_3< d\leq \alpha_3+\min\{\alpha_1,\alpha_2\}}}
\zeta(\vec{n})\;\omega(\alpha_1,\alpha_2,d{-}d_1,d{-}\alpha_3)-\zeta(\alpha_2+\alpha_3+1,\{1\}^{\alpha_1-1})\ ,
\end{multline}
with $\alpha_1,\alpha_2\geq 1$, $\alpha_3\geq0$. This identity contains a hypergeometric function ${}_2F_1$ through the function $\omega$, given in \req{5om0}. Examples are
\begin{align}
\begin{split}
\zeta(2) \zeta(2, 1)&=\zeta(2, 3) + \zeta(3, 2) + \zeta(4, 1) + \zeta(2, 1, 2) + \zeta(2, 2, 1)\ ,\\
\zeta(3) \zeta(2, 1) + \zeta(2) \zeta(3, 1) &= \zeta(2, 4) + \zeta(3, 3) + \zeta(4, 2) + 3 \zeta(5, 1) + \zeta(2,1,3)\\ &+ \zeta(2, 3, 1) + \zeta(3, 1, 2) + \zeta(3, 2, 1) - \zeta(4, 1, 1)\ ,
\end{split}
\end{align}
for $(\alpha_1,\alpha_2,\alpha_3)=(2,2,1)$ and $(\alpha_1,\alpha_2,\alpha_3)=(3,2,1)$, respectively. From the results \req{6124}, \req{5sub4} and eqs. \req{62b} follows the family of MVZ identities:
\begin{multline}\label{6mzvid2}
\sum\limits_{l_1,l_2\geq1}\zeta(l_1+1,\{1\}^{l_2-1})\sum\limits_{\substack{w=\alpha_1+\alpha_2+\alpha_3-l_1-l_2\\
n_1\geq \alpha_3+2;~n_i\geq2}}\zeta(\vec{n})\binom{w-\alpha_3-2d}{\alpha_1-l_1-d}\\
=\sum\limits_{\substack{w=\alpha_1+\alpha_2+\alpha_3\\
n_1\geq \alpha_3+2;~n_i\geq2}}\zeta(\vec{n})\binom{w-\alpha_3-2d}{\alpha_1-d}-\sum\limits_{\substack{w=\alpha_1+\alpha_2+\alpha_3\\ n_1>\alpha_2;~d=\alpha_1}}
\zeta(\vec{n})\ ,
\end{multline}
with $\alpha_1,\alpha_2\geq1$, $\alpha_3\geq0$. Two examples with $(\alpha_1,\alpha_2,\alpha_3)=(2,2,1)$ and $(\alpha_1,\alpha_2,\alpha_3)=(2,2,1)$, respectively, are
\begin{align}\begin{split}
\zeta(2) \zeta(3)&=2 \zeta(5) -\zeta(4, 1)\ ,\\
\zeta(2) \zeta(4) + \zeta(3) \zeta(2, 1)&= 3 \zeta(6) + \zeta(3, 3) - \zeta(5, 1) \ .
\end{split}\end{align}
The transformation $\alpha_1\leftrightarrow \alpha_2$ in \req{6mzvid2} changes only the second sum on the r.h.s. Thus we can write:
\begin{align}\label{gensth}
  \sum\limits_{\substack{w=\alpha_3\\d=\alpha_1;~n_1>\alpha_2}}\zeta(\vec{n})
=\sum\limits_{\substack{w=\alpha_3\\d=\alpha_2;~n_1>\alpha_1}}\zeta(\vec{n})\ ,\ \ \  \alpha_1>\alpha_2\geq1\ , \ \ \alpha_3\geq \alpha_1+\alpha_2\ .
\end{align}
This is an interesting generalization of the sum theorem \req{6sth}, which arises for $\alpha_2=1$. For MZVs of weights $w=5$ and $w=6$ additionally to eqs. \req{6ex2} the independent relations 
\begin{align}
 \begin{split}
  \zeta(4,1)&=\zeta(3,1,1)\ ,\\
  \zeta(5,1)&=\zeta(3,1,1,1)\ ,\\
 \zeta(5,1)+\zeta(4,2)&=\zeta(4,1,1)+\zeta(3,2,1)+\zeta(3,1,2)\ ,
 \end{split}
\end{align}
are generated. The identity \req{4mzvid} arises from \req{gensth} for $\alpha_3=\alpha_1+\alpha_2$. All relations following from \req{gensth} are contained in \req{6mzvid2}, when using the regions for the parameters $\alpha_1,\alpha_2,\alpha_3$ given below that identity. 
Alternatively we could use the additional condition $\alpha_1\geq \alpha_2$ in \req{6mzvid2}
and generate the remaining identities with \req{gensth}. 

\paragraph[One $j_i$ set to zero:]{One $\bm{j_i}$ set to zero:} Finally we discuss the identities generated through
\begin{align}
f(j_1,j_2,0,j_4,j_5)=f(j_2,0,j_4,j_5,j_1)\ .
\end{align} 
Using the symmetry \req{6sym2} and eq. \req{61245} we obtain
\begin{multline}
\sum\limits_{l_1,l_2\geq1}\zeta(l_1+1,\{1\}^{l_2-1})\sum\limits_{\substack{w=\alpha_1+\alpha_2+\alpha_3+\alpha_4-l_1-l_2;~ n_1\geq \alpha_3+2\\
 \alpha_4< d\leq \alpha_4+\min\{\alpha_1{-}l_1,\alpha_2{-}l_2\}}}
 \zeta(\vec{n})\;\omega(\alpha_1{-}l_1,\alpha_2{-}l_2,d{-}d_1,d{-}\alpha_4)\\-
 \sum\limits_{\substack{w=\alpha_1+\alpha_2+\alpha_3+\alpha_4;~ n_1\geq \alpha_3+2\\
 \alpha_4< d\leq \alpha_4+\min\{\alpha_1,\alpha_2\}}}
 \zeta(\vec{n})\;\omega(\alpha_1,\alpha_2,d{-}d_1,d{-}\alpha_4)\ ,
\end{multline}
with $\alpha_1,\alpha_2\geq1$ and $\alpha_3,\alpha_4\geq0$. This combination of MZVs appears on one side of the general identity and the other one can be obtained through the transformation 
$$(\alpha_1,\alpha_2,\alpha_3,\alpha_4)\rightarrow (\alpha_4+1,\alpha_3+1,\alpha_2-1,\alpha_1-1)\ .$$
For example the parameters $(\alpha_1,\alpha_2,\alpha_3,\alpha_4)=(3,1,1,1)$ lead to
\begin{multline}
-\zeta(5, 1) =  \zeta(2) \zeta(2, 1, 1) - \zeta(2, 1, 3) - \zeta(2, 3, 1) -  \zeta(3, 1, 2) - \zeta(3, 2, 1) - 2 \zeta(4, 1, 1)\\ -\zeta(2, 1, 1, 2) - \zeta(2, 1, 2, 1) - \zeta(2, 2, 1, 1)\ .
\end{multline}
Let us emphasize, that identities related to limits, where less arguments $j_i$ equal zero, include those with more $j_i$ set
to zero as special cases. All identities presented in this section can be generated using the single expression \req{mzvgen}.
Moreover, additional identities arise, which are not given explicitly here. 

\vskip0.5cm
\goodbreak
\centerline{\noindent{\bf Acknowledgments} }\vskip 1mm
We wish to thank Johannes Broedel and Herbert Gangl for useful discussions.
St.St. is grateful to the Mainz Institute for Theoretical Physics (MITP) for its hospitality and its partial support during the completion of this work.

\appendix
\section[Limits of $v(j_1,j_2,j_3,j_4,j_5)$]{Limits of $\bm{v(j_1,j_2,j_3,j_4,j_5)}$}

The results for limits of the function $v(j_1,j_2,j_3,j_4,j_5)$, which involve weightings with multi--index sums are listed below. Note that $(j_2\neq0) \land (j_3\neq0)$ holds for all of them. They follow directly from \req{5g} and eqs. \req{5om6limit}. However, we calculate these limits starting from \req{5op} and using proper identities of section 6.1 to check their consistency.

The case $j_1=j_4=j_5=0$ uses identity \req{5sumex3}:
\begin{align}
 &v(0,j_2,j_3,0,0)\nonumber\\
 ={}&(-1)^{j_2+j_3}\sum\limits_{\delta_3,\beta_3}(-1)^{\delta_3+\beta_3}I(0)\{I(0)^{j_2-\delta_3-\beta_3}
 ,\left.I(0)^{j_3-\delta_3-\beta_3},I(0,0)^{\beta_3},I(1,0)^{\delta_3}\}I(1)\right|_{z=1}\nonumber\\
 ={}&(-1)^{j_2+j_3} \sum\limits_{\delta_3}(-1)^{\delta_3}\sum\limits_{\substack{w=j_2+j_3+2\\d=\delta_3+1}}
 \zeta(\vec{n})\;\omega_2'(\vec{n}{-}\vec{2};j_2{-}\delta_3)\nonumber\\
 ={}&(-1)^{j_2+j_3}\sum\limits_{w=j_2+j_3+2}\zeta(\vec{n})(-1)^{d-1}\omega_2'(\vec{n}{-}\vec{2};j_2{-}d{+}1)\ .\label{a23}
\end{align}
Setting $j_4=j_5=0$ and applying identity \req{5sumex5} yields:
\begin{align}
 &v(j_1,j_2,j_3,0,0)\nonumber\\
 ={}&(-1)^{j_1+j_2+j_3}\sum\limits_{\delta_1,\delta_3}(-1)^{\delta_3}\binom{\delta_1+\delta_3}
 {\delta_3}\sum\limits_{\beta_1,\beta_3}(-1)^{\beta_1+\beta_3}I(0)\{I(0)^{j_3-\delta_3-\beta_1-\beta_3},
 I(0)^{j_1-\delta_1-\beta_1},\nonumber\\
 & I(0,0)^{\beta_1},I(0)^{j_2-\delta_1-\delta_3-\beta_3},I(0,0)^{\beta_3},
 \left.I(1,0)^{\delta_1+\delta_3}\}I(1)\right|_{z=1}\nonumber\\
 ={}&(-1)^{j_1+j_2+j_3}\sum\limits_{\delta_1,\delta_3}(-1)^{\delta_3}\binom{\delta_1+\delta_3}{\delta_3}\sum
 \limits_{\substack{w=j_1+j_2+j_3+2\\d=\delta_1+\delta_3+1}}\zeta(\vec{n})\;
 \omega_3'(\vec{n}{-}\vec{2};j_1{-}\delta_1,j_2{-}\delta_1{-}\delta_3)\nonumber\\
 ={}&(-1)^{j_1+j_2+j_3}\sum\limits_{w=j_1+j_2+j_3+2}\zeta(\vec{n})\sum\limits_{\delta_3}(-1)^{\delta_3}\binom{d-1}{\delta_3}
 \omega_3'(\vec{n}{-}\vec{2};j_1{+}\delta_3{-}d{+}1,j_2{-}d{+}1)\ .\label{a123}
\end{align}
Identity \req{5sumex4} is useful for the limit $j_1=j_4=0$:
\begin{align}
 &v(0,j_2,j_3,0,j_5)\nonumber\\
 ={}&(-1)^{j_2+j_3+j_5}\sum\limits_{\delta_3,\beta_3}(-1)^{\delta_3+\beta_3}I(0)\{I(0)^{j_2-\delta_3-\beta_3},\left.
 I(0)^{j_3-\delta_3-\beta_3},I(0,0)^{\beta_3},I(1)^{j_5},I(1,0)^{\delta_3}\}I(1)\right|_{z=1}\nonumber\\
 ={}&(-1)^{j_2+j_3+j_5}\sum\limits_{\delta_3}(-1)^{\delta_3}\sum\limits_{\substack{w=j_2+j_3+j_5+2\\d=1+j_5+\delta_3}}\zeta(\vec{n})\;
  \omega_{2a}'(\vec{n}{-}\vec{2};j_2{-}\delta_3,j_5)\nonumber\\
 ={}&(-1)^{j_2+j_3+j_5}\sum\limits_{\substack{w=j_2+j_3+j_5+2\\d\geq1+j_5}}\zeta(\vec{n})(-1)^{d-1-j_5}
 \omega_{2a}'(\vec{n}{-}\vec{2};j_2{+}j_5{+}1{-}d,j_5)\ .\label{a235}
\end{align}
For $j_5=0$ we use identity \req{5sumex8} to obtain:
\begin{align}
 &v(j_1,j_2,j_3,j_4,0)\nonumber\\
 ={}&(-1)^{j_1+j_2+j_3+j_4}\sum\limits_{\delta_1,\delta_3,\delta_4,\vec{\beta},\epsilon_1,\epsilon_2}
 (-1)^{\delta_3+|\vec{\beta}|+\epsilon_1+\epsilon_2}I(0)\{I(0)^{j_1-\delta_1-\beta_1-\beta_2-\epsilon_2},
 I(0)^{j_2-\delta_1-\delta_3-\beta_3-\beta_4},\nonumber\\
 &I(0)^{j_3-\delta_3-\delta_4-\beta_1-\beta_3},I(0)^{j_4-\delta_4-\beta_2-\beta_4-\epsilon_1},I(0,0)^{\beta_1},
I(0,0)^{\beta_2},I(0,0)^{\beta_3},I(0,0)^{\beta_4},\nonumber\\
&I(1,0)^{\delta_4-\epsilon_2},I(1,0,0)^{\epsilon_2},I(1,0)^{\delta_1-\epsilon_1},\left.I(1,0,0)^{\epsilon_1},
I(1,0)^{\delta_3}\}I(1)\right|_{z=1}\nonumber\\
={}&(-1)^{j_1+j_2+j_3+j_4}\sum\limits_{\delta_1,\delta_3,\delta_4}(-1)^{\delta_3}
\sum\limits_{\substack{w=j_1+j_2+j_3+j_4+2\\ d=\delta_1+\delta_3+\delta_4+1}}
 \zeta(\vec{n})\nonumber\\
 \times&\;\omega_5'\left(\vec{n}-\vec{2};
\begin{matrix} 
\delta_4,~\delta_1\\ 
j_1{-}\delta_1,~j_2{-}\delta_1{-}\delta_3,~j_3{-}\delta_3{-}\delta_4~,j_4{-}\delta_4
\end{matrix}
\right)\nonumber\\
 ={}&(-1)^{j_1+j_2+j_3+j_4}\sum\limits_{w=j_1+j_2+j_3+j_4+2}\zeta(\vec{n})\sum\limits_{\delta_1,\delta_4}
 (-1)^{d-1+\delta_1+\delta_4}\nonumber\\
 \times&\;\omega_5'\left(\vec{n}-\vec{2};
\begin{matrix} 
\delta_4,~\delta_1\\ 
j_1{-}\delta_1,{~}j_2{+}\delta_4{-}d{+}1,~j_3{+}\delta_1{-}d{+}1,~j_4{-}\delta_4
\end{matrix}
\right)\ .\label{a1234}
\end{align}
Identity \req{5sumex7} allows to determine the MZV representation for the limit $j_4=0$:
\begin{align}
&v(j_1,j_2,j_3,0,j_5)\nonumber\\
={}&(-1)^{j_1+j_2+j_3+j_5}\sum\limits_{\delta_1,\delta_3}(-1)^{\delta_3}\binom{\delta_1+\delta_3}{\delta_1}
\sum\limits_{\beta_1,\beta_3,\delta_2}(-1)^{\beta_1+\beta_3+\delta_2}I(0)\{I(0)^{j_3-\delta_3-\beta_1-\beta_3},\nonumber\\
&I(0)^{j_1-\delta_1-\delta_2-\beta_1},I(0,0)^{\beta_1},I(0)^{j_2-\delta_1-\delta_3-\beta_3},I(0,0)^{\beta_3},
I(1)^{j_5-\delta_2},I(1,0)^{\delta_2},\nonumber\\
&\left.I(1,0)^{\delta_1+\delta_3}\}I(1)\right|_{z=1}\nonumber\\
={}&(-1)^{j_1+j_2+j_3+j_5}\sum\limits_{\delta_1,\delta_3}(-1)^{\delta_3}\binom{\delta_1+\delta_3}{\delta_1}
\sum\limits_{\substack{w=j_1+j_2+j_3+j_5+2\\d=j_5+\delta_1+\delta_3+1}}\zeta(\vec{n})\nonumber\\
\times&\;\omega_4'\left(\vec{n}-\vec{2};
\begin{matrix} 
j_5\\ 
j_1{-}\delta_1,~j_2{-}\delta_1{-}\delta_3
\end{matrix}
\right)\nonumber\\
={}&(-1)^{j_1+j_2+j_3+j_5}\sum\limits_{w=j_1+j_2+j_3+j_5+2}\zeta(\vec{n})
\sum\limits_{\delta_1}(-1)^{\delta_1+d-1-j_5}\binom{d-1-j_5}{\delta_1}\nonumber\\
\times&\;\omega_4'\left(\vec{n}-\vec{2};
\begin{matrix} 
j_5\\ 
j_1{-}\delta_1,~j_2{+}j_5{-}d{+}1
\end{matrix}
\right)\ .\label{a1235}
\end{align}


\begin{thebibliography}{10}


\bibitem{Kotikov:1991pm} 
  A.V.~Kotikov,
``Differential equations method: New technique for massive Feynman diagrams calculation,''
  Phys.\ Lett.\ B {\bf 254}, 158 (1991);
 ``Differential equation method: The Calculation of N point Feynman diagrams,''
  Phys.\ Lett.\ B {\bf 267}, 123 (1991);\\
  E.~Remiddi,
``Differential equations for Feynman graph amplitudes,''
  Nuovo Cim.\ A {\bf 110}, 1435 (1997)
  [hep-th/9711188].

\bibitem{Knizhnik:1984nr} 
  V.G.~Knizhnik and A.B.~Zamolodchikov,
``Current Algebra and Wess-Zumino Model in Two-Dimensions,''
  Nucl.\ Phys.\ B {\bf 247}, 83 (1984).

\bibitem{Bernard:1987df} 
  D.~Bernard,
``On the Wess-Zumino-Witten Models on the Torus,''
  Nucl.\ Phys.\ B {\bf 303}, 77 (1988);
``On the Wess-Zumino-Witten Models on Riemann Surfaces,''
  Nucl.\ Phys.\ B {\bf 309}, 145 (1988).


\bibitem{Motivic} 
  O.~Schlotterer and S.~Stieberger,
 ``Motivic Multiple Zeta Values and Superstring Amplitudes,''
  J.\ Phys.\ A {\bf 46}, 475401 (2013)
  [arXiv:1205.1516 [hep-th]].


\bibitem{Broedel:2014vla} 
  J.~Broedel, C.R.~Mafra, N.~Matthes and O.~Schlotterer,
``Elliptic multiple zeta values and one-loop superstring amplitudes,''
  arXiv:1412.5535 [hep-th].




\bibitem{Oprisa:2005wu} 
  D.~Oprisa and S.~Stieberger,
``Six gluon open superstring disk amplitude, multiple hypergeometric series and Euler-Zagier sums,''
  hep-th/0509042.
  
\bibitem{Smirnov}
V.A. Smirnov, 
``Evaluating Feynman Integrals,''
Springer Berlin Heidelberg, November 2010. 
  



\bibitem{Rumaenien}
M.A.  Jivulescu, A. Napoli, A. Messina,
``General solution of a second order non-homogenous linear difference equation with 
noncommutative coefficients,''
Applied Mathematics \& Information Sciences {\bf 4} (2010), 1-–14.


\bibitem{Slater} 
L.J. Slater,  ``Generalized hypergeometric functions,''  
 Cambridge University Press (2008).









\bibitem{Takayama} 
N. Takayama, ``Gr\"obner basis and the problem of contiguous relations,'' 
Japan J. Appl. Math. {\bf 6} (1989) 147.






\bibitem{Kalmykov:2006hu} 
  M.Y.~Kalmykov, B.F.~L.~Ward and S.~Yost,
``All order epsilon-expansion of Gauss hypergeometric functions with integer and half/integer values of parameters,''
  JHEP {\bf 0702}, 040 (2007)
  [hep-th/0612240].











\bibitem{Kalmykov} 
  M.Y.~Kalmykov, B.F.L.~Ward and S.A.~Yost,
``On the all-order epsilon-expansion of generalized hypergeometric functions with integer values of parameter,''
  JHEP {\bf 0711}, 009 (2007)
  [arXiv:0708.0803 [hep-th]].


\bibitem{Boels} 
  R.H.~Boels,
``On the field theory expansion of superstring five point amplitudes,''
  Nucl.\ Phys.\ B {\bf 876}, 215 (2013)
  [arXiv:1304.7918 [hep-th]].







\bibitem{Granville} A. Granville, 
``A decomposition of Riemanns zeta-function,'' 
London Mathematical Society Lecture Notes (Proceedings of the Kyoto Conference) 
vol 247 (1997) pages 95-101.


\bibitem{D} 
  J.M.~Borwein, D.M.~Bradley, D.J.~Broadhurst and P.~Lisonek,
``Special values of multiple polylogarithms,''
  Trans.\ Am.\ Math.\ Soc.\  {\bf 353}, 907 (2001)
  [math/9910045 [math-ca]].
 
\bibitem{BrownPoly}
F. Brown, 
``Single-valued multiple polylogarithms in one variable,''
C.R. Acad. Sci. Paris, Ser. I {\bf 338}, 527-532 (2004).
  
  
\bibitem{LeMurakami} T.Q.T. Le and J. Murakami,
``Kontsevich's integral for the Kauffman polynomial,''
Nagoya Math. J. {\bf 142} (1996), 39--65.


\bibitem{Furusho}
H. Furusho, 
``The multiple zeta value algebra and the stable derivation algebra,'' 
Publ. Res. Inst. Math. Sci. {\bf 39} no 4. (2003). 695-720.

 
  
  
\bibitem{Fleischer:1998nc}
  J.~Fleischer, A.V.~Kotikov and O.L.~Veretin,
  ``Applications of the large mass expansion,''
  Acta Phys.\ Polon.\ B {\bf 29} (1998) 2611
  [hep-ph/9808243];
``Analytic two loop results for selfenergy type and vertex type diagrams with one nonzero mass,''
  Nucl.\ Phys.\ B {\bf 547}, 343 (1999)
  [hep-ph/9808242];\\
A.V.~Kotikov, L.N.~Lipatov, A.I.~Onishchenko and V.N.~Velizhanin,
``Three loop universal anomalous dimension of the Wilson operators in N=4 SUSY Yang-Mills model,''
  Phys.\ Lett.\ B {\bf 595}, 521 (2004)
  [Erratum-ibid.\ B {\bf 632}, 754 (2006)]
  [hep-th/0404092].
  
 
 \bibitem{Mafra:2011nw} 
  C.R.~Mafra, O.~Schlotterer and S.~Stieberger,
``Complete N-Point Superstring Disk Amplitude II. Amplitude and Hypergeometric Function Structure,''
  Nucl.\ Phys.\ B {\bf 873}, 461 (2013)
  [arXiv:1106.2646 [hep-th]].
 
 \bibitem{Stieberger:2012rq} 
  S.~Stieberger and T.R.~Taylor,
``Maximally Helicity Violating Disk Amplitudes, Twistors and Transcendental Integrals,''
  Phys.\ Lett.\ B {\bf 716}, 236 (2012)
  [arXiv:1204.3848 [hep-th]].




\bibitem{Goncharov} A.B. Goncharov, 
``Multiple $\zeta$-values, hyperlogarithms and mixed Tate motives,''
preprint, 1993. 


\bibitem{goncharov}
A.B. Goncharov,
``Multiple polylogarithms and mixed Tate motives,''
	[arXiv:math/ 0103059v4 [math.AG]];
``Multiple polylogarithms, cyclotomy and modular complexes,''
Math. Res. Letters 5, (1998) 497-516  [arXiv:1105.2076v1 [math.AG]]

  
\bibitem{BrownFuchs}
F. Brown, 
``Single-valued hyperlogarithms and unipotent differential equations,''  preprint.
  
\bibitem{GPSS} G.~Puhlf\"urst and S.~Stieberger,
``A Feynman Integral and its Recurrences and Associators,''
  arXiv:1511.03630 [hep-th].
    
    \bibitem{Henn:2014qga} 
  J.M.~Henn,
``Lectures on differential equations for Feynman integrals,''
  J.\ Phys.\ A {\bf 48}, no. 15, 153001 (2015)
  [arXiv:1412.2296 [hep-ph]].
  
\bibitem{Lust:2008qc} 
  D.~L\"ust, S.~Stieberger and T.R.~Taylor,
 ``The LHC String Hunter's Companion,''
  Nucl.\ Phys.\ B {\bf 808}, 1 (2009)
  [arXiv:0807.3333 [hep-th]].



\bibitem{Stieberger:2006te} 
  S.~Stieberger and T.R.~Taylor,
``Multi-Gluon Scattering in Open Superstring Theory,''
  Phys.\ Rev.\ D {\bf 74}, 126007 (2006)
  [hep-th/0609175].


\bibitem{Broedel:2013tta} 
  J.~Broedel, O.~Schlotterer and S.~Stieberger,
``Polylogarithms, Multiple Zeta Values and Superstring Amplitudes,''
  Fortsch.\ Phys.\  {\bf 61}, 812 (2013)
  [arXiv:1304.7267 [hep-th]].
  


\bibitem{Broedel:2013aza} 
  J.~Broedel, O.~Schlotterer, S.~Stieberger and T.~Terasoma,
 ``All order $\alpha^{\prime}$-expansion of superstring trees from the Drinfeld associator,''
  Phys.\ Rev.\ D {\bf 89}, no. 6, 066014 (2014)
  [arXiv:1304.7304 [hep-th]].


\bibitem{Barreiro:2013dpa} 
  L.A.~Barreiro and R.~Medina,
``RNS derivation of N-point disk amplitudes from the revisited S-matrix approach,''
  Nucl.\ Phys.\ B {\bf 886}, 870 (2014)
  [arXiv:1310.5942 [hep-th]].





\bibitem{Green:1981xx} 
  M.B.~Green and J.H.~Schwarz,
``Supersymmetrical Dual String Theory. 2. Vertices and Trees,''
  Nucl.\ Phys.\ B {\bf 198}, 252 (1982);\\
J.H.~Schwarz,
``Superstring Theory,''
  Phys.\ Rept.\  {\bf 89}, 223 (1982);\\
A.A.~Tseytlin,
``Vector Field Effective Action in the Open Superstring Theory,''
  Nucl.\ Phys.\ B {\bf 276}, 391 (1986).


\bibitem{Barreiro:2005hv} 
L.A.~Barreiro and R.~Medina,
``5-field terms in the open superstring effective action,''
  JHEP {\bf 0503}, 055 (2005)
  [hep-th/0503182];\\
S.~Stieberger and T.R.~Taylor,
``Amplitude for N-Gluon Superstring Scattering,''
  Phys.\ Rev.\ Lett.\  {\bf 97}, 211601 (2006)
  [hep-th/0607184].


 \bibitem{Mafra:2011nv} 
  C.R.~Mafra, O.~Schlotterer and S.~Stieberger,
``Complete N-Point Superstring Disk Amplitude I. Pure Spinor Computation,''
  Nucl.\ Phys.\ B {\bf 873}, 419 (2013)
  [arXiv:1106.2645 [hep-th]];



 \bibitem{Drummond:2013vz} 
  J.M.~Drummond and E.~Ragoucy,
``Superstring amplitudes and the associator,''
  JHEP {\bf 1308}, 135 (2013)
  [arXiv:1301.0794 [hep-th]].

\bibitem{work} Work to appear.



\bibitem{C} M.E. Hoffman,
``The Algebra of Multiple Harmonic Series,''
Journal of Algebra, {\bf  194} (1997)  477Ð-495.




\end{thebibliography}
\end{document}